\newcommand*{\QUANTUM}{}
\ifdefined\QUANTUM
\documentclass[a4paper,onecolumn,11pt]{quantumarticle}
\pdfoutput=1
\else
\documentclass[11pt]{article}
\fi

\usepackage{color, xcolor, colortbl}
\usepackage{graphicx,epstopdf}
\usepackage{amsmath}
\usepackage{amssymb}
\usepackage{amsthm}
\usepackage{bm}
\usepackage{geometry}
\usepackage{appendix}
\usepackage{multirow}
\usepackage{braket}
\usepackage[english]{babel}
\usepackage{hyperref}
\usepackage[capitalize,noabbrev]{cleveref}
\usepackage{caption}
\usepackage{enumerate}
\usepackage{enumitem}
\crefrangeformat{enumi}{#3#1#4--#5#2#6}

\usepackage{algorithm}
\usepackage{algorithmic}
\usepackage{xspace}
\usepackage[qm]{qcircuit}
\usepackage{subfigure}
\usepackage{multirow}

\usepackage{newfloat}
\usepackage{etoolbox}
\usepackage{dsfont}
\usepackage{booktabs}

\usepackage{tikz-cd}
\usepackage{stmaryrd}

\newtheorem{theorem}{Theorem}
\newtheorem{definition}[theorem]{Definition}

\newtheorem{lemma}[theorem]{Lemma}
\newtheorem{corollary}[theorem]{Corollary}

\newtheorem{problem}[theorem]{Problem}

\newcommand{\bP}{\mathds{P}}

\newcommand{\bI}{\mathbb{I}}

\newcommand{\rd}{\mathrm{d}}

\newcommand{\bvec}[1]{\mathbf{#1}}
\DeclareMathAlphabet{\mymathbb}{U}{BOONDOX-ds}{m}{n}

\newcommand{\vc}{\bvec{c}}

\newcommand{\vk}{\bvec{k}}

\newcommand{\vp}{\bvec{p}}
\newcommand{\vq}{\bvec{q}}

\renewcommand{\Re}{\mathrm{Re}}
\renewcommand{\Im}{\mathrm{Im}}
\newcommand{\I}{\mathrm{i}}

\newcommand{\mc}[1]{\mathcal{#1}}
\newcommand{\mf}[1]{\mathfrak{#1}}
\newcommand{\wt}[1]{\widetilde{#1}}

\newcommand{\abs}[1]{\left\lvert#1\right\rvert}
\newcommand{\norm}[1]{\left\lVert#1\right\rVert}

\newcommand{\Or}{\mathcal{O}}

\newcommand{\NN}{\mathbb{N}}
\newcommand{\RR}{\mathbb{R}}
\newcommand{\CC}{\mathbb{C}}

\newcommand{\bPP}{\mathbb{P}}

\newcommand{\expt}[1]{\mathds{E}\left( #1 \right)}
\newcommand{\expl}{\mathrm{exp}}

\newcommand{\myargmin}{\mathop{\mathrm{argmin}}}

\newcommand{\ceil}[1]{\left\lceil#1\right\rceil}
\newcommand{\spans}{\mathrm{span}}
\newcommand{\matrepwrt}[2]{\left[#1\right]_{#2}}

\newcommand{\fsim}{\text{FsimGate}}
\newcommand{\scp}[1]{{(#1)}}
\newcommand{\var}{\mathrm{Var}}

\usepackage[normalem]{ulem}

\newcommand{\DeptMath}{Department of Mathematics, University of California, Berkeley, California 94720 USA}
\newcommand{\Google}{Google Quantum AI, Venice, CA 90291, USA}

\begin{document}
	\title{Beyond Heisenberg Limit Quantum Metrology through Quantum Signal Processing}
	\author{Yulong Dong}
	\email[Electronic address: ]{dongyl@berkeley.edu}
	\affiliation{\DeptMath}
	\affiliation{\Google}
	\orcid{0000-0003-0577-2475}
	
	\author{Jonathan A. Gross}
	
	\affiliation{\Google}
	\orcid{}
	
	\author{Murphy Yuezhen Niu}
	\email[Electronic address: ]{murphyniu@google.com}
	\affiliation{\Google}
	\orcid{}
	
	\begin{abstract}
Leveraging quantum effects in metrology such as entanglement and coherence allows one to measure parameters with enhanced sensitivity~\cite{Lloyd2006}.
However, time-dependent noise
can disrupt such Heisenberg-limited amplification.
We propose a quantum-metrology method based on the quantum-signal-processing framework to overcome these realistic noise-induced limitations in practical quantum metrology.
Our algorithm separates the gate parameter $\varphi$~(single-qubit Z phase) that is susceptible to time-dependent error from the target gate parameter $\theta$~(swap-angle between $\ket{10}$ and $\ket{01}$ states) that is largely free of time-dependent error.
Our method achieves an accuracy of $10^{-4}$ radians in standard deviation for learning $\theta$ in superconducting-qubit experiments, outperforming existing alternative schemes by two orders of magnitude.
We also demonstrate the increased robustness in learning time-dependent gate parameters through fast Fourier transformation and sequential phase difference. We show both theoretically and numerically that there is an interesting transition of the optimal metrology variance   scaling as a function of circuit depth $d$ from the pre-asymptotic regime $d \ll 1/\theta$  to Heisenberg limit $d \to \infty$. Remarkably,  in the pre-asymptotic regime our method's estimation variance on time-sensitive parameter $\varphi$ scales faster than the asymptotic Heisenberg limit as a function of depth, $\text{Var}(\hat{\varphi})\approx 1/d^4$.
Our work is the first quantum-signal-processing algorithm that demonstrates practical application in laboratory quantum computers. 

	\end{abstract}
	
	\maketitle
	
	\makeatletter
	\xydef@\rparenthesized{\xy@@{\setboxz@h{%
				\A@=\X@c \advance\A@\R@c \B@=\Y@c \advance\B@-\D@c
				\setboxz@h{$\m@th\bracecr$}\dimen@ii=\dp\z@ \advance\A@-.5\wdz@
				\setboxz@h{$\m@th\bracec$}\dimen@=\dp\z@
				\ifdim\U@c<.5\dimen@ \U@c=.5\dimen@ \fi
				\ifdim\D@c<.5\dimen@ \advance\B@-.5\dimen@ \advance\B@\D@c \D@c=.5\dimen@ \fi
				\advance\U@c.6\p@ \advance\D@c.6\p@ \advance\B@-.6\p@
				\dimen@ii\U@c \advance\dimen@ii\D@c
				\kern\A@\raise\B@\vbox to \dimen@ii{%
					\nointerlineskip\hbox{$\m@th\braceul\smash{\kern-4pt{}_\gatesup}$}%
					\kern-.61\dimen@ \cleaders\copy\z@\vfil \kern-.61\dimen@
					\nointerlineskip\hbox{$\m@th\bracedl$}\kern\z@}}%
			\ht\z@=\z@ \dp\z@=\z@ \wd\z@=\z@ \boxz@}}
	\makeatother
	
	\def\gatesup{{}}
	
	\newpage
	\tableofcontents
	
	\section{Introduction}
	
One of the leading applications for a quantum computer is to simulate non-trivial quantum systems that are formidable to simulate classically. Quantum Signal Processing~(QSP) is a framework that allows us to treat the inherent quantum dynamics as quantum signals, and perform universal transformations on the input to realize targeted quantum dynamics. Despite being one of the leading algorithms in achieving the highest accuracy and  efficiency for simulating non-trivial quantum  systems in the fault-tolerant regime, no near-term application is known to-date with QSP. Such lack of application comes from the mismatch between the large amount of quantum noise in near-term quantum devices and the low noise tolerance of QSP algorithms. Instead of working against noise, we utilize QSP to amplify quantum gate parameters in the presence of time-dependent noise that  breaks  existing Heisenberg-limit achieiving metrology methods.
We provide to our knowledge the first near-term application of QSP that has been realized on laboratory quantum computers: realizing quantum metrology with  efficiency and accuracy beyond what's achievable with naive quantum amplification in the presence of realistic noise.
The analytic structure of QSP circuits provides us a powerful tool set to transform quantum dynamics to separate parameters with different dependence on environmental noise.
Such separation is crucial in allowing us to achieve the highest accuracy in experimentally measuring population swapping during a two-qubit gate.

The efficiency of a quantum metrology algorithm is measured by the total amount of physical resources~(in term of number of qubits and the number of measurements/physical runtime) needed to achieve a given learning accuracy.
The optimal efficiency is reached when the standard deviation of the parameter estimate scales inversely proportionally to the physical resources~\cite{Lloyd2006}, which typically corresponds to the number of applications of the gate being characterized.
Characterization protocols exhibiting this precision scaling are said to achieve the Heisenberg limit. Assuming a procedure achieving the  Heisenberg limit, the accuracy of  quantum gate characterization  in practice is bottle-necked by finite coherence time and time-dependent errors caused by low-frequency noise and other control imperfections.
The later prevents the Heisenberg-limited amplification of the targeted parameter by introducing unwanted disturbance to the measured signal over the course of the physical deployment.
For this reason, existing quantum metrology protocols are limited to an accuracy of $10^{-2}$--$10^{-3}$ radians when estimating quantum gate angles.
This falls short of many error-threshold requirements for scalable fault-tolerant quantum computation, in addition to various near-term applications.

	Currently, there are two types of characterization schemes.
	The first kind achieves the Heisenberg limit and the robustness against state preparation and measurement errors for estimating single-qubit gate parameters~\cite{kimmel}.
	This method has since been generalized to multi-qubit settings~\cite{neill_accurately_2021,arute_observation_2020}.
	This first kind of method learns the parameter with standard deviation scaling inversely proportional to the circuit depth $d$.
	For the circuit depths realizable  with near-term quantum computers, when considering drift error, this first family of methods can only achieve a standard deviation of $10^{-2}$ to $10^{-3}$ radians for gate parameters~\cite{arute_observation_2020}.  
	The second kind, which is the most widely used so far, includes randomized benchmarking, cross-entropy benchmarking and related techniques that share similar mechanisms, see \cite{PhysRevLett.106.180504,PhysRevA.77.012307,PhysRevA.85.042311,GoogleQuantumSupremacy2019}.
	On the high-level, these protocols involve running a random sequence of quantum operations and measuring the output probabilities, which can then be fit for the unknown model parameters that represent the single or two-qubit gate.  The primary issue with this technique is that the randomization in the sequence prevents the control angles from adding up coherently. As a result, this family of methods fails to achieve optimal efficiency for quantum parameter estimation.  This limits the accuracy with which parameters can be determined (typically on the order of a few degrees) given realistic resources and runtime.

We study the analytical structure of a class of quantum-metrology circuits using the theoretical framework of QSP~\cite{LowChuang2017,GilyenSuLowEtAl2019,WangDongLin2021,DongLinNiEtAl2022}.  The analysis enables us to propose a metrology tool that is robust against realistic time-dependent Z-phase errors which occur in superconducting qubit systems. As an example, we focus on learning a two-qubit gate parameter, the swap angle $\theta$ between basis states $\ket{10}$ and  $\ket{01}$, and the phase difference $\varphi$ between the same two basis states.
Our QSP metrology algorithm separates the estimation of $\theta$ from that of $\varphi$. Moreover, it offers a faster than Heisenberg limit convergence as a function of circuit depth in the learning of parameter $\varphi$ in a pre-asymptotic regime of experimental interest. Such faster convergence further improves our metrology accuracy on $\varphi$, which is directly impacted by the time-dependent Z-phase errors.
We also analyze the stability of the metrology in the presence of other experimental noise and sampling errors.
We prove that our method achieves the Cram\'{e}r-Rao lower bound in the presence of sampling errors, and achieves up to $10^{-4}$ STD accuracy in learning swap angle $\theta$ in both simulation and experimental deployments on superconducting qubits.
Furthermore, we show with  theoretical analysis and numerical simulation that there is an interesting transition of the optimal metrology variance scaling as a function of circuit depth $d$ from the pre-asymptotic regime $d \ll 1/\theta$ to Heisenberg limit $d \to \infty$.
	
	\subsection{Main results}
	In this section, we summarize the main results of our metrology algorithm. We start by defining the a metrology problem, \fsim\ calibration, followed by analytical derivation of a designed QSP circuit using \fsim s. Building upon these closed-form results, we propose a calibration method combining Fourier analysis with QSP to separate the two gate parameters of interests in their functional forms. 
This enables fast and deterministic data post-processing using only direct algebraic operations rather than iterative blackbox optimization. Furthermore, the separation of inference problems improves the robustness of the calibration method against dominantly time-dependent error on one of the gate parameters. The analysis and modeling of Monte Carlo sampling error also indicate that the calibration method achieves the fundamental quantum metrology optimality in a practical regime with experimentally affordable resources.
	
		Defining the logical basis states as $\ket{0_\ell} := \ket{01}$ and $\ket{1_\ell} := \ket{10}$, the single-excitation subspace spanned by $\mc{B}_2 := \{ \ket{0_\ell}, \ket{1_\ell} \}$ is isomorphic to the state space of a single qubit, on which the \fsim\ can represent any unitary. Gauging out the global phase, the matrix representation of a generic \fsim\ is parametrized as 
	\begin{equation}
	\label{eq:single-excite-fsim-gate}
	U_\fsim^{\mc{B}_2}(\theta, \varphi, \chi) = \left(\begin{array}{cc}
	e^{-\I\varphi}\cos\theta & -\I e^{\I\chi}\sin\theta \\
	-\I e^{-\I\chi}\sin\theta & e^{\I\varphi}\cos\theta
	\end{array}
	\right) = e^{-\I (\varphi - \chi - \pi)/2 Z} e^{\I \theta X} e^{-\I (\varphi + \chi + \pi)/2 Z}.
	\end{equation}
	Here, $X$ and $Z$ are Pauli matrices defined by the basis states $\mc{B}_2$. The parameter $\theta$ is referred to as the \textit{swap angle}, and the $\varphi$ is referred to as the \textit{single-qubit phase}. Note that in the physical basis,  $\varphi$ amounts to  the difference  of Z basis phase accumulation during the two-qubit gate between the two physical qubits~\cite{foxen2020}. The problem of \fsim\ calibration is to infer $\theta$ and $\varphi$ against realistic noise given the access to the \fsim\ and basic quantum operations. 
To measure extremely small swap angle $\theta \le 10^{-3}$, previous methods~(see Sec.~\ref{sec:prior-art}) require an implementation of deep quantum circuits in which many \fsim's are applied. Moreover, the accuracy of  $\theta$ estimation depends on the stability and accuracy of inferring $\varphi$. Consequently, prior methods are bottled-necked by time-dependent drift error in  $\varphi$. We design a quantum metrology method based on QSP to satisfy practical constraints in reaching fault-tolerant threshold level accuracy in realistic experimental metrology:  (1) the depth of the quantum circuit is as shallow as possible, (2) the accuracy of the inference does not deviate heavily in the presence of realistic quantum noise.   
The calibration of an \fsim\ boils down to the following problem.
	\begin{problem}[Calibrating \fsim]\label{prob:qspc}
		Given the access to an unknown \fsim\
		   and arbitray single-qubit quantum gates and projective measurements, the problem is to infer $\theta$ and $\varphi$ of the \fsim\ with bounded error and finite amount of measurement repetitions.
	\end{problem}
	Previous methods~\cite{kimmel,neill_accurately_2021,arute_observation_2020} based on optimal measurements~\cite{Caves1994} for achieving Heisenberg limit fall short at providing sufficient accuracy in $\theta$ when  $\theta \ll 1$. Two major factors are limiting these traditionally regarded ``optimal'' metrology schemes.
	First, the accuracy in $\theta$ depends on the amplification factor, i.e. maximum circuit depth that is limited to below $100$ given the state-of-the-art qubit coherence time.
	Second, time-dependent  unitary error in $\varphi$ is prevalent in our system of concern~\cite{google2020hartree} which invalidates basic assumptions in traditionally optimal and Heisenberg-limit-achieving metrology schemes.

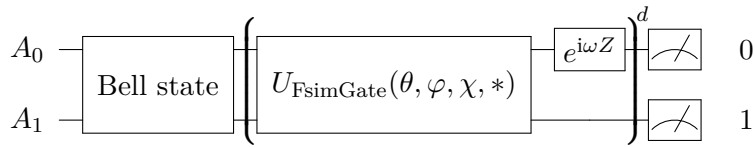
\begin{figure}[htbp]
		\begin{center}
			\[\scalebox{1}{
				\Qcircuit @C=0.8em @R=1.em {
					\lstick{A_0} & \multigate{1}{\text{Bell state}} & \multigate{1}{U_\fsim(\theta,\varphi,\chi,*)} & \gate{e^{\I\omega Z}} &\meter & \rstick{0} \\
					\lstick{A_1} &\ghost{\text{Bell state}}	& \ghost{U_\fsim(\theta,\varphi,\chi,*)} &\qw&\meter & \rstick{1} \gategroup{1}{3}{2}{4}{.7em}{(}\gdef\gatesup{d}\gategroup{1}{3}{2}{4}{.7em}{)}
			}}
			\]
		\end{center}
		\caption{Quantum Circuit for QSPC. The input quantum state is prepared to be Bell state in either $\ket{+_\ell}$ or $\ket{\I_\ell}$ according to the type of experiment. The quantum circuit enjoys a periodic structure of the unknown \fsim\ and a tunable $Z$ rotation.\label{fig:qsp-pc-circuit}}
	\end{figure}

	In this work, we provide a calibration method inferring the angles in an unknown \fsim\ when the swap angle is extremely small of order below $10^{-3}$ while facing  realistic time-dependent phase errors in $\varphi$. The calibration method leverages the structure of periodic circuits analyzed by the theory of QSP, and provides a framework to understand quantum calibration from the prospective of quantum algorithms. We call this new type of calibration method \textit{QSP calibration} (QSPC). Let $\omega$ be a tunable phase parameter, then QSPC measures the transition probability of the quantum circuit in \cref{fig:qsp-pc-circuit}. The transition probability corresponding to the Bell state $\ket{+_\ell} := \frac{1}{\sqrt{2}}\left( \ket{0_\ell} + \ket{1_\ell} \right)$ is denoted as $p_X(\omega; \theta, \varphi, \chi)$, and that corresponding to the Bell state $\ket{\I_\ell} := \frac{1}{\sqrt{2}}\left( \ket{0_\ell} + \I \ket{1_\ell} \right)$ is denoted as $p_Y(\omega; \theta, \varphi, \chi)$. 
	
	Our first main result leverages the theory of QSP to unveil the analytical structure of the periodic circuit in \cref{fig:qsp-pc-circuit}. 
	\begin{theorem}[Structure of QSPC]\label{thm:structure-of-qsp-pc}
		Let $d \in \NN$ be the number of \fsim\ applications in the QSPC circuit, and 
		\begin{equation}\label{qspc-h-eq}
		    \mf{h}(\omega; \theta, \varphi, \chi) := p_X(\omega; \theta, \varphi, \chi) - \frac{1}{2} + \I \left(p_Y(\omega; \theta, \varphi, \chi) - \frac{1}{2}\right)
		\end{equation}
		be the reconstructed function derived from the measurement probability. Then, it admits a finite Fourier series expansion
		\begin{equation}
		\mf{h}(\omega; \theta, \varphi, \chi) = \sum_{-d+1}^{d-1} c_k(\theta, \varphi, \chi) e^{2\I k \omega}.
		\end{equation}
		Furthermore, for nonnegative indices $k = 0, 1, \cdots, d-1$, the Fourier coefficients take the form
		\begin{equation}
		c_k(\theta,\chi,\varphi) = \I e^{-\I\chi} e^{-\I(2k+1)\varphi} \theta + \mathrm{max} \left\{ \Or\left(\theta^3\right), \Or\left((d\theta)^5\right) \right\}.
		\end{equation}
	\end{theorem}
	As a remark, the defined quantities using the measurement probability can be viewed as the expectation value of the logical Pauli operators. That is
	\begin{equation}
	\begin{split}
	    & \langle X_\ell \rangle(\omega; \theta, \varphi, \chi) = 2 p_X(\omega; \theta, \varphi, \chi) - 1,\quad \langle Y_\ell \rangle(\omega; \theta, \varphi, \chi) = 2 p_Y(\omega; \theta, \varphi, \chi) - 1,\\
	    &\text{and } \mf{h}(\omega; \theta, \varphi, \chi) = \frac{1}{2} \big( \langle X_\ell \rangle(\omega; \theta, \varphi, \chi) + \I \langle Y_\ell \rangle(\omega; \theta, \varphi, \chi) \big).
	\end{split}
	\end{equation}
	\cref{thm:structure-of-qsp-pc} provides the intuition behind QSPC.
	The first implication is that the number of degrees of freedom of the calibration problem is finite. The finiteness of the degree of the Fourier series implies that sampling the reconstructed function on $(2d-1)$ distinct $\omega$-points is sufficient to completely characterize its information. The second implication is that the dependencies on $\theta$ and $\varphi$ are completely factored in the amplitude and the phase of the Fourier coefficients, respectively.
	
	Let the sample points be equally spaced $\omega_j = \frac{j\pi}{2d-1}$ where $j = 0, 1, \cdots, 2d-2$. If accurate access to the reconstructed function is assumed, and the data vector is denoted as $\vec{\mf{h}} := \left(\mf{h}(\omega_0), \mf{h}(\omega_1), \cdots, \mf{h}(\omega_{2d-2})\right)^\top$, then performing Fast Fourier Transformation (FFT) of the data vector explicitly gives the Fourier coefficients $\vec{c} = \mathsf{FFT}\left(\vec{\mf{h}}\right)$. Furthermore, $\theta$ and $\varphi$ can be read from the amplitude and the phase of the Fourier coefficients respectively. Hence, by fixing the data sampling process from quantum circuits, we formally write the inference problem of QSPC as an instance of \cref{prob:qspc} as follows.
	\begin{problem}[Calibrating \fsim\ using QSPC]\label{prob:inference-Fourier}
	(1) QSPC: Given experimentally measured probabilities of QSPC circuits in \cref{fig:qsp-pc-circuit} on $\{\omega_j: j = 0, \cdots, 2d-2\}$, the problem is to infer $\theta$ and $\varphi$ accurately.
	
	(2) QSPC in Fourier space, or QSPC-F: Given experimentally measured Fourier coefficients of nonnegative indices, the problem is to infer $\theta$ and $\varphi$ accurately.
	\end{problem}
	
	We remark that the Fourier coefficients of negative indices are discarded in the modeled problem because their magnitudes are nearly vanishing (see \cref{thm:approx-coef-first-order}). Consequently, because of the almost vanishing Fourier coefficients of negative indices, QSPC-F does not loose too much information comparing with that of QSPC.
	
	The finite number of measurement samples induces the Monte Carlo sampling error to the experimentally measured probability, which is denoted as $p_{X(Y)}^\expl$ for distinction. In the presence of Monte Carlo sampling error, the experimentally measured probability is randomly distributed around the exact measured probability and the statistical fluctuation decreases when the sample size increases. An immediate implication of the characterization of the Monte Carlo sampling error is the signal-to-noise ratio (SNR) of \cref{prob:inference-Fourier}. Below we provide a lower bound on the SNR in presence of Monte Carlo sampling errors for QSPC-F.
	
	\begin{theorem}[SNR of QSPC-F]\label{thm:snr-qsp-pc}
		Let $M$ be the number of measurement samples, and $v_k$ be the additive Monte Carlo sampling error on the $k$-th Fourier coefficient, namely, $c_k^\expl = c_k(\theta, \varphi, \chi) + v_k$. When $d^5 \theta^4 \ll 1$, the SNR, defined as the lower bound on the elementwise SNR, satisfies
		\begin{equation}
		\mathrm{SNR}_k := \frac{\abs{c_k(\theta,\varphi,\chi)}^2}{\expt{\abs{v_k}^2}} \ge \mathrm{SNR} := 2(2d-1)M\sin^2\theta \left(1 - \frac{4}{3}(d\theta)^2\left(1 + 3d^3\theta^2\right)\right).
		\end{equation}
	\end{theorem}
	Remarkably, in the regime $1 \ll d \ll \theta^{-4/5}$, the SNR is approximately equal to
	\begin{equation}
	    \mathrm{SNR} \approx 4 d M \theta^2
	\end{equation}
	up to leading order. We will use this approximate SNR in the results below to capture the main scaling dependence on circuit depth $d$, sample size $M$ and gate angle $\theta$. 

To achieve the optimal inference accuracy,  we design the statistical estimators solving QSPC-F in \cref{prob:inference-Fourier}, and prove their optimality against  Monte Carlo sampling error. We define these statistical estimators in \cref{def:estimator-qsp-pc}, and derive  their performance  in \cref{prop:variance-qsp-pc-fourier}. Lastly, in \cref{subsec:CRLB-pre-asymptotic}, we prove  that our statistical estimators are optimal and attain the Cram\'{e}r-Rao lower bound of QSPC (\cref{prob:inference-Fourier}) in a practical regime $d \theta \ll 1$ in which the experimental resource is affordable.
	
	\begin{definition}[QSPC-F estimators]\label{def:estimator-qsp-pc}
		For any $k = 0, \ldots, d-2$, the sequential phase difference is defined as 
		\begin{equation}
		\Delta_k := \mathsf{phase}\left(c_k^\expl \overline{c_{k+1}^\expl}\right),\ \text{ and }\ \vec{\Delta} := \left(\Delta_0, \Delta_1, \ldots, \Delta_{d-2}\right)^\top.
		\end{equation}
		Let the all-one vector be $\vec{\mymathbb{1}} = (\underbrace{1,\ldots,1}_{d-1})^\top$ and the discrete Laplacian matrix be
		\begin{equation*}
		\mf{D} = \left(\begin{array}{rrrrr}
		2 & -1 & 0 & \cdots & 0\\
		-1 & 2 & -1 & \cdots & 0\\
		0 & -1 & 2 & \cdots & 0\\
		\vdots & \vdots & \vdots & & \vdots\\
		0 & 0 & 0 & \cdots & 2
		\end{array}\right) \in \RR^{(d-1)\times (d-1)}.
		\end{equation*}
		The statistical estimators solving QSPC-F are
		\begin{equation}\label{estimator_eq}
		\hat{\theta} = \frac{1}{d} \sum_{k=0}^{d-1} \abs{c_k^\expl} \quad \text{ and } \quad  \hat{\varphi} = \frac{1}{2} \frac{\vec{\mymathbb{1}}^\top \mf{D}^{-1} \vec{\Delta}}{\vec{\mymathbb{1}}^\top \mf{D}^{-1} \vec{\mymathbb{1}}}.
		\end{equation}
	\end{definition}
	We remark that the above estimators do not depend on unknown parameters and are fully deterministic functions of the measurements values, i.e. $\{p_X^\expl(\omega_j), p_Y^\expl(\omega_j)\}$ and the Fourier coefficients $\{c_k^\expl \}$ derived from measurement probabilities.
	Furthermore, the computation of the estimators only need direct algebraic operations. As a consequence, our calibration schemes avoid the black-box  optimization step in conventional methods to achieve Heisenberg limit~\cite{neill_accurately_2021}.
	This not only prevents the decreased performance due to the sub-optimality of the adopted solver, it also significantly speeds up the inference process.
	Moreover, once realistic quantum noise is introduced, the cost function landscape for conventional inference can be highly oscillatory~\cite{neill_accurately_2021} making global optimization ever more challenging.
	In comparison, our estimator in \cref{estimator_eq} is deterministic and offers fast and stable inference without the need of black-box optimization solver.
	Another salient feature of our estimators is the independence between the two parameters: $\theta$ depends on the amplitude of the Fourier coefficients from QSPC output, and $\varphi$ depends only on the differential phase of the Fourier coefficients from different moments. Such orthogonality provides another level of stability in estimating swap angle  $\theta$ in face of realistic time-dependent phase errors in $\varphi$. This is the first quantum metrology method to our knowledge to explicitly make such separation thanks to the powerful analytic forms given by the analysis and the theory of QSP~\cite{LowChuang2017,GilyenSuLowEtAl2019,WangDongLin2021}. We provide more comprehensive analysis of such stability in \cref{sec:realistic-error}. For completeness, we summarize the inference of QSPC-F in \cref{alg:qsp-pc}.

	\begin{algorithm}[htbp]
		\caption{Inferring unknown angles in \fsim\ with extremely small swap angle using QSPC-F estimators}
		\label{alg:qsp-pc}
		\begin{algorithmic}
			\STATE{\textbf{Input:} A \fsim\ $U_\fsim(\theta,\varphi,\chi,*)$, an integer $d$ (the number of applications of \fsim).}
			\STATE{\textbf{Output:} Estimators $\hat{\theta}, \hat{\varphi}$}
			\STATE{}
			\STATE{Initiate a complex-valued data vector $\vec{\mf{h}}^\expl \in \CC^{2d-1}$.}
			\FOR{$j = 0, 1, \cdots, 2d-2$}
			\STATE{Set the tunable $Z$-phase modulation angle as $\omega_j = \frac{j}{2d-1} \pi$.}
			\STATE{Perform the quantum circuit in \cref{fig:qsp-pc-circuit} and measure the transition probabilities $p_X^\expl(\omega_j)$ and $p_Y^\expl(\omega_j)$.}
			\STATE{Set $\vec{\mf{h}}^\expl_j \leftarrow p_X^\expl(\omega_j) - \frac{1}{2} + \I\left(p_Y^\expl(\omega_j) - \frac{1}{2}\right)$.}
			\ENDFOR
			\STATE{Compute the Fourier coefficients $\vec{c}^\expl = \mathsf{FFT}\left(\vec{\mf{h}}^\expl\right)$.}
			\STATE{Compute estimators $\hat{\theta}$ and $\hat{\varphi}$ according to \cref{def:estimator-qsp-pc}.}
		\end{algorithmic}
	\end{algorithm}
	
	The performance of the staitical estimators is measured by their unbiasness and variance. In \cref{subsec:stat-estimator-MC}, we derive the performance of QSPC-F estimators with the following theorem by treating  QSPC-F (\cref{prob:inference-Fourier}) as linear statistical models. Furthermore, in \cref{sec:lower-bound-qspc-metrology}, we show that QSPC-F estimators in \cref{def:estimator-qsp-pc} are optimal by saturating the Cram\'{e}r-Rao lower bound of the inference problem.
    \begin{theorem}[Variance of QSPC-F estimators]\label{prop:variance-qsp-pc-fourier}
    	If the higher order remainders are neglected, in the regime $d \ll 1/\theta$, QSPC-F estimators are unbiased and the variances are analytically given as follows
    	\begin{equation}
    		\begin{split}
    		& \mathrm{Var}\left(\hat{\theta}\right) \approx \frac{1}{8 d^2 M} \quad \text{ and } \quad \mathrm{Var}\left(\hat{\varphi}\right) \approx \frac{3}{8 d^4\theta^2 M}.
    		\end{split}
    	\end{equation}
    \end{theorem}
    
\begin{figure}[htbp]
        \centering
        \subfigure[\label{fig:preasym-hl-regime}]{
            \includegraphics[width=.65\textwidth]{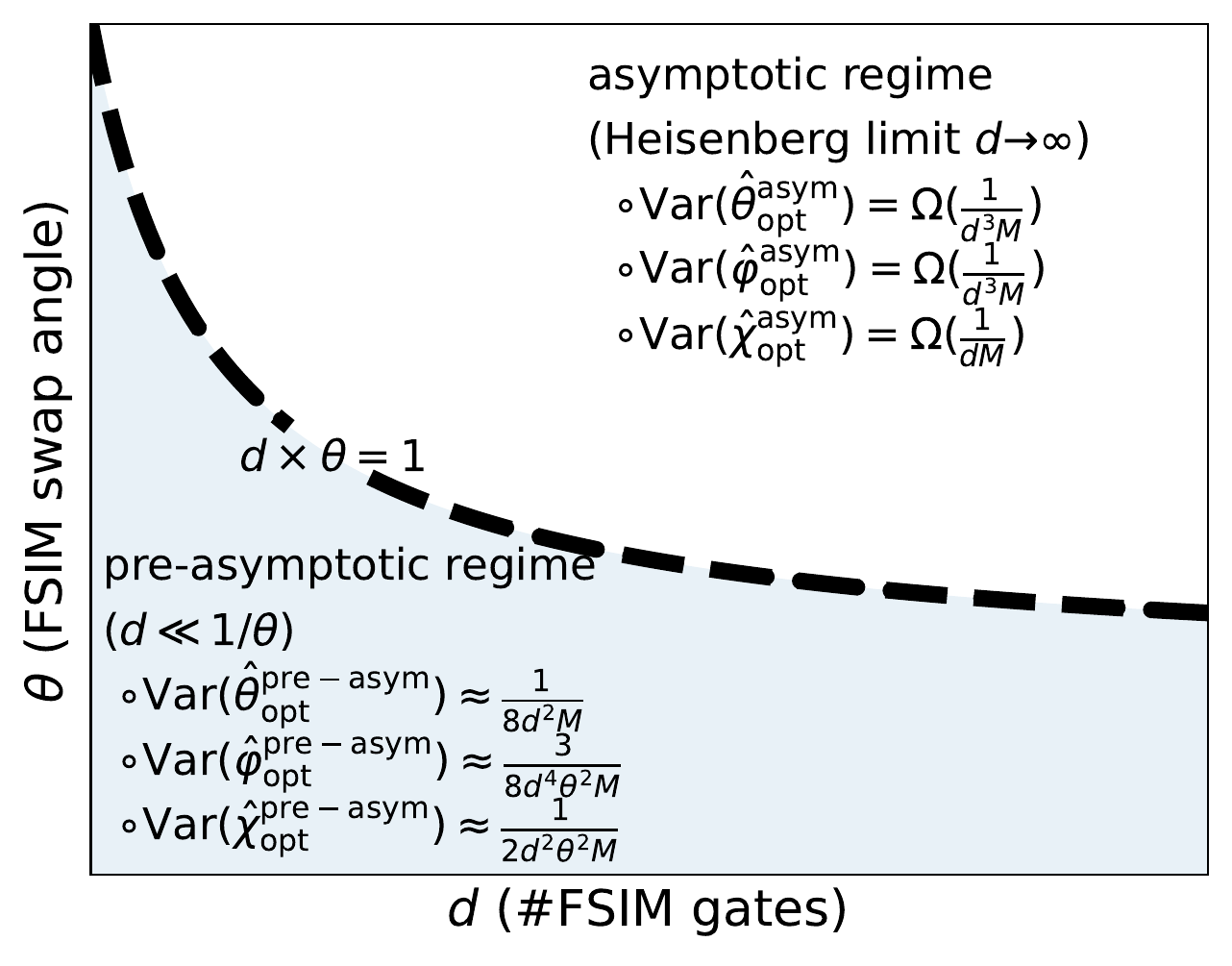}
        }
        \subfigure[\label{fig:exact_crlb-qspcf}]{
            \includegraphics[width=\textwidth]{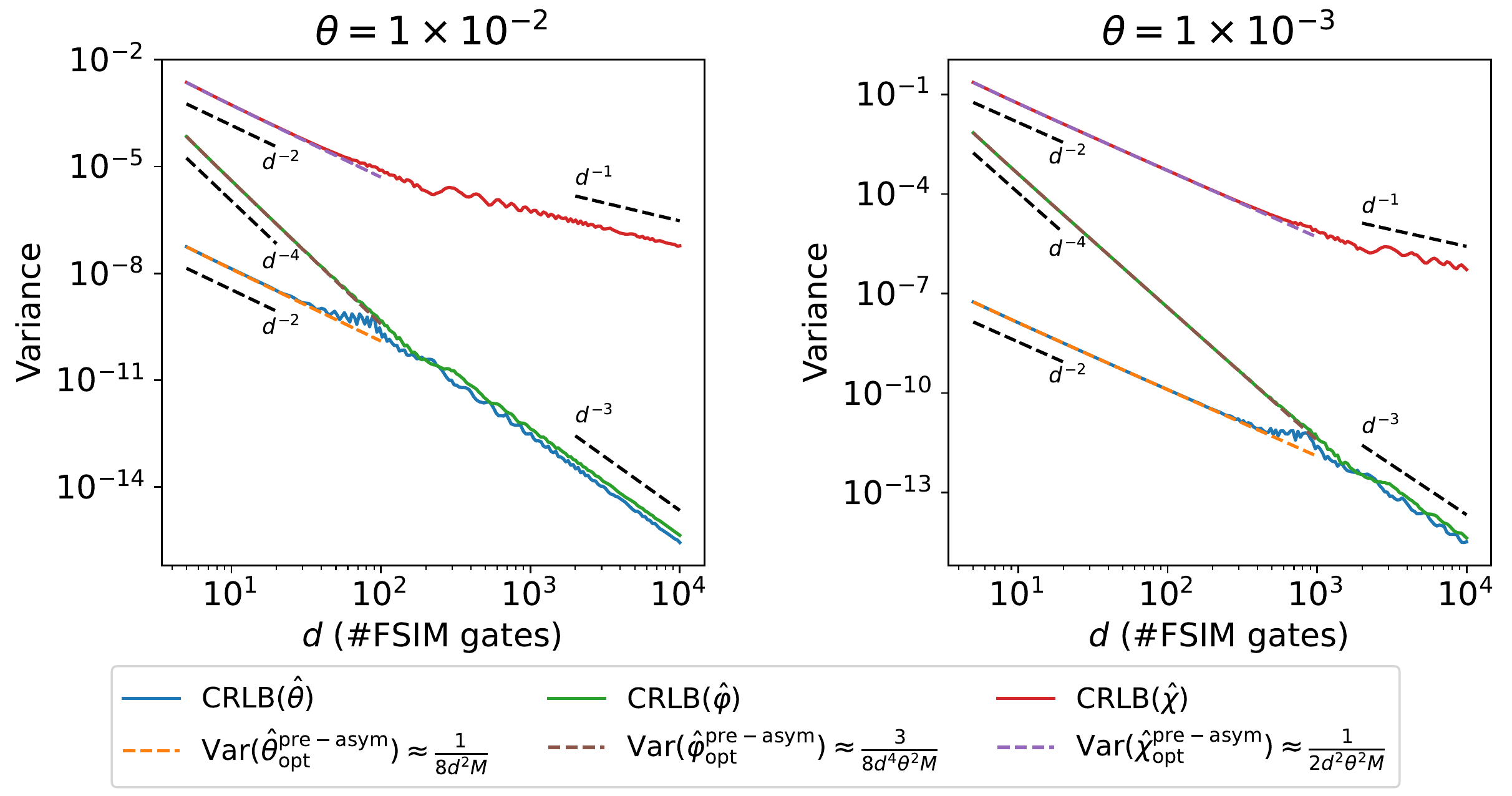}
        }
        \caption{A nontrivial transition of the optimal variance in solving QSPC. The theoretical analysis of the transition is in \cref{sec:lower-bound-qspc-metrology}. (a) Phase diagram showing the nontrivial transition of the optimal variance in solving QSPC. The optimal variance in the pre-asymptotic regime is attained by QSPC-F estimators. (b) Cram\'{e}r-Rao lower bound (CRLB) and the approximately derived optimal variance in the pre-asymptotic regime. The single-qubit phases are set to $\varphi = \pi/16$ and $\chi = 5\pi/32$. The number of measurement samples is set to $M = 1\times10^5$.}
    \end{figure}

  According to the framework developed in Ref. \cite{Lloyd2006}, the variance of any quantum metrology is lower bounded by the Heisenberg limit. It indicates that when $d$ is large enough, Heisenberg limit expects the optimal variance scales as $1/(d^3 M)$. This seemingly contradicts \cref{prop:variance-qsp-pc-fourier}, where the variance of QSPC-F $\varphi$-estimator can achieve $1/(d^4 M)$. We remark that this counterintuitive conclusion is due to the pre-asymptotic regime $d \ll 1/\theta$. In \cref{sec:lower-bound-qspc-metrology}, we analyze the Cram\'{e}r-Rao lower bound (CRLB) of QSPC (\cref{prob:inference-Fourier}). The optimal variance which is given by CRLB is exactly solvable in the pre-asymptotic regime $d \ll 1/\theta$. The exact optimal variance exhibits some nontrivial pre-asymptotic behaviours. The optimal variance in $\varphi$-estimator scales as $1/(d^4M)$ although circuits are not entirely run coherently, and the optimal variance of $\chi$-estimator scales as $1/(d^2M)$ although $\chi$ is completely not amplified in quantum circuits. The key reason in the analysis is that measurement probabilities are very close to $1/2$ in the pre-asymptotic regime. Yet when $d$ is large enough to pass to the asymptotic regime, measurement probabilities might arbitrarily take values. Furthermore, the analysis of the CRLB suggests that the optimal variance agrees with the Heisenberg limit. This nontrivial transition of optimal variance is theoretically analyzed and numerically justified in \cref{sec:lower-bound-qspc-metrology}. We summarize this nontrivial transition of the optimal variance scaling of QSPC as a phase diagram in \cref{fig:preasym-hl-regime}. To numerically justify the transition, we  compute the exact CRLB of QSPC when $\theta = 1\times 10^{-2}$ and $\theta = 1\times 10^{-3}$. In \cref{fig:exact_crlb-qspcf}, the slope of the curve in log-log scale exhibits a clear transition before and after $d = 1/\theta$ which supports the phase diagram in \cref{fig:preasym-hl-regime}. Furthermore, the numerical CRLB agrees with our theoretical derived optimal variance in the pre-asymptotic regime.
  Detailed theoretical and numerical discussions of the transition is carried out in \cref{sec:lower-bound-qspc-metrology}.
  Such pre-asymptotic features harness the unique structure of QSP circuit: the measurement outcome~Eq.~(\ref{qspc-h-eq})  concentrates around a constant value regardless of the gate parameter values, to achieve faster convergence than what is allowed in the asymptotic regime.
    
    Exploiting the analysis in the Fourier space can also provide fruitful structure for mitigating decoherence. To illustrate, we propose a mitigation scheme for the globally depolarizing error in \cref{sec:depolarizing}. Numerical simulation shows that the scheme can accurately mitigate the depolarizing error and can drastically improve the performance of QSPC-F estimators. Furthermore, we also numerically investigate the robustness of the QSPC-F estimators against  low frequency qubit frequency-drift error~\cite{wudarski2022characterizing} based on the observation from real experiments. The numerical results in \cref{sec:additional-numerical} suggests that the QSPC-F estimators give reasonable estimations with acceptable accuracy in the presence of complex realistic error. In \cref{sec:readout}, we make an explicit resource estimation for sufficiently accurately mitigating the readout error. Consequentially, we use those techniques to deploy QSPC-F on real quantum device. The calibration results are given and discussed in \cref{sec:calibrate-experiment}.

\begin{figure}[htbp]
		\centering
		\includegraphics[width=\textwidth]{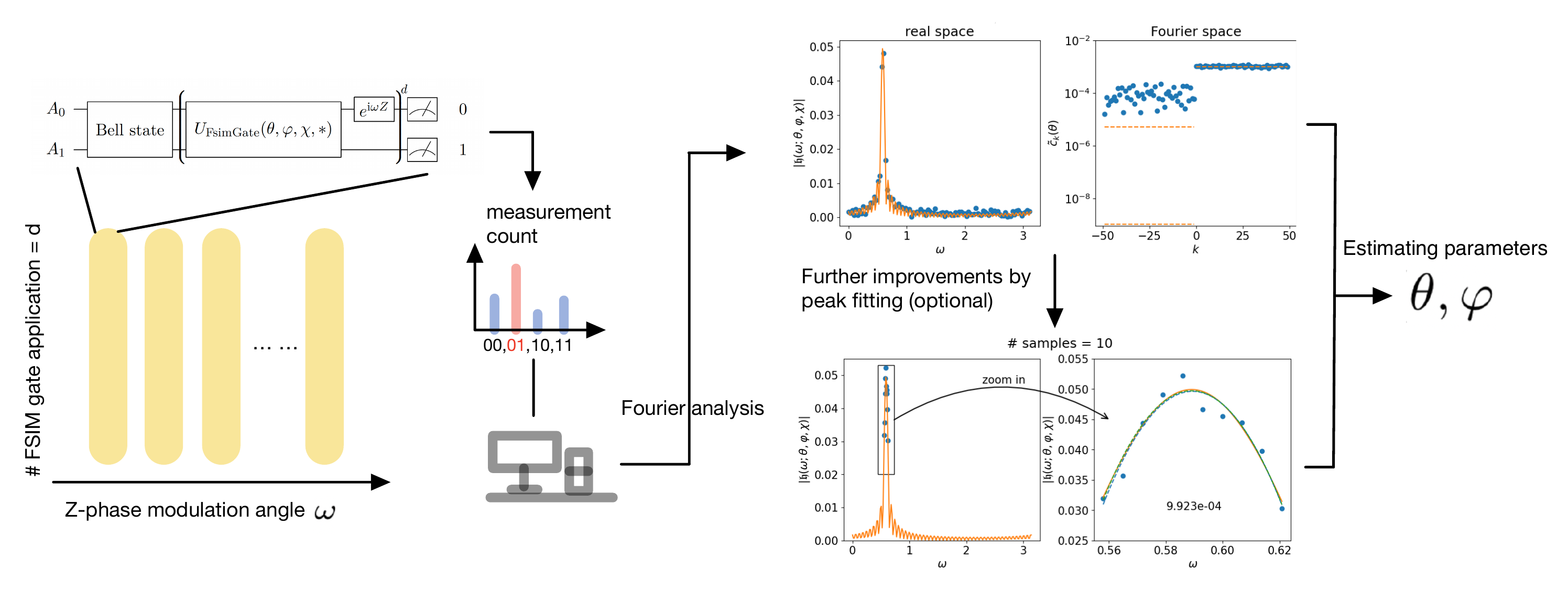}
		\caption{Flowchart of main procedures for solving QSPC.}
		\label{fig:qspc}
	\end{figure}

	\cref{prop:variance-qsp-pc-fourier} implies that QSPC-F estimator $\hat{\varphi}$ gives an accurate estimation of the phase angle $\varphi$. Furthermore, \cref{cor:modulus-magnitude-sin2theta} indicates that the amplitude $\abs{\mf{h}_d(\omega;\theta,\varphi,\chi)}$ attains maximum when phase matching condition $\omega = \varphi$ is satisfied. We explicitly write the dependence on the degree $d$ as the subscript. The analytical results derived in \cref{sec:analytical-result-qspc} indicates that the degree parameter $d$ controls the maximum height of the  amplitude function $\abs{\mf{h}_d(\omega;\theta,\varphi,\chi)}$ and the angle parameter $\omega$ determines the sampling location. With this interpretation, QSPC-F provides an algorithm using the information scanning over the polynomial of a given degree parameter $d$. Hence, it is natural to ask whether unleashing the constraint of fixed $d$ can yield other calibration methods. We plot the amplitude as a function of $\omega$ as an example in \cref{fig:qspc}. In accordance with the numerical demonstration and the analytical expression given in \cref{cor:modulus-magnitude-sin2theta}, we can see there is a sharp and dominated peak around the matched phase $\omega = \varphi$, which yields more robustness of the sampled signal against possible noise. Assuming some a priori estimator $\hat{\varphi}_\mathrm{pri}$ to the single-qubit phase $\varphi$, it is more preferable to sample near the estimated peak to boost the robustness of the samples against noise. One immediate consideration is sampling data with fixed $\omega=\hat{\varphi}_\mathrm{pri}$ but varying $d$. To analyze the signal, we might trust $\hat{\varphi}_\mathrm{pri}$ as the location of the peak. Then, unless $\hat{\varphi}_\mathrm{pri} = \varphi$ exactly holds, the estimator on $\theta$ is always biased. We quantify this effect explicitly in \cref{{thm:bias-prog-diff}}. At the cost of introducing bounded bias, the variance of the estimator to $\theta$ is improved to $3/(4d^3M)$ using additional $\Or(d^2)$ \fsim's, which gets an additional $d$ dependence in the denominator comparing with that of QSPC-F. Specifically, one might take $\hat{\varphi}$ from QSPC-F as an input of the algorithm, the performance guarantee of the induced estimator on $\theta$ is given in \cref{cor:bias-prog-diff-QSPC-F}. Remarkably, QSPC-F already uses $\Or(d^2)$ \fsim's and hence the additional improvement on the estimation on $\theta$ does not asymptotically affect the amount of gates.
	
	By leveraging the ability to sample data with variable degree $d$ and $\omega$, we can consider regressing the data on $\theta$ and other unknown angles with respect to its analytical formula. Suppose $n$ samples are made, M-estimation theory \cite{KeenerTheoreticalStatistics2010} gives that there exists an unbiased estimator on $\theta$ so that the variance scales asymptotically as $\Or\left(1/(d^2nM)\right)$. Assuming the amount of \fsim's is $\Or(d^2)$, the variance could be improved to $\Or\left(1/(d^3M)\right)$ when $d$ is large enough. This agrees with Heisenberg limit. In practice, the estimator is approximated by minimizing some cost function. The complex landscape of nonlinear minimization and the sample signal with small magnitude could largely contaminate the estimation via black-box minimization. To overcome issue on the vanishing signal, we can first perform QSPC-F to get $\hat{\theta}, \hat{\varphi}$ and then sample on the interval $\mc{I} = \left[ \hat{\varphi} - \frac{\pi}{2d}, \hat{\varphi} + \frac{\pi}{2d} \right]$. It can be shown that this interval contains the highest peak with high probability which gives relatively high magnitude of the signal against noise. Furthermore, given that QSPC-F provides a reliable estimation, $\hat{\theta}$ and $\hat{\varphi}$ are close to the true values which can be used as the initial guess of the minimization to improve the performance.
	
	The hardness of the regression around the peak also comes from the complex landscape of numerically solving the nonlinear regression problem. At the same time, the additional bias of a previously discussed improvement is because the location of the peak is over-confidently assumed to be the a priori value. To address these issues and improve the performance of estimation, we propose a heuristic algorithm called \textit{peak fitting}. The proposal follows an observation that the highest peak within half width can be well approximated by a parabola. Regressing the data with respect to a parabola instead, the problem boils down to an ordinary least square problem which can be solved directly using simple algebraic operations. Hence, the complexity in the optimization landscape is circumvented while the tradeoff is a further parabolic approximation and possible induced bias. On the other hand, the a priori $\hat{\varphi}_\mathrm{pri}$ is used to determine the sampling interval $\mc{I}$ and for post-selection.
	Trusting $\hat{\varphi}_\mathrm{pri}$ as a good estimation to the peak location $\varphi$, we accept the fitted parabola if its peak location does not deviate much from $\hat{\varphi}_\mathrm{pri}$.
	According to \cref{cor:modulus-magnitude-sin2theta}, dividing the fitted peak magnitude by $d$ yields an estimation to $\theta$. Although there is no theoretical performance guarantee of the peak fitting, a significant improvement against Monte Carlo sampling error can be found in numerical results in \cref{subsec:num-result-MC}. In \cref{alg:qspc-peak-fitting}, the algorithm of peak fitting is presented for completeness.
	
		\begin{algorithm}[htbp]
		\caption{Improving $\theta$ estimation using peak fitting}
		\label{alg:qspc-peak-fitting}
		\begin{algorithmic}
			\STATE{\textbf{Input:} A \fsim\ $U_\fsim(\theta,\varphi,\chi,*)$, an integer $d$ (the number of applications of \fsim), an integer $n$ (the number of sampled angles), a priori $\hat{\varphi}_\mathrm{pri}$ (can be generated by QSPC-F), a threshold $\beta^\mathrm{thr} \in [0, 1]$.}
			\STATE{\textbf{Output:} Estimators $\hat{\theta}_\mathrm{pf}$}
			\STATE{}
			\STATE{Initiate real-valued data vectors $\vec{\mf{p}}^\expl, \vec{\mf{w}} \in \RR^n$.}
			\FOR{$j = 0, 1, \cdots, n-1$}
			\STATE{Set the tunable $Z$-phase modulation angle as $\omega_j = \hat{\varphi}_\mathrm{pri} + \frac{\pi}{d}\left(\frac{j}{n-1}-\frac{1}{2}\right)$.}
			\STATE{Peform the quantum circuit in \cref{fig:qsp-pc-circuit} and measure the transition probabilities $p_X^\expl(\omega_j)$ and $p_Y^\expl(\omega_j)$.}
			\STATE{Set $\vec{\mf{p}}^\expl_j \leftarrow \sqrt{\left(p_X^\expl(\omega_j) - \frac{1}{2}\right)^2 + \left(p_Y^\expl(\omega_j) - \frac{1}{2}\right)^2}$ and $\vec{\mf{w}}_j \leftarrow \omega_j$.}
			\ENDFOR
			\STATE{Fit $\left(\vec{\mf{w}}, \vec{\mf{p}}^\expl\right)$ with respect to to parabolic model $\mf{p} = \beta_0\left(\mf{w} - \beta_1\right)^2 + \beta_2$.}
            \IF{$\beta_0 < 0$ (concavity) and $\abs{\beta_1 - \hat{\varphi}_\mathrm{pri}} < \beta^\mathrm{thr}$ (small deviation from a priori)}
			\STATE{Set $\hat{\theta}_\mathrm{pf} \leftarrow \beta_2 / d$. The improvement is accepted.}
			\ELSE
			\STATE{Set $\hat{\theta}_\mathrm{pf} \leftarrow \mathrm{None}$. The improvement is rejected.}
			\ENDIF
		\end{algorithmic}
	\end{algorithm}
	
	We give a flowchart in \cref{fig:qspc} which summarizes and illustrates the main procedures of QSPC.

	\subsection{Background and related works}
	Quantum computing is a promising computational resource for accelerating many problems arising from physics, material science, and scientific computing. To build an accurate quantum computer, one needs high-fidelity quantum gates. The controlled-Z gate (CZ) is widely used in quantum computing for a variety of tasks, such as demonstrating quantum supremacy~\cite{GoogleQuantumSupremacy2019}, accurately computing electronic structure properties~\cite{neill_accurately_2021}, and performing error correction~\cite{chen_exponential_2021,krinner_realizing_2022,zhao_realization_2022}. Some physical implementations of the CZ gate use pulse protocols capable of realizing a large class of excitation-preserving two-qubit quantum gates,  a.k.a., \fsim. Despite the demand for a high-fidelity gate, in practice the physical implementation of the \fsim is always noisy and the resulted implementation slightly deviates from the exact operation. In order to characterize extremely small gate angle deviation, coherent phase amplification  are used to infer the parameters of an unknown quantum gate. Because the swap angle of CZ is $\theta_\mathrm{CZ} = 0$, calibrating noisy CZ boils down to the calibration of \fsim\ with extremely small swap angle. Several standard tools for performing this characterization are Periodic/Floquet calibration and cross-entropy benchmarking (XEB) characterization, both of which we summarize in~\cref{sec:prior-art}.

	\subsection{Discussion and open questions}
Our proposed QSPC-F estimators leverages the polynomial structure of periodic circuits derived from the theory of QSP and the Fourier analysis. Consequentially, the inference of the swap angle $\theta$ is largely decoupled with that the single-qubit phase $\varphi$. When some constant phase drift is imposed to the system, the inference is not affected thanks to the robustness of discrete Fourier transform and sequential phase difference to small phase drift errors. Furthermore, the QSPC-F estimators exhibit robustness against realistic error in numerical simulations and the deployment on quantum devices. We developed an error mitigation method against globally depolarizing error using the difference in the Fourier coefficients. To  further mitigate more generic quantum errors, we have to investigate case by case different realistic noise effects on the structure of Fourier coefficients. 

We design the optimal QSPC-F estimators based on error analysis of Monte Carlo sampling error. In \cref{sec:Monte-Carlo-sampling-error}, the inference problem in the presence of Monte Carlo sampling error is reduced to linear statistical models whose optimal ordinary least square estimators give the QSPC-F estimators. Although we show through both simulation and experimental deployments that QSPC-F estimators are robust against realistic error,  the optimality of QSPC-F agasint realistic errors remains unknown. To fully optimize the design of statistical estimators, we need to model and study the behaviour and statistics of the realistic error using tools from classical statistics, Bayesian inference and statistical machine learning. Our future work will try to addrsss this important problem with a deepened understanding of a wider range of realistic errors.

An important caveat of our QSP based metrology scheme is that we picked a given set of state initialization and measurements. This specific choice defined in \cref{prob:inference-Fourier}   is an instance of a more generic setting in \cref{prob:qspc}. Although theoretical analysis and numerical simulation justify that QSPC-F estimators are optimal in the given parameter regime and the given state preparation and measurement scheme, it remains an open question whether we can derive the optimal estimators in the most genric setting in \cref{prob:qspc} by optimizing circuit structure, initialization and measurement schemes.

The QSPC-F estimators are only reliable in the pre-asymptotic regime in which $d$ is moderate so that experiments can afford the resource requirements. Such non-asymptotic performance gaurantee is tied in with our main objective of mitigating detrimental effect of time-dependent noise. As next step, one can consider the optimal estimators which is fast and efficiently derivable from experimental data in the asymptotic regime with sufficiently large $d$. The analysis of QSPC provides fruitful toolbox for designing new quantum metrology protocols that leverage and transform the unwanted quantum dynamics from environmental noise. Generalizing a deterministic estimator from our work to a variational one can offer greater flexibility and optimality, but requires deeper understanding of the  landscape inherited from the QSPC structure. This can also guide the design of MLE with fast local convergence, and hence push the optimality and robustness to the asymptotic regime.

Lastly, the structure of QSPC circuit enjoys a periodic circuit which can be viewed as a QSP circuit with fixed modulation angle $\omega$ in each layer. However, the theory of QSP allows the modulation angle of each layer being independent. Unleashing the constraint of fixed modulation angle, the structure of the polynomial becomes more complicated and can be multivariate. It remains an open question whether this generalization could help the inference in the presence of inhomogeneous phase drift error.
	
	\subsection*{Acknowledgments:}
	This work is partially supported by the NSF Quantum Leap Challenge Institute (QLCI) program through grant number OMA-2016245 (Y.D.). The authors thank discussions with Lin Lin, Vadim Smelyanskiy,   K. Birgitta Whaley, Ryan Babbush and Zhang Jiang.

	\section{Preliminaries}
	\subsection{Fermionic simulation gate (\fsim)}
	Fermionic simulation gate (\fsim) is a class of two-qubit quantum gates preserving the excitation. Acting on two qubits $A_0$ and $A_1$, the \fsim\ is parametrized by a few parameters and the quantum gate is denoted graphically as follows.
	\begin{center}
		\[\scalebox{1}{
			\Qcircuit @C=0.8em @R=1.em {
				\lstick{A_0: \ket{a_0}} & \multigate{1}{U_\fsim(\theta,\varphi,\chi,\psi,\phi)} &\qw \\
				\lstick{A_1: \ket{a_1}} &\ghost{U_\fsim(\theta,\varphi,\chi,\psi,\phi)} & \qw 
		}} 
		\]
	\end{center}
	Ordering the basis as $\mc{B} := \{ \ket{00}, \ket{01}, \ket{10}, \ket{11} \}$ where the qubits are ordered as $\ket{a_0a_1} := \ket{a_0}_{A_0}\ket{a_1}_{A_1}$, the unitary matrix representation of the \fsim\ is given by
	\begin{equation}
	U_\fsim(\theta,\varphi,\chi,\psi,\phi) = \left(\begin{array}{*{4}c}
	1 & 0 & 0 & 0 \\
	0 & e^{-\I \varphi-\I\psi}\cos\theta & -\I e^{\I \chi - \I \psi} \sin\theta & 0\\
	0 & -\I e^{-\I\chi-\I\psi}\sin\theta & e^{\I\varphi-\I\psi}\cos\theta & 0\\
	0 & 0 & 0 & e^{-\I(\phi + 2 \psi)}
	\end{array}\right).
	\end{equation}
	As a consequence of the preservation of excitation, there is a two-dimensional invariant subspace of the \fsim, which is referred to as the single-excitation subspace spanned by basis states $\mc{B}_2 = \{ \ket{01}, \ket{10} \}$. Restricted on the single-excitation subspace $\mc{E}_2 := \spans\ \mc{B}_2$, the matrix representation of the \fsim\ is (up to a global phase)
	\begin{equation}
	\begin{split}
	& \matrepwrt{U_\fsim(\theta, \varphi, \chi, \phi, \psi)}{\mc{B}_2} =: U_\fsim^{\mc{B}_2}(\theta, \varphi, \chi)\\
	&= \left(\begin{array}{cc}
	e^{-\I\varphi}\cos\theta & -\I e^{\I\chi}\sin\theta \\
	-\I e^{-\I\chi}\sin\theta & e^{\I\varphi}\cos\theta
	\end{array}
	\right) = e^{-\I \frac{\varphi - \chi - \pi}{2} Z} e^{\I \theta X} e^{-\I \frac{\varphi + \chi + \pi}{2} Z}.
	\end{split}
	\end{equation}
	Here, $X$ and $Z$ are logical Pauli operators by identifying logical quantum states $\ket{0}_\ell := \ket{01}$ and $\ket{1}_\ell := \ket{10}$. As a remark, it provides a parametrization of any general $\mathrm{SU}(2)$ matrix. 
	
	One of the most important two-qubit quantum gates is controlled-Z gate (CZ). It forms universal gate sets with several single-qubit gates and it is a pivotal building block for demonstrating surface code\cite{acharya2022suppressing}. CZ is in the gate class of \fsim 's, which can be generated by setting $\theta = \varphi = \chi = \psi = 0$ and $\phi = \pi$. Due to the noisy implementation of CZ, the resulting quantum gate is an \fsim\ slightly deviating the perfect CZ. In order to perform high-fidelity quantum computation, one has to characterize the erroneous parameters of an \fsim\ which include CZ as a special case. The characterization of gate parameters relies on quantum calibration techniques.
	
	\subsection{Prior art}
	\label{sec:prior-art}
	
	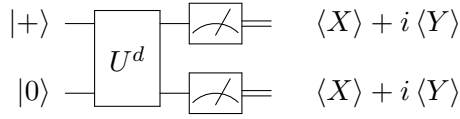
\begin{figure}
    \centerline{
    \Qcircuit @C=1em @R=1em {
        \lstick{\ket{+}}
        & \multigate{1}{U^d}
        & \meter & \cw
        & \rstick{\expval{X}+i\expval{Y}}
        \\
        \lstick{\ket{0}}
        & \ghost{U^d}
        & \meter & \cw
        & \rstick{\expval{X}+i\expval{Y}}
    }
    }
    \caption{Phase-method style of Floquet-characterization circuit.
    One of the qubits (in the diagram above, the top qubit) is prepared in the superposition state $\ket{+}$, and the other qubit is left in the ground state.
    After repeating the unitary we're characterizing a number of times, we measure either the expectation value of $X$ or $Y$ on each qubit to give us complex numbers from which the matrix elements in the single-excitation subspace can be inferred.}
    \label{fig:phase-method-circuit}
\end{figure}
	
    \fsim s have been calibrated at the Heisenberg limit using a technique called
	Periodic or Floquet calibration~\cite{neill_accurately_2021,arute_observation_2020}, which is an extension of robust phase estimation~\cite{kimmel} to multi-qubit gates.
	It leverages the excitation-preserving structure of the \fsim\ to measure the parameters using a restricted set of circuits (compared to full process tomography).
	This technique amplifies unitary errors in the gate through repeated applications between measurements, leading to variance in the estimated parameters that scales inversely with the square of the number of gate applications instead of simply scaling inversely with the number of gate applications, thus achieving the Heisenberg limit.
	An important style of Floquet calibration is called the phase method, which uses circuits of the form shown in~\cref{fig:phase-method-circuit}.

	One difficulty with these techniques is that small values of the swap angle $\theta$ are difficult to  be amplified in the presence of larger single-qubit phases.
	This can be addressed adaptively, by first measuring the unwanted single-qubit phases and applying compensating pulses, but this strategy is limited by the precision with which one can compensate, and the speed with which these single-qubit phases drift relative to the experiment time.
	For these reasons, in practice estimation of the swap angle is often done with the depth-1 circuits from the phase method, commonly referred to as unitary tomography~\cite{foxen2020}.

    An alternative characterization scheme using cross-entropy-benchmarking (XEB) circuits was described in Sec. C. 2. of the supplemental material for~\cite{GoogleQuantumSupremacy2019}.
    This characterization tool randomizes various noise sources into an effective depolarizing channel, allowing noise to be simply characterized along with unitary parameters.
    Randomization comes at a cost, though, requiring a large number of random circuits to get a representative sample of the distribution.
    Also, randomization interferes with the ability of unitary errors to build up coherently, keeping this method from achieving the Heisenberg limit.
    This makes it difficult for XEB characterization to resolve angles below $10^{-2}$ radians in practice.
	
	\subsection{Quantum signal processing (QSP)}
	The quantum circuit used in QSPC (see \cref{fig:qsp-pc-circuit}) contains a periodic circuit structure in which the \fsim\ and a Z-rotation are interleaved. This circuit structure coincides with a quantum algorithm called quantum signal processing (QSP) \cite{LowChuang2017,GilyenSuLowEtAl2019}. QSP is an useful quantum algorithm for solving numerical linear algebra problems such as quantum linear system problems and Hamiltonian simulation by properly choosing a set of phase factors~\cite{DongMengWhaleyEtAl2020,martyn2021grand}. Specifically, in this paper, we will use the polynomial structure induced by the theory of QSP \cite{LowChuang2017,GilyenSuLowEtAl2019,WangDongLin2021,DongLinNiEtAl2022}. The following theorem is a simplified version of \cite[Theorem 1]{WangDongLin2021}.
	\begin{theorem}[Polynomial structure of symmetric QSP]\label{thm:qsp}
		Let $d\in \NN$ and $\varOmega := (\omega_0, \cdots, \omega_d) \in \RR^{d+1}$ be a set of phase factors. Then, for any $x \in [-1,1]$, the following product of $\mathrm{SU}(2)$-matrices admits a representation
		\begin{equation}
		\label{eqn:qsp-gslw}
		U(x, \varOmega) = e^{\I \omega_0 Z} \prod_{j=1}^{d} \left( e^{\I \arccos(x) X} e^{\I \omega_j Z} \right) = \left( \begin{array}{cc}
		P(x) & \I Q(x) \sqrt{1 - x^2}\\
		\I Q^*(x) \sqrt{1 - x^2} & P^*(x)
		\end{array} \right)
		\end{equation}
		for some $P,Q\in \CC[x]$ satisfying that
		\begin{enumerate}[label=(\arabic*)]
			\item \label{itm:1} $\deg(P) \leq d, \deg(Q) \leq d-1$,
			\item \label{itm:2} $P(x)$ has parity $(d\mod2)$ and $Q(x)$ has parity $(d-1 \mod 2)$,
			\item  \label{itm:3} $|P(x)|^2 + (1-x^2) |Q(x)|^2 = 1, \forall x \in [-1, 1]$.
		\end{enumerate}
		Here, the superscript $*$ denotes the complex conjugate of a polynomial, namely $P^*(x) = \sum_i \overline{p_i} x^i$ if $P(x) = \sum_i p_i x^i$ with $p_i \in \CC$. Furthermore, if $\varOmega$ is chosen to be symmetric, namely $\omega_j = \omega_{d-j}$ for any $j$, then $Q \in \RR[x]$ is a real polynomial.
	\end{theorem}
	\begin{proof}
	    We will give a straightforward proof for completeness. 
	    
	    ``\textit{Condition (1)}'': Note that $\mathrm{SU}(2)$ matrices satisfy
	    \begin{equation*}
	        e^{\I \arccos(x) X} = x I + \I \sqrt{1-x^2} X \quad \mathrm{and} \quad X e^{\I \omega Z} = e^{-\I \omega Z} X.
	    \end{equation*}
	    The polynomial representation in \cref{eqn:qsp-gslw} follows the expansion and rearranging Pauli $X$ matrices. The condition (1) follows the observation that the leading term is at most $x^d$ when there are even number of Pauli $X$ matrices in the expansion while it is at most $x^{d-1}$ when the number of Pauli $X$ matrices is odd. 
	    
	    ``\textit{Condition (2)}'': To see condition (2), we note that under the transformation $x \mapsto -x$, we have
	    \begin{equation*}
	        e^{\I \arccos(-x) X} = e^{\left(\pi - \arccos(x)\right)X} = - e^{-\I \arccos(x) X} = - Z e^{\I\arccos(x) X} Z.
	    \end{equation*}
	    Therefore
	    \begin{equation*}
	       U(-x,\varOmega) = (-1)^d Z U(x,\varOmega) Z = \left( \begin{array}{cc}
		(-1)^d P(x) & \I (-1)^{d-1} Q(x) \sqrt{1 - x^2}\\
		\I (-1)^{d-1} Q^*(x) \sqrt{1 - x^2} & (-1)^d P^*(x)
		\end{array} \right)
	    \end{equation*}
	    which implies that
	    \begin{equation*}
	        P(-x) = (-1)^d P(x) \quad \mathrm{and} \quad Q(-x) = (-1)^{d-1} Q(x)
	    \end{equation*}
	    which is the parity condition.
	    
	    ``\textit{Condition (3)}'': Condition (3), which is equivalent to $\det U(x,\varOmega) = 1$, directly follows the special unitarity.
	    
	    ``\textit{Symmetric QSP}'': Note that when $\Phi$ is symmetric, $U(x, \Phi)$ is invariant under the matrix transpose which reverses the order of phase factors. Using $U(x, \Phi) = U(x,\Phi)^\top$, the condition on the polynomial $Q(x) = Q^*(x)$ follows the transformation of the off-diagonal element. Therefore, $Q \in \RR[x]$ is a real polynomial.
	\end{proof}
	
	The previous theorem bridges the gap between the periodic circuits and the analysis of polynomial. In the paper, we will frequently invoke some important inequalities of polynomials, which are stated in \cref{app:poly-ineq} for completeness.

	\subsection{Notation}
	Throughout the paper, $M$ refers to the number of measurement samples unless otherwise noted. For a matrix $A\in\CC^{m\times n}$, the transpose, Hermitian conjugate and complex conjugate are denoted by $A^{\top}$, $A^{\dag}$, $\overline{A}$, respectively. The same notations are also used for the operations on a vector. The complex conjugate of a complex number $a$ is denoted as $\overline{a}$. We define the basis kets of the state space of a qubit as follows
\[
\ket{0} := \begin{pmatrix}
1\\0
\end{pmatrix}, \quad \ket{1} := 
\begin{pmatrix}
0\\ 1
\end{pmatrix}.
\]
	
	\section{Analytical structure of periodic circuit}\label{sec:analytical-result-qspc}
	The QSPC circuit in \cref{fig:qsp-pc-circuit} enjoys a periodic structure by interleaving \fsim\ and $Z$-rotation. This periodic structure is studied by the theory of QSP (\cref{thm:qsp}). Consequentially, the QSPC circuit admits some polynomial representation. In this section, we will derive the analytical form of the structure of the QSPC circuit. We start from the exact closed-form results of the QSPC circuit in \cref{sec:exact-pc}. In \cref{sec:app-pc}, we derive a good approximation to the closed-form exact results. The analysis in this section proves \cref{thm:structure-of-qsp-pc}.
	\subsection{Exact representation of the periodic circuit}\label{sec:exact-pc}
	We abstract a simple $\mathrm{SU}(2)$-product model which can be shown as the building block of the QSPC circuit in \cref{fig:qsp-pc-circuit}. It turns out that the model admits a polynomial representation. 
	\begin{definition}[Building block of QSPC]\label{def:qspc-unitary}
		Let $\theta, \omega \in \RR$ be any angles, $d \in \NN$ by any positive integer. Then, the matrix representation of a periodic circuit with $d$ repetitions and $Z$-phase modulation angle $\omega$ is
		\begin{equation}
		U^\scp{d}(\omega,\theta) = \left(e^{\I \omega Z} e^{\I \theta X}\right)^{d} e^{\I\omega Z}.
		\end{equation}
	\end{definition}
	In the quantum circuit defined above, the $X$- and $Z$-rotations are interleaved, which agrees with the structure of QSP in \cref{thm:qsp}. The theory of QSP implies that the $\mathrm{SU}(2)$-product model enjoys a structure representing by polynomials which is given by the following lemma.
	\begin{lemma}\label{lma:qsp-unitary-polynomial}
		Let $x = \cos(\theta) \in [-1,1]$. There exists a complex polynomial $P_\omega^\scp{d} \in \CC_d[x]$ and a real polynomial $Q_\omega^\scp{d} \in \RR_{d-1}[x]$ so that
		\begin{equation}\label{eqn:lma:qsp-unitary-polynomial}
		U^\scp{d}\left(\omega, \arccos(x)\right) = \left(\begin{array}{cc}
		P^\scp{d}_\omega(x) & \I \sqrt{1-x^2} Q^\scp{d}_\omega(x) \\
		\I \sqrt{1-x^2} Q^{\scp{d} *}_\omega(x) & P^{\scp{d} *}_\omega(x)
		\end{array}\right).
		\end{equation}
		Furthermore, the special unitarity of $U^\scp{d}(\omega, \arccos(x))$ yields 
		\begin{equation}\label{eqn:P2+Q2=1}
		P_\omega^\scp{d}(x) P_\omega^{\scp{d} *}(x) + (1-x^2) \left(Q_\omega^\scp{d}(x)\right)^2 = 1.
		\end{equation}
	\end{lemma}
	\begin{proof}
		Following \cite[Theorem 4]{GilyenSuLowEtAl2019}, there exists two polynomials $P^\scp{d}_\omega, Q^\scp{d}_\omega \in \CC[x]$ so that \cref{eqn:lma:qsp-unitary-polynomial} holds. Because $U^\scp{d}(\omega, \arccos(x))$ is a QSP unitary with a set of symmetric phase factors, $Q_\omega^\scp{d} \in \RR_{d-1}[x]$ is a real polynomial according to \cite[Theorem 2]{DongMengWhaleyEtAl2020}. \cref{eqn:P2+Q2=1} holds by taking the determinant of \cref{eqn:lma:qsp-unitary-polynomial}.
	\end{proof}
	The exact presentation of the pair of polynomials $(P_\omega^\scp{d}, Q_\omega^\scp{d})$ can be determined via recurrence on a special set of points $d = 2^j, j \in \NN$ (see \cref{lma:poly-rep-special-pts} in Appendix). Based on it, we prove the generalized result to any positive integer $d$ by using induction. This gives a complete characterization of the structure of the $\mathrm{SU}(2)$-product model in \cref{def:qspc-unitary}.
	
	\begin{theorem}\label{thm:QSP-PC-P-Q}
		Let $d = 1, 2, \ldots$ be any positive integer. Then 
		\begin{equation}
		P_\omega^\scp{d}(x) = e^{\I\omega} \left(\cos\left(d \sigma\right) + \I \frac{\sin\left(d \sigma\right)}{\sin\sigma} \left(\sin \omega\right) x\right) \text{ and } Q_\omega^\scp{d}(x) = \frac{\sin\left(d \sigma\right)}{\sin\sigma}
		\end{equation}
		where $\sigma = \arccos\left(\left(\cos \omega\right) x \right)$.
	\end{theorem}
	\begin{proof}
		Let us prove the theorem by induction. The base case is $d=1$, where $P^\scp{1}_\omega(x) = e^{2\I\omega} x$ and $Q^\scp{1}_\omega(x) = 1$. Assuming that the induction hypothesis holds for $d$, we will prove it also holds for $d+1$. Using \cref{def:qspc-unitary,lma:qsp-unitary-polynomial}, the polynomials can be determined by a recurrence relation
		\begin{equation}
		U^\scp{d+1}(\omega,\theta) = U^\scp{d}(\omega,\theta) e^{\I\theta X} e^{\I \omega Z} \Rightarrow \left\{
		\begin{array}{l}
		P^\scp{d+1}_\omega(x) = e^{\I\omega}\left(x P^\scp{d}_\omega(x) - (1-x^2) Q^\scp{d}_\omega(x)\right),\\
		Q^\scp{d+1}_\omega(x) = e^{-\I\omega} \left(P^\scp{d}_\omega(x) + x Q^\scp{d}_\omega(x)\right).
		\end{array}
		\right.
		\end{equation}
		Using the induction hypothesis, we have
		\begin{equation}
		\begin{split}
		P_\omega^\scp{d+1}(x) &= e^{\I\omega} \left(\cos\sigma\cos(d\sigma) - \left(1-\left(1-\sin^2\omega\right)x^2\right) \frac{\sin(d\sigma)}{\sin\sigma} \right.\\
		&\quad\quad\quad\quad\quad\quad \left.+ \I \left(\sin\omega\right)x \frac{\sin\sigma\cos(d\sigma)+\cos\sigma\sin(d\sigma)}{\sin\sigma}\right)\\
		&= e^{\I\omega} \left(\cos\left((d+1) \sigma\right) + \I \frac{\sin\left((d+1) \sigma\right)}{\sin\sigma} \left(\sin \omega\right) x\right)
		\end{split}
		\end{equation}
		and
		\begin{equation}
		Q_\omega^\scp{d+1}(x) = \cos(d\sigma) + \cos\sigma\frac{\sin(d\sigma)}{\sin\sigma} = \frac{\sin\left((d+1)\sigma\right)}{\sin\sigma}.
		\end{equation}
		Therefore, the theorem follows induction.
	\end{proof}
The closed-form results above help us to analyze the dynamics of \cref{fig:qsp-pc-circuit} where we  apply  a Pauli $Z$ modulation $e^{\I\omega Z_{A_0}}$ to the periodic circuit. Restricted to the single-excitation subspace, the matrix representation of the QSPC circuit in \cref{fig:qsp-pc-circuit} is
	\begin{equation}\label{eqn:circuit-rep-of-qspc-with-building-block}
	\mc{U}^\scp{d}(\omega; \theta, \varphi) = \matrepwrt{\left(e^{\I \omega Z_{A_0}} U_\fsim\left(\theta,\varphi,\chi,*\right)\right)^d}{\mc{B}_2} = e^{\I\frac{\chi+\pi+\varphi}{2} Z} U^\scp{d}(\omega-\varphi, \theta) e^{-\I\left(\omega + \frac{\chi+\pi-\varphi}{2} \right)Z}.
	\end{equation}
	The initial two-qubit state of the QSP circuit can be prepared as Bell states $\ket{+_\ell}$ or $\ket{\I_\ell}$ by using Hadamard gate, phase gate and CNOT gate. Recall that we denote the probability by measuring qubits $A_0A_1$ with $01$ as
	\begin{equation}
	p_X(\omega; \theta, \varphi) = \abs{\braket{0_\ell|\mc{U}^\scp{d}(\omega; \theta, \varphi)|+_\ell}}^2
	\end{equation}
	when the initial state is $\ket{+_\ell}$, and
	\begin{equation}
	p_Y(\omega; \theta, \varphi) = \abs{\braket{0_\ell|\mc{U}^\scp{d}(\omega; \theta, \varphi)|\I_\ell}}^2
	\end{equation}
	when the initial state is $\ket{\I_\ell}$ respectively. These bridge the gap between the analytical results derived based on \cref{def:qspc-unitary} and the measurement probabilities from the experimental setting. We are ready to prove the first half of \cref{thm:structure-of-qsp-pc}. 
	
	\begin{theorem}\label{thm:reconstruction-h-Fourier-expansion}
		The function reconstructed from the measurement probability admits the following Fourier series expansion:
		\begin{equation}
		\mf{h}(\omega; \theta, \varphi, \chi) := p_X(\omega; \theta, \varphi, \chi) + \I p_Y(\omega; \theta, \varphi, \chi) - \frac{1+ \I}{2} = \sum_{k = -d+1}^{d-1} c_{k}(\theta, \chi, \varphi) e^{2 \I k \omega}
		\end{equation}
		where
		\begin{equation}
		c_k(\theta,\chi,\varphi) = \I e^{-\I\chi} e^{-\I(2k+1)\varphi} \wt{c}_k(\theta)\quad \text{and}\quad \wt{c}_k(\theta) \in \RR.
		\end{equation}
	\end{theorem}
	\begin{proof}
		For simplicity, let $\beta \in \mathrm{U}(1)$ and $\ket{\beta} := \frac{1}{\sqrt{2}}\left(\ket{0_\ell} + \beta \ket{1_\ell}\right)$. Then, $\ket{\beta=1} = \ket{+_\ell}$ and $\ket{\beta = \I} = \ket{\I_\ell}$. Given the input quantum state is $\ket{\beta}$, we have the measurement probability
		\begin{equation}
		\begin{split}
		p_\beta(\omega;\theta,\varphi) &= \abs{\braket{0_\ell | \mc{U}^\scp{d}(\omega; \theta, \varphi)| \beta}}^2\\
		&= \abs{\frac{1}{\sqrt{2}} e^{\I\frac{\varphi+\chi+\pi}{2}} \bra{0_\ell} U^\scp{d}(\omega-\varphi, \theta) \left(e^{-\I\left(\omega + \frac{\chi+\pi-\varphi}{2} \right)}\ket{0_\ell} + \beta e^{\I\left(\omega + \frac{\chi+\pi-\varphi}{2} \right)}\ket{1_\ell}\right)}^2\\
		&= \frac{1}{2} + \Re\left( \overline{\beta} e^{\I(\varphi-\chi-2\omega)} P_{\omega-\varphi}^\scp{d}(\cos\theta) \I \sin\theta Q_{\omega-\varphi}^\scp{d}(\cos\theta) \right).
		\end{split}
		\end{equation}
		Then, $p_X = p_{\beta=1}$ and $p_Y = p_{\beta = \I}$. Furthermore, it holds that
		\begin{equation}\label{eqn:pX-Re-pY-Im}
		\begin{split}
		& p_X(\omega; \theta, \varphi) - \frac{1}{2} = \Re\left( e^{\I(\varphi-\chi-2\omega)} P_{\omega-\varphi}^\scp{d}(\cos\theta) \I \sin\theta Q_{\omega-\varphi}^\scp{d}(\cos\theta) \right),\\
		& p_Y(\omega; \theta, \varphi) - \frac{1}{2} = \Im\left( e^{\I(\varphi-\chi-2\omega)} P_{\omega-\varphi}^\scp{d}(\cos\theta) \I \sin\theta Q_{\omega-\varphi}^\scp{d}(\cos\theta) \right).
		\end{split}
		\end{equation}
		Therefore, the reconstructed function is
		\begin{equation}
		\mf{h}(\omega; \theta, \varphi, \chi) = \I e^{-\I(\chi+\varphi)} \sin\theta e^{-2\I(\omega - \varphi)} P_{\omega-\varphi}^\scp{d}(\cos\theta) Q_{\omega-\varphi}^\scp{d}(\cos\theta) =: \I e^{-\I(\chi+\varphi)} \wt{\mf{h}}(\omega-\varphi,\theta).
		\end{equation}
		Note that following \cref{thm:QSP-PC-P-Q}
		\begin{equation}
		\begin{split}
		&P_{\omega+\pi-\varphi}^\scp{d}(\cos\theta) = (-1)^{d+1} P_{\omega-\varphi}^\scp{d}(\cos\theta),\ Q_{\omega+\pi-\varphi}^\scp{d}(\cos\theta) = (-1)^{d-1} Q_{\omega-\varphi}^\scp{d}(\cos\theta)\\
		& \Rightarrow\ \wt{\mf{h}}(\omega+\pi-\varphi,\theta) = \wt{\mf{h}}(\omega-\varphi,\theta).
		\end{split}
		\end{equation}
		That means $\wt{\mf{h}}(\omega-\varphi,\theta)$ is $\pi$-periodic in the first argument. Furthermore, $\wt{\mf{h}}(\omega-\varphi,\theta)$ is a trigonometric polynomial in $(\omega-\varphi)$. Thus, it admits the Fourier series expansion:
		\begin{equation}
		\wt{\mf{h}}(\omega-\varphi,\theta) = \sum_{k=-d+1}^{d-1} \wt{c}_k(\theta) e^{2\I k (\omega-\varphi)}
		\end{equation}
		with coefficients
		\begin{equation}
		    \wt{c}_k(\theta) = \frac{\sin\theta}{\pi} \int_0^\pi e^{-2\I(k+1)\omega} P_\omega^\scp{d}(\cos\theta) Q_\omega^\scp{d}(\cos\theta) \rd \omega.
		\end{equation}
		The upper limit and lower limit of the summation index $\pm (d-1)$ can be verified by straightforward computation. According to \cref{thm:QSP-PC-P-Q}, we also have
		\begin{equation}
		P_{-\omega}^\scp{d}(\cos\theta) = \overline{P_\omega^\scp{d}(\cos\theta)},\ Q_\omega^\scp{d}(\cos\theta) \in \RR\ \Rightarrow\ \wt{\mf{h}}(\omega,\theta) = \overline{\wt{\mf{h}}(-\omega,\theta)}\ \Rightarrow\ \wt{c}_k(\theta) \in \RR.
		\end{equation}
		The proof is completed.
	\end{proof}
	It is also useful to study the magnitude of the reconstructed function. It gives the intuition of the distribution of the magnitude over different modulation angle $\omega$. The following corollary indicated that the magnitude of the reconstructed function attains its maximum $d\theta$ when the phase matching condition $\omega = \varphi$ is achieved.
	\begin{corollary}\label{cor:modulus-magnitude-sin2theta}
		The magnitude of $p_X(\omega; \theta, \varphi)-\frac{1}{2}$ and $p_Y(\omega; \theta, \varphi)-\frac{1}{2}$ are of order $\sin\theta$. Furthermore
		\begin{equation}\label{eqn:px2+py2-calibration}
		\begin{split}
		\mf{p}(\omega-\varphi, \theta) &:= \abs{\mf{h}(\omega;\theta,\varphi,\chi)}^2 =  \left(p_X(\omega; \theta, \varphi) - \frac{1}{2}\right)^2 + \left(p_Y(\omega; \theta, \varphi) - \frac{1}{2}\right)^2\\
		&= \sin^2(\theta) \frac{\sin^2(d\sigma)}{\sin^2(\sigma)} \left(1 - \sin^2(\theta) \frac{\sin^2(d\sigma)}{\sin^2(\sigma)}\right).
		\end{split}
		\end{equation}
		Here $\sigma = \arccos\left(\cos(\omega-\varphi)\cos(\theta)\right)$.
	\end{corollary}
	\begin{proof}
		Using \cref{thm:QSP-PC-P-Q,eqn:P2+Q2=1,eqn:pX-Re-pY-Im} as intermediate steps, we have
		\begin{equation}
		\begin{split}
		\mf{p}(\omega-\varphi, \theta) &= \abs{e^{\I(\varphi-\chi-2\omega)} P_{\omega-\varphi}^\scp{d}(\cos\theta) \I \sin\theta Q_{\omega-\varphi}^\scp{d}(\cos\theta)}^2\\
		&= \sin^2(\theta) \abs{Q_{\omega-\varphi}^\scp{d}(\cos\theta)}^2 \abs{P_{\omega-\varphi}^\scp{d}(\cos\theta)}^2\\
		&= \sin^2(\theta) \abs{Q_{\omega-\varphi}^\scp{d}(\cos\theta)}^2 \left(1 - \sin^2(\theta) \abs{Q_{\omega-\varphi}^\scp{d}(\cos\theta)}^2\right)\\
		&= \sin^2(\theta) \frac{\sin^2(d\sigma)}{\sin^2(\sigma)} \left(1 - \sin^2(\theta) \frac{\sin^2(d\sigma)}{\sin^2(\sigma)}\right)
		\end{split}
		\end{equation}
		which completes the proof.
	\end{proof}
	
	As a remark, if the transition probability between tensor-product states is measured, the magnitude of the signal (the nontrivial $\theta$ dependence in the transition probability) is $\Or\left(\sin^2\theta\right)$. Nonetheless, by preparing the input quantum state as Bell states, \cref{cor:modulus-magnitude-sin2theta} reveals that the magnitude of the signal is lifted to $\Or(\sin\theta)$ instead. Therefore, when $\theta$ is extremely small, it is a significant improvement of the SNR especially in the presence of realistic errors.
	
	\subsection{Approximate Fourier coefficients}\label{sec:app-pc}
	\cref{thm:reconstruction-h-Fourier-expansion} shows that the $\theta$ and $\varphi$ dependence are factored completely in the amplitude and the phase of the Fourier coefficients of the reconstructed function $\mf{h}$ respectively. Given the angle $\omega$ of the $Z$-rotation is tunable, we can sample the data point by performing the QSPC circuit in \cref{fig:qsp-pc-circuit} with equally spaced angles $\omega_j = \frac{j}{2d-1} \pi$ where $j = 0, \cdots, 2d-2$. These $2(2d-1)$ quantum experiments yield two sequences of measurement probabilities $\vp_X^\expl := \left(p^\expl_X(\omega_0), p^\expl_X(\omega_1), \cdots, p^\expl_X(\omega_{2d-2})\right)$ and $\vp_Y^\expl := \left(p^\expl_Y(\omega_0), p^\expl_Y(\omega_1), \cdots, p^\expl_Y(\omega_{2d-2})\right)$. Therefore, we can compute $\mf{h}^\expl = \vp_X^\expl + \vp_Y^\expl - \frac{1+\I}{2}$ from experimental data. The Fourier coefficients of $\mf{h}$ can be computed by fast Fourier transform (FFT). The Fourier coefficients $\vc^\expl := \left(c_{-d+1}^{(\expl)}, c_{-d+2}^{(\expl)}, \cdots, c_{d-1}^{(\expl)}\right) = \mathsf{FFT}(\mf{h}^\expl)$ can be computed efficiently using FFT. In order to infer $\theta$ and $\varphi$ accurately and efficiently from the data, we need to study the approximate structure of the Fourier coefficients first.
	
	\begin{theorem}\label{thm:approx-coef-first-order}
		Let $\hat{\mf{h}}(\omega, \cos\theta) := \wt{\mf{h}}(\omega,\theta)/(\sin\theta e^{-\I \omega})$. There is an approximation to it: 
		\begin{equation}\label{eqn:hat_c_k_star_expr}
		\begin{split}
		& \hat{\mf{h}}^\star(\omega,\cos\theta) = \sum_{k=-d+1}^{d-1} \hat{c}_k^\star(\theta) e^{\I(2k+1)\omega},\quad \text{where }\\
		 & \hat{c}_k^\star(\theta) = \left\{
		\begin{array}{ll}
		1 - \frac{1}{2}\left(3d^2-k^2-(k+1)^2-\left(d-(2k+1)\right)^2\right) (1-\cos\theta) \ \text{ if } 0 \le k \le d-1, \\
		- \frac{1}{2} \left(d^2 + (d+2k+1)^2 - k^2 - (k+1)^2\right) (1-\cos\theta) \quad \text{ if } -d+1 \le k \le -1.
		\end{array} \right.
		\end{split}
		\end{equation}
		The approximation error is upper bounded as
		\begin{equation}
		\max_{\omega \in [0,\pi]} \abs{\hat{\mf{h}}(\omega,\cos\theta) - \hat{\mf{h}}^\star(\omega,\cos\theta)} \le 2 d^5 \theta^4
		\end{equation}
		and for any $k$
		\begin{equation}
		\abs{\wt{c}_k(\theta) - \sin\theta \hat{c}_k^\star(\theta)} \le 2(d\theta)^5.
		\end{equation}
	\end{theorem}
	\begin{proof}
		Following \cref{thm:QSP-PC-P-Q}, we have
		\begin{equation}
		\begin{split}
		\wt{\mf{h}}(\omega,\theta) &= \sin\theta e^{-2\I \omega} P_\omega^\scp{d}(\cos\theta) Q_\omega^\scp{d}(\cos\theta) = \sin\theta e^{-\I\omega} \left(\cos(d\sigma) + \I \sin\omega \cos\theta \frac{\sin(d\sigma)}{\sin\sigma}\right)\frac{\sin(d\sigma)}{\sin\sigma}\\
		&= \sin\theta e^{-\I\omega} \left(T_d(\cos\sigma) + \I \sin\omega \cos\theta U_{d-1}(\cos\sigma)\right)U_{d-1}(\cos\sigma)
		\end{split}
		\end{equation}
		where $\cos\sigma = \cos\omega \cos\theta$ and $T_d \in \RR_d[x], U_{d-1} \in \RR_{d-1}[x]$ are Chebyshev polynomials of the first and second kind respectively. Then
		\begin{equation}
		\begin{split}
		\hat{\mf{h}}(\omega,\cos\theta) &:= \left(T_d(\cos\sigma) + \I \sin\omega \cos\theta U_{d-1}(\cos\sigma)\right)U_{d-1}(\cos\sigma) \\
		&= \frac{1}{2} U_{2d-1}(\cos\sigma) + \I \sin\omega\cos\theta U_{d-1}^2(\cos\sigma).
		\end{split}
		\end{equation}
		Therefore, for a given $\omega$, $\hat{\mf{h}}(\omega, \cos\theta)$ is a polynomial in $\cos\theta$ of degree at most $2d-1$. According to \cref{cor:modulus-magnitude-sin2theta}, we have for any $\omega$
		\begin{equation}\label{eqn:h-re-im-upper-bound}
		\max_{\theta \in [0,\pi]} \max\left\{ \abs{\Re\left(\hat{\mf{h}}(\omega,\cos\theta)\right)}, \abs{\Im\left(\hat{\mf{h}}(\omega,\cos\theta)\right)} \right\} \le \max_{\theta \in [0,\pi]} \abs{\hat{\mf{h}}(\omega,\cos\theta)} \le \max_{\theta \in [0,\pi]} \abs{\frac{\sin(d\sigma)}{\sin\sigma}} \le d.
		\end{equation}
		Applying Taylor's theorem and expanding $\hat{\mf{h}}(\omega,\cos\theta)$ with respect to $1-\cos\theta$, there exists $\xi \in (\cos\theta, 1)$ so that
		\begin{equation}\label{eqn:taylor-remainder}
		\hat{\mf{h}}(\omega,\cos\theta) = \hat{\mf{h}}(\omega,1) + \frac{\partial \hat{\mf{h}}(\omega,x)}{\partial x}\bigg|_{x=1} (\cos\theta - 1) + \frac{1}{2} \frac{\partial^2 \hat{\mf{h}}(\omega,x)}{\partial x^2}\bigg|_{x=\xi}(\cos\theta - 1)^2.
		\end{equation}
		Here
		\begin{equation}
		\hat{\mf{h}}(\omega,1) = e^{\I d \omega} U_{d-1}(\cos\omega) = e^{\I \omega} \sum_{k=0}^{d-1} e^{2\I k \omega}.
		\end{equation}
		Furthermore,
		\begin{equation}
		\begin{split}
		& \Re\left(\frac{\partial \hat{\mf{h}}(\omega,x)}{\partial x}\bigg|_{x=1} \right) = \frac{1}{2} \frac{\partial U_{2d-1}(x\cos\omega)}{\partial x}\bigg|_{x=1} = \frac{\cos\omega}{2} U_{2d-1}^\prime(\cos\omega)\\
		&= \cos\omega \sum_{j=0}^{d-1} (2j+1) U_{2j}(\cos\omega) = \cos\omega \sum_{j=0}^{d-1} (2j+1) \sum_{k=-j}^j e^{2\I k \omega}\\
		&= \cos\omega \sum_{k=-(d-1)}^{d-1} (d^2-k^2) e^{2\I k \omega}
		\end{split}
		\end{equation}
		and
		\begin{equation}
		\begin{split}
		& \Im\left(\frac{\partial \hat{\mf{h}}(\omega,x)}{\partial x}\bigg|_{x=1} \right) = \sin\omega U_{d-1}(\cos\omega) \left(U_{d-1}(\cos\omega) + 2\cos\omega U_{d-1}^\prime(\cos\omega)\right)\\
		&= \sin(d\omega)\left(\sum_{\substack{k=-(d-1)\\ \text{stepsize 2}}}^{d-1} e^{\I k \omega} + \frac{1}{2} \left(e^{\I\omega} + e^{-\I\omega}\right) \sum_{\substack{k=-(d-2)\\ \text{stepsize 2}}}^{d-2} (d^2-k^2) e^{\I k \omega}\right)\\
		&= \sin(d\omega) \sum_{\substack{k=-(d-1)\\ \text{stepsize 2}}}^{d-1} (d^2-k^2) e^{\I k \omega} = \frac{1}{2\I} \left(\sum_{\substack{k=-2d+1\\ \text{stepsize 2}}}^{-1} (2d+k)k e^{\I k \omega} + \sum_{\substack{k=1\\ \text{stepsize 2}}}^{2d-1} (2d-k)k e^{\I k \omega}\right).
		\end{split}
		\end{equation}
		Let the approximation of $\hat{\mf{h}}(\omega,\cos\theta)$ be
		\begin{equation}
		\hat{\mf{h}}^\star(\omega,\cos\theta) := \hat{\mf{h}}(\omega,1) + \frac{\partial \hat{\mf{h}}(\omega,x)}{\partial x}\bigg|_{x=1} (\cos\theta - 1).
		\end{equation}
		Then, the previous computation shows it admits a Fourier series expansion:
		\begin{equation}
		\begin{split}
		& \hat{\mf{h}}^\star(\omega,\cos\theta) = \sum_{k=-d+1}^{d-1} \hat{c}_k^\star(\theta) e^{\I(2k+1)\omega},\ \text{where } \\
		& \hat{c}_k^\star(\theta) = \left\{
		\begin{array}{ll}
		1 + \frac{1}{2}\left(3d^2-k^2-(k+1)^2-\left(d-(2k+1)\right)^2\right) (\cos\theta - 1) \ \text{ if } 0 \le k \le d-1, \\
		\frac{1}{2} \left(d^2 + (d+2k+1)^2 - k^2 - (k+1)^2\right) (\cos\theta - 1) \quad \text{ if } -d+1 \le k \le -1.
		\end{array} \right.
		\end{split}
		\end{equation}
		The approximation error can be bounded by using \cref{eqn:taylor-remainder}. For any $\omega \in [0,\pi]$, we have
		\begin{equation}
		\begin{split}
		& \abs{\hat{\mf{h}}(\omega,\cos\theta) - \hat{\mf{h}}^\star(\omega,\cos\theta)} \le \frac{(1-\cos\theta)^2}{2} \max_{x \in [-1,1]} \abs{\frac{\partial^2 \hat{\mf{h}}(\omega,x)}{\partial x^2}}\\
		& \le \frac{\theta^4}{8} \sqrt{\left(\max_{x \in [-1,1]}\abs{\frac{\partial^2 \Re\left(\hat{\mf{h}}(\omega,x)\right)}{\partial x^2}}\right)^2 + \left(\max_{x \in [-1,1]}\abs{\frac{\partial^2 \Im\left(\hat{\mf{h}}(\omega,x)\right)}{\partial x^2}}\right)^2}.
		\end{split}
		\end{equation}
		Note that $\Re\left(\hat{\mf{h}}(\omega,x)\right)$ and $\Im\left(\hat{\mf{h}}(\omega,x)\right)$ are real polynomials in $x$ of degree at most $2d-1$. Invoking the Markov brothers' inequality (\cref{thm:Markovs-ineq}), we further get
		\begin{equation}
		\begin{split}
		& \abs{\hat{\mf{h}}(\omega,\cos\theta) - \hat{\mf{h}}^\star(\omega,\cos\theta)}\\
		& \le \frac{\sqrt{2} (2d-1)^2 d(d-1)}{6} \theta^4 \max_{x \in [-1,1]} \max\left\{ \abs{\Re\left(\hat{\mf{h}}(\omega,x)\right)}, \abs{\Im\left(\hat{\mf{h}}(\omega,x)\right)} \right\}\\
		& \le \frac{\sqrt{2} (2d-1)^2 d^2(d-1)}{6} \theta^4 \le 2 d^5 \theta^4,
		\end{split}
		\end{equation}
		where \cref{eqn:h-re-im-upper-bound} is used. The error bound can be transferred to that of the Fourier coefficients. Using the previous result and triangle inequality, one has
		\begin{equation}
		\begin{split}
		\abs{\wt{c}_k(\theta) - \sin\theta \hat{c}_k^\star(\theta)} &= \abs{\frac{\sin\theta}{\pi} \int_0^\pi e^{-\I (2k+1) \omega} \left(\hat{\mf{h}}(\omega,\cos\theta) - \hat{\mf{h}}^\star(\omega,\cos\theta)\right) \rd \omega}\\
		& \le \sin\theta \frac{1}{\pi} \int_0^\pi \abs{\hat{\mf{h}}(\omega,\cos\theta) - \hat{\mf{h}}^\star(\omega,\cos\theta)} \rd \omega \le 2 \left(d\theta\right)^5. 
		\end{split}
		\end{equation}
		The proof is completed.
	\end{proof}
	
	There are two implications of the previous theorem. First, it suggests that the magnitude of the Fourier coefficients of negative indices are $\Or\left(\sin^3\theta\right)$. If they are included in the formalism of the inference problem in the Fourier space, the accuracy of inference may be heavily contaminated because of the nearly vanishing SNR when $\theta \ll 1$. On the other hand, the amplitude of the Fourier coefficients of nonnegative indices tightly concentrate at $\sin\theta$ when $\theta \ll 1$. Therefore, a nice linear approximation of the Fourier coefficients holds in the case of extremely small swap angle: for any $k = 0, \cdots, d-1$
	\begin{equation}
	c_k(\theta,\chi,\varphi) = \I e^{-\I\chi} e^{-\I(2k+1)\varphi} \theta + \mathrm{max} \left\{ \Or\left(\theta^3\right), \Or\left((d\theta)^5\right) \right\}.
	\end{equation}
	This proves the second half of \cref{thm:structure-of-qsp-pc}.
	
	\section{Robust estimator against Monte Carlo sampling error}\label{sec:Monte-Carlo-sampling-error}
A dominant and unavoidable source of errors in quantum metrology is Monte Carlo sampling error due to the finite sample size in quantum measurements. Such limitation derives from both practical concerns of the efficiency of quantum metrology, and realistic constraints where some system parameters can drift over time and can only be monitored by sufficiently fast protocols. In this section, we  analyze the effect of Monte Carlo sampling error in our proposed metrology algorithm by characterizing the sampling error as a function of quantum circuit depth, \fsim\ parameters and sample size. The result will also be used in \cref{sec:lower-bound-qspc-metrology} to prove that our estimator based on QSPC is optimal. In the following analysis, we annotate  with superscript ``$\expl$'' to represent experimentally measured probability as oppose to expected probability from theory.
	\subsection{Modeling the Monte Carlo sampling error}\label{subsec:model_MC}
	We start the analysis by statistically modeling the Monte Carlo sampling error on the measurement probabilities. Furthermore, we also derive the sampling error induced on the Fourier coefficients derived from experimental data. The result is summarized in the following lemma.
	\begin{lemma}\label{lma:monte-carlo-error-magnitude}
		Let $M$ be the number of measurement samples in each experiment. When $M$ is large enough, the measurement probability $p_X^\expl(\omega_j)$ is approximately normal distributed
		\begin{equation}
		p_X^\expl(\omega_j) = p_X(\omega_j;\theta,\varphi,\chi) + \Sigma_{X,j} u_{X,j},\text{ where } u_{X,j} \sim N(0,1) \text{ and } \frac{1-4(d\theta)^2}{4M} \le \Sigma_{X,j}^2 \le \frac{1}{4M}.
		\end{equation}
		The same conclusion holds for $p_Y^\expl(\omega_j)$. Furthermore, by computing the Fourier coefficients via FFT, the Fourier coefficients are approximately complex normal distributed
		\begin{equation}
		c_k^\expl = c_k(\theta,\varphi,\chi) + \left\{ \begin{array}{ll}
		v_k &, k = 0,\cdots, d-1,   \\
		v_{2d-1+k} &, k = -d+1,\cdots, -1. 
		\end{array}\right.
		\end{equation}
		where $v_k$'s are complex normal distributed random variables so that
		\begin{equation}
		\begin{split}
		&\text{for any }k:\ \expt{v_k} = 0,\ \frac{1 - 2(d\theta)^2}{2M(2d-1)} \le \expt{\abs{v_k}^2} \le \frac{1}{2M(2d-1)},\\
		&\text{ and for any } k \ne k^\prime:\ \abs{\expt{v_k \overline{v_{k^\prime}}}} \le \frac{(d\theta)^2}{M(2d-1)}.
		\end{split}
		\end{equation}
		Consequentially, when $d\theta \ll 1$, these random variables $v_k$'s can be approximately assumed to be uncorrelated.
	\end{lemma}
	\begin{proof}
		Given a quantum experiment with angle $\omega_j$, the measurement generates i.i.d. Bernoulli distributed outcomes $b_i$'s, namely $\bP(b_i=0) = 1-\bP(b_i=1) = p_X(\omega_j;\theta,\varphi,\chi)$. Then, the measurement probability is estimated by $p_X^\expl(\omega_j) = \frac{1}{M}\sum_{i=1}^M (1-b_i)$. When the sample size $M$ is large enough, $p^\expl_X(\omega_j)$ is approximately normal distributed following the central limit theorem where the mean is $\expt{p_X^\expl(\omega_j)} = p_X(\omega_j;\theta,\varphi,\chi)$ and the variance is 
		\begin{equation}\label{eqn:variance-mc-error-concentration}
		\begin{split}
		\Sigma_{X,j}^2 &:= \var\left(p^\expl_X(\omega_j)\right) = \frac{p_X(\omega_j;\theta,\varphi,\chi)\left(1-p_X(\omega_j;\theta,\varphi,\chi)\right)}{M}\\
		& = \frac{1}{4M} - \frac{\left(p_X(\omega_j;\theta,\varphi,\chi) - \frac{1}{2}\right)^2}{M} \le \frac{1}{4M}.
		\end{split}
		\end{equation}
		The other side of the inequality $\Sigma_{X,j}^2 \ge \frac{1-4(d\theta)^2}{4M}$ follows that $(p_X-\frac{1}{2})^2 \le (d\theta)^2$ from \cref{cor:modulus-magnitude-sin2theta}. The same analysis is applicable to $p_Y^\expl(\omega_j)$. 
		
		To compute the Fourier coefficients from the experimental data, we perform FFT on $\mf{h}^\expl_j := p^\expl_X(\omega_j) + \I p_Y^\expl(\omega_j) - \frac{1+\I}{2}$ reconstructed from the experimental data. We have $\expt{\mf{h}^\expl_j} = \mf{h}(\omega_j;\theta,\varphi,\chi)$. Furthermore, let $\wt{u}_j = \mf{h}^\expl_j - \expt{\mf{h}^\expl_j} = \Sigma_{X,j}u_{X,j} + \I \Sigma_{Y,j}u_{Y,j}$, then it holds that
		\begin{equation}
		\begin{split}
		& \text{for any }j,\ \expt{\wt{u}_j} = 0,\ \expt{\abs{\wt{u}_j}^2} = \Sigma_{X,j}^2 + \Sigma_{Y,j}^2 = \frac{1}{2M} - \frac{\mf{p}(\omega_j-\varphi,\theta)}{M}, \\
		&\text{and for any }j \ne j^\prime, \expt{\wt{u}_j \overline{\wt{u}_{j^\prime}}} = 0.
		\end{split}
		\end{equation}
		The FFT gives the Fourier coefficients as
		\begin{equation}
		\left(\begin{array}{c}
		c_0^\expl\\ \vdots \\ c_{d-1}^\expl \\ c_{-d+1}^\expl \\ \vdots \\ c_{-1}^\expl
		\end{array}\right) = \frac{1}{2d-1} \Omega^\dagger \left(\begin{array}{l}
		\mf{h}^\expl_0 \\ \vdots  \\ \mf{h}^\expl_{2d-2}
		\end{array}\right), \text{ where } \Omega_{jk} = e^{\I \frac{2\pi jk}{2d-1}}.
		\end{equation}
		Using the linearity, we get 
		\begin{equation}\label{eqn:monte-carlo-error-in-Fourier-coef}
		v_k = \frac{1}{2d-1}\left(\Omega^\dagger \mf{h}^\expl\right)_k = \frac{1}{2d-1} \sum_{j=0}^{2d-2} \overline{\Omega_{kj}} \wt{u}_j.
		\end{equation}
		The mean is $\expt{v_k} = \frac{1}{2d-1} \sum_{j=0}^{2d-2} \overline{\Omega_{kj}} \expt{\wt{u}_j} = 0$. The covariance is
		\begin{equation}
		\expt{v_k \overline{v_{k^\prime}}} = \frac{1}{(2d-1)^2} \sum_{j, j^\prime = 0}^{2d-2} \overline{\Omega_{kj}} \Omega_{k^\prime j^\prime} \expt{\wt{u}_j \overline{\wt{u}_{j^\prime}}} = \frac{1}{(2d-1)^2} \sum_{j=0}^{2d-2} e^{\I \frac{2\pi}{2d-1}(k^\prime-k)j} \expt{\abs{\wt{u}_j}^2}.
		\end{equation}
		When $k = k^\prime$, it gives
		\begin{equation}
		\begin{split}
		& \expt{\abs{v_k}^2} = \frac{1}{2M(2d-1)} - \frac{1}{M (2d-1)^2} \sum_{j=0}^{2d-2} \mf{p}(\omega_j-\varphi, \theta)\\
		& \Rightarrow \frac{1 - 2(d\theta)^2}{2M(2d-1)} \le \expt{\abs{v_k}^2} \le \frac{1}{2M(2d-1)}
		\end{split}
		\end{equation}
		where $0 \le \mf{p}(\omega_j-\varphi, \theta) \le (d\theta)^2$ is used which follows \cref{cor:modulus-magnitude-sin2theta}.
		
		On the other hand, when $k \ne k^\prime$, the constant term $\frac{1}{2M}$ in $\expt{\abs{\wt{u}_j}^2}$ vanishes because $\sum_{j=0}^{2d-2} e^{\I \frac{2\pi}{2d-1}(k^\prime-k)j} = (2d-1) \delta_{k k^\prime}$. Then, using triangle inequality and \cref{cor:modulus-magnitude-sin2theta}, we get
		\begin{equation}
		\begin{split}
		\abs{\expt{v_k \overline{v_{k^\prime}}}} &= \abs{\frac{1}{(2d-1)^2} \sum_{j=0}^{2d-2} e^{\I \frac{2\pi}{2d-1}(k^\prime-k)j} \mf{p}(\omega_j-\varphi, \theta)}\\
		& \le \frac{1}{M (2d-1)^2} \sum_{j=0}^{2d-2} \abs{\mf{p}(\omega_j-\varphi, \theta)} \le \frac{(d\theta)^2}{M(2d-1)}.
		\end{split}
		\end{equation}
		The proof is completed.
	\end{proof}
	With a characterization of Monte Carlo sampling error, we are able to measure the robustness of the signal against error by the signal-to-noise ratio (SNR). The SNR of each Fourier coefficient is defined as the ratio between the squared Fourier coefficient and the variance of its associated additive sampling error. We define the SNR of QSPC-F in \cref{prob:inference-Fourier} by the minimal component-wise SNR. The following theorem gives a characterization of the SNR.
	\begin{theorem}\label{thm:snr-qspc}
		When $d^{\frac{5}{4}} \theta \ll 1$, the signal-to-noise ratio satisfies
		\begin{equation}
		\mathrm{SNR}_k := \frac{\abs{c_k(\theta,\varphi,\chi)}^2}{\expt{\abs{v_k}^2}} \ge \mathrm{SNR} := 2(2d-1)M\sin^2\theta \left(1 - \frac{4}{3}(d\theta)^2\left(1 + 3d^3\theta^2\right)\right).
		\end{equation}
	\end{theorem}
	\begin{proof}
		According to \cref{eqn:hat_c_k_star_expr}, for any $k = 0, \cdots, d-1$
		\begin{equation}
		1 - \frac{2}{3} (d\theta)^2 \le \hat{c}_k^\star(\theta) \le 1.
		\end{equation}
		Applying \cref{thm:approx-coef-first-order}, we have
		\begin{equation}
		c_k(\theta,\varphi,\chi) \ge \sin\theta\left(\hat{c}_k^\star(\theta) - 2d^5\theta^4 \right) \ge \sin\theta \left(1 - \frac{2}{3}(d\theta)^2\left(1 + 3d^3\theta^2\right) \right).
		\end{equation}
		Furthermore, by Bernoulli's inequality,
		\begin{equation}
		\abs{c_k(\theta,\varphi,\chi)}^2 \ge \sin^2\theta \left(1 - \frac{4}{3}(d\theta)^2\left(1 + 3d^3\theta^2\right) \right).
		\end{equation}
		Combing the derived inequality with \cref{lma:monte-carlo-error-magnitude}, it gives
		\begin{equation}
		\mathrm{SNR}_k = \frac{\abs{c_k(\theta,\varphi,\chi)}^2}{\expt{\abs{v_k}^2}} \ge 2(2d-1)M\sin^2\theta \left(1 - \frac{4}{3}(d\theta)^2\left(1 + 3d^3\theta^2\right)\right),
		\end{equation}
		which completes the proof.
	\end{proof}
	
	\subsection{Statistical estimator against Monte Carlo sampling error}\label{subsec:stat-estimator-MC}
	As a consequence of \cref{thm:snr-qspc}, when SNR is high, namely $d M \theta^2 \gg 1$, the noise modeling in Ref. \cite{Tretter1985} suggests that a linear model with normal distributed noise can well approximate the problem in which the $\theta$- and $(\varphi,\chi)$-dependence are decoupled following \cref{thm:reconstruction-h-Fourier-expansion}.
	When $k = 0, \cdots, d-1$ and $d^5 \theta^4 \ll 1$, we have 
	\begin{equation}\label{eqn:modeled-problem-Fourier-space}
	\begin{split}
	& \mathsf{amplitude}\left(c^\expl_k\right) = \wt{c}_k(\theta) + v_k^\scp{\mathrm{amp}} \approx \theta + v_k^\scp{\mathrm{amp}},\\
	& \mathsf{phase}(c^\expl_k) = \frac{\pi}{2} - \chi - (2k+1) \varphi + v_k^\scp{\mathrm{pha}} \text{ (up to } 2\pi\text{-periodicity)}
	\end{split}
	\end{equation}
	where $v_k^\scp{\mathrm{amp}}$ and $v_k^\scp{\mathrm{pha}}$ are normal distributed and are approximately $v_k^\scp{\mathrm{amp}} = \Re(v_k)$, $v_k^\scp{\mathrm{pha}} = \Im(v_k)/\wt{c}_k(\theta)$ according to Ref. \cite{Tretter1985}. Let the covariance matrices be $\mc{C}^\scp{\mathrm{amp}}$ and $\mc{C}^\scp{\mathrm{pha}}$. For any $k$ and $k^\prime$
	\begin{equation}
	\mc{C}_{k,k^\prime}^\scp{\mathrm{amp}} := \expt{v_k^\scp{\mathrm{amp}}v_{k^\prime}^\scp{\mathrm{amp}}} \text{ and } \mc{C}_{k,k^\prime}^\scp{\mathrm{pha}} := \expt{v_k^\scp{\mathrm{pha}}v_{k^\prime}^\scp{\mathrm{pha}}}.
	\end{equation}
	Let the data vectors be
	\begin{equation}
	\abs{\vec{c}^\expl} := \left(\mathsf{amplitude}(c_0^\expl), \cdots, \mathsf{amplitude}(c_{d-1}^\expl)\right)^\top,\ \vec{\wt{\mymathbb{1}}} = (\underbrace{1,\cdots,1}_d)^\top.
	\end{equation}
	The maximum likelihood estimator (MLE) is found by minimizing the negated log-likelihood function
	\begin{equation}
	\hat{\theta} = \myargmin_\theta \left(\abs{\vec{c}^\expl} - \theta \vec{\wt{\mymathbb{1}}}\right)^\top \left(\mc{C}^\scp{\mathrm{amp}}\right)^{-1} \left(\abs{\vec{c}^\expl} - \theta \vec{\wt{\mymathbb{1}}}\right).
	\end{equation}
	which follows the normality in \cref{lma:monte-carlo-error-magnitude}.
	
	In order to estimate $\varphi$, we can apply the Kay's phase unwrapping estimator in Ref.\cite{Kay1989}, a.k.a. weighted phase average estimator (WPA). The estimator is based on the sequential phase difference of the successive coefficients:
	\begin{equation}\label{eqn:sequential-phase-difference}
	\mathsf{phase}\left(c_k^\expl \overline{c_{k+1}^\expl}\right) = 2 \varphi + v_k^\scp{\mathrm{pha}} - v_{k+1}^\scp{\mathrm{pha}},\ k = 0, 1, \cdots, d-2.
	\end{equation}
	Remarkably, by computing the sequential phase difference, the troublesome $(2\pi)$-periodicity in \cref{eqn:modeled-problem-Fourier-space} can be overcome. According to this equation, the noise is turned to a colored noise process. Let the covariance be
	\begin{equation}\label{eqn:covariance-WPA}
	\mc{D}_{k,k^\prime} := \expt{\left(v_k^\scp{\mathrm{pha}} - v_{k+1}^\scp{\mathrm{pha}}\right)\left(v_{k^\prime}^\scp{\mathrm{pha}} - v_{k^\prime+1}^\scp{\mathrm{pha}}\right)} = \mc{C}_{k,k^\prime}^\scp{\mathrm{pha}} + \mc{C}_{k+1,k^\prime+1}^\scp{\mathrm{pha}} - \mc{C}_{k,k^\prime+1}^\scp{\mathrm{pha}} - \mc{C}_{k+1,k^\prime}^\scp{\mathrm{pha}}.
	\end{equation}
	Then, the WPA estimator is derived by the following MLE:
	\begin{equation}
	\hat{\varphi} = \myargmin_{\varphi} \left(\vec{\Delta} - 2 \varphi \vec{\mymathbb{1}}\right)^\top \mc{D}^{-1} \left(\vec{\Delta} - 2 \varphi \vec{\mymathbb{1}}\right)
	\end{equation}
	where the data vectors are
	\begin{equation}
	\vec{\Delta} = \left( \mathsf{phase}\left(c_0^\expl \overline{c_{1}^\expl}\right), \cdots, \mathsf{phase}\left(c_{d-2}^\expl \overline{c_{d-1}^\expl}\right) \right)^\top, \text{ and } \vec{\mymathbb{1}} = (\underbrace{1,\cdots,1}_{d-1})^\top.
	\end{equation}
	To solve the MLE, we need to study the structure of covariance matrices, which is given by the following lemma.
	\begin{lemma}\label{lma:covariance-mod-pha-estimation}
		When $d\theta \le \frac{1}{5}$ and $d^3\theta^2 \le 1$, for any $k \ne k^\prime$
		\begin{equation}
		\abs{\mc{C}_{k,k^\prime}^\scp{\mathrm{amp}}} \le \frac{(d\sin\theta)^2}{M(2d-1)}, \text{ and } \abs{\mc{C}_{k,k^\prime}^\scp{\mathrm{pha}}} \le \frac{4 d^2}{3 M(2d-1)}.
		\end{equation}
		For any $k$
		\begin{equation}
		\abs{\mc{C}_{k,k}^\scp{\mathrm{amp}} - \frac{1}{4M(2d-1)}} \le \frac{2(d\sin\theta)^2}{M(2d-1)}, \text{ and } \abs{\mc{C}_{k,k}^\scp{\mathrm{pha}} - \frac{1}{4M(2d-1)\sin^2\theta}} \le \frac{10 d^2}{3 M(2d-1)}.
		\end{equation}
	\end{lemma}
	\begin{proof}
		We first estimate the covariance of the real and imaginary components of the Monte Carlo sampling error in Fourier coefficients. Following \cref{eqn:monte-carlo-error-in-Fourier-coef,lma:monte-carlo-error-magnitude}, for any $k \ne k^\prime$,
		\begin{equation}
		\abs{\expt{\Re(v_k)\Re(v_{k^\prime}) + \Im(v_k)\Im(v_{k^\prime})}} \le \abs{\expt{v_k \overline{v_{k^\prime}}}} \le \frac{(d\sin\theta)^2}{M(2d-1)},
		\end{equation}
		and for any $k$,
		\begin{equation}
		\frac{1-2(d\sin\theta)^2}{2M(2d-1)} \le \expt{\Re^2(v_k) + \Im^2(v_k)} = \expt{\abs{v_k}^2} \le \frac{1}{2M(2d-1)}.
		\end{equation}
		For any $k\ne k^\prime$,
		\begin{equation}
		\begin{split}
		&\abs{\expt{\Re(v_k)\Re(v_{k^\prime}) - \Im(v_k)\Im(v_{k^\prime})}}  \le \abs{\expt{v_k v_{k^\prime}}} = \abs{\frac{1}{(2d-1)^2} \sum_{j, j^\prime = 0}^{2d-2} \overline{\Omega_{kj} \Omega_{k^\prime j^\prime}} \expt{\wt{u}_j \wt{u}_{j^\prime}}}\\
		&\le  \frac{1}{(2d-1)^2} \sum_{j= 0}^{2d-2} \abs{\expt{\wt{u}^2_j }} = \frac{1}{(2d-1)^2} \sum_{j= 0}^{2d-2} \abs{\Sigma_{X,j}^2 - \Sigma_{Y,j}^2} \\
		&\le \frac{1}{M (2d-1)^2} \sum_{j= 0}^{2d-2} \mf{p}(\omega_j-\varphi,\theta) \le \frac{(d\sin\theta)^2}{M(2d-1)}.
		\end{split}
		\end{equation}
		Similarly for any $k$,
		\begin{equation}
		\begin{split}
		\abs{\expt{\Re^2(v_k) - \Im^2(v_k)}} &\le \abs{\expt{\Re^2(v_k) - \Im^2(v_k) + 2\I \Re(v_k)\Im(v_k)}}\\
		&= \abs{\expt{v_k^2}} \le \frac{1}{(2d-1)^2}\sum_{j=0}^{2d-2} \abs{\expt{\wt{u}_j^2}} \le \frac{(d\sin\theta)^2}{M(2d-1)}.
		\end{split}
		\end{equation}
		Using triangle inequality and the derived results, we have for any $k$
		\begin{equation}
		\begin{split}
		\abs{\expt{\Re^2(v_k)} - \frac{1}{4M(2d-1)}} \le& \frac{1}{2}\abs{\expt{\Re^2(v_k) + \Im^2(v_k)} - \frac{1}{2M(2d-1)}} \\
		&\quad + \frac{1}{2} \abs{\expt{\Re^2(v_k) - \Im^2(v_k)}} \le \frac{(d\sin\theta)^2}{M(2d-1)}.
		\end{split}
		\end{equation}
		The same argument is applicable to the imaginary component
		\begin{equation}
		\abs{\expt{\Im^2(v_k)} - \frac{1}{4M(2d-1)}} \le \frac{(d\sin\theta)^2}{M(2d-1)}.
		\end{equation}
		When $k \ne k^\prime$,
		\begin{equation}
		\begin{split}
		\abs{\expt{\Re(v_k)\Re(v_{k^\prime}}} \le& \frac{1}{2}\abs{\expt{\Re(v_k)\Re(v_{k^\prime}) + \Im(v_k)\Im(v_{k^\prime})}} \\
		&\quad + \frac{1}{2} \abs{\expt{\Re(v_k)\Re(v_{k^\prime}) - \Im(v_k)\Im(v_{k^\prime})}} \le \frac{(d\sin\theta)^2}{M(2d-1)}
		\end{split}
		\end{equation}
		and
		\begin{equation}
		\begin{split}
		\abs{\expt{\Im(v_k)\Im(v_{k^\prime}}} \le& \frac{1}{2}\abs{\expt{\Re(v_k)\Re(v_{k^\prime}) + \Im(v_k)\Im(v_{k^\prime})}} \\
		&\quad+ \frac{1}{2} \abs{\expt{\Re(v_k)\Re(v_{k^\prime}) - \Im(v_k)\Im(v_{k^\prime})}} \le \frac{(d\sin\theta)^2}{M(2d-1)}.
		\end{split}
		\end{equation}
		Estimating \cref{eqn:hat_c_k_star_expr} gives for any $k \ge 0$
		\begin{equation}
		1 - \frac{2}{3} (d\theta)^2 \le \hat{c}_k^\star(\theta) \le 1.
		\end{equation}
		Assuming that $d^3\theta^2 \le 1$, and applying \cref{thm:approx-coef-first-order}, it holds that for any $k$
		\begin{equation}
		\sin\theta\left(1 - \frac{8}{3} (d\theta)^2\right) \le \wt{c}_k(\theta) \le\sin\theta + 2 (d\theta)^5.
		\end{equation}
		Furthermore, if $d\theta \le \frac{1}{5}$, it holds that
		\begin{equation}
		\abs{\frac{\sin\theta}{\wt{c}_k(\theta)}} \le \frac{1}{1 - \frac{8}{3}(d\theta)^2} < \sqrt{\frac{4}{3}}.
		\end{equation}
		Then, for any $k \ne k^\prime$
		\begin{equation}
		\abs{\expt{v_k^\scp{\mathrm{pha}}v_{k^\prime}^\scp{\mathrm{pha}}}} = \frac{1}{\abs{\wt{c}_k(\theta)\wt{c}_{k^\prime}(\theta)}} \abs{\expt{\Im(v_k)\Im(v_{k^\prime}}} \le \frac{4 d^2}{3 M(2d-1)}.
		\end{equation}
		For any $k$, applying triangle inequality, it yields that 
		\begin{equation}
		\begin{split}
		& \abs{\expt{\left(v_k^\scp{\mathrm{pha}}\right)^2} - \frac{1}{4M(2d-1) \sin^2\theta}} \\
		&\le \frac{1}{\wt{c}_k^2(\theta)} \abs{\expt{\Im^2(v_k)} - \frac{1}{4M(2d-1)}} + \frac{1}{4M(2d-1)\sin^2\theta} \frac{\abs{\wt{c}_k^2(\theta) - \sin^2\theta}}{\wt{c}_k^2(\theta)}\\
		&\le \frac{\sin^2\theta}{\wt{c}_k^2(\theta)} \frac{d^2}{M(2d-1)} \left( 1 + \frac{4}{3} \frac{\theta^2}{\sin^2\theta} \left(1 + d^5\theta^4\right)\right) \le \frac{5}{2} \frac{\sin^2\theta}{\wt{c}_k^2(\theta)} \frac{d^2}{M(2d-1)} \le \frac{10 d^2}{3 M(2d-1)},
		\end{split}
		\end{equation}
		where the inequality $\frac{\theta^2}{\sin^2\theta} \le \frac{1}{25 \sin^2(1/5)} < \frac{9}{8}$ when $\theta \le \frac{1}{5d} \le \frac{1}{5}$ is used to simplify the constant.
		The proof is completed.
	\end{proof}
	
	Because of the sequential phase difference, we also need to study the structure of the covariance matrix of the colored noise in \cref{eqn:sequential-phase-difference}. It is given by the following corollary.
	\begin{corollary}\label{cor:elementwise-bound-discrete-Laplacian}
		Let
		\begin{equation}
		\wt{D} := \frac{1}{4M(2d-1)\sin^2\theta} \mf{D}, \text{ where } \mf{D}_{k, k^\prime} = \left\{\begin{array}{ll}
		2 & ,\ k = k^\prime, \\
		-1 & ,\ \abs{k-k^\prime} = 1,\\
		0 & ,\ \text{otherwise}.
		\end{array}\right.
		\end{equation}
		Then, when $d\theta \le \frac{1}{5}$ and $d^3\theta^2 \le 1$,
		\begin{equation}\label{eqn:covariance-WPA-elementwise-bound}
		\abs{D_{k,k^\prime} - \wt{D}_{k,k^\prime}} \le \frac{d^2}{M(2d-1)}  \times \left\{
		\begin{array}{ll}
		\frac{28}{3} & , k = k^\prime, \\
		\frac{22}{3} & , \abs{k - k^\prime} = 1, \\
		\frac{16}{3} & , \text{otherwise}.
		\end{array}\right.
		\end{equation}
	\end{corollary}
	\begin{proof}
		The element-wise bound \cref{eqn:covariance-WPA-elementwise-bound} follows immediately by applying triangle inequality with \cref{lma:covariance-mod-pha-estimation} and the defining equation \cref{eqn:covariance-WPA}. 
	\end{proof}
	
	Consequentially, the log-likelihood functions are well approximated by quadratic forms in terms constant matrices. The approximate forms yield the MLEs of QSPC-F in \cref{prob:inference-Fourier}:
	\begin{equation}
	\hat{\theta} = \frac{1}{d} \sum_{k=0}^{d-1} \abs{c_k^\expl} \quad \text{ and } \quad  \hat{\varphi} = \frac{1}{2} \frac{\vec{\mymathbb{1}}^\top \mf{D}^{-1} \vec{\Delta}}{\vec{\mymathbb{1}}^\top \mf{D}^{-1} \vec{\mymathbb{1}}}.
	\end{equation}
	Their variances can also be computed by using approximate covariance matrices, which gives
	\begin{equation}\label{eqn:var-qspc-estimator}
	\begin{split}
	& \mathrm{Var}\left(\hat{\theta}\right) \approx \frac{1}{4Md(2d-1)} \approx \frac{1}{8 M d^2} \quad \text{ and } \quad \mathrm{Var}\left(\hat{\varphi}\right) \approx \frac{3}{4Md(2d-1)(d^2-1)\theta^2} \approx \frac{3}{8Md^4\theta^2}.
	\end{split}
	\end{equation}
	In practice, an additional moving average filter in Ref. \cite{ShenLiu2019} can be applied to the data to further numerically boost the SNR. As a remark, considering the inference problem as linear statistical models, the estimators derived from MLEs have variances matching the Cram\'{e}r-Rao lower bound \cite{RifeBoorstyn1974}. It means the derived estimators are optimal in solving QSPC-F (\cref{prob:inference-Fourier}). For completeness, we exactly compute the optimal variance from Cram\'{e}r-Rao lower bound and discuss the optimality of the estimators in \cref{sec:lower-bound-qspc-metrology}.
	
	\subsection{Improving the estimator of swap angle using the peak information provided by $\hat{\varphi}$}
	In this subsection, we explicitly write down the dependence on $d$ as the subscript of relevant functions because $d$ is variable in the analysis.
	
	Once we have a priori $\hat{\varphi}_\mathrm{pri}$, it gives an accurate estimation of phase making $\abs{\mf{h}}$ attain its maximum, which is often referred to as the phase matching condition. The a priori phase $\hat{\varphi}_\mathrm{pri}$ can be some statistical estimator from other subroutines. For example, it can be the QSPC-F $\varphi$-estimator. By setting the phase modulation angle to $\omega = \hat{\varphi}_\mathrm{pri}$ in the QSPC circuit, we compute the amplitude of the reconstructed function for variable degrees and compute the differential signal by
	\begin{equation}
	\begin{split}
	&\{ \abs{\mf{h}_j^\expl} : j = d, d+2, d+4, \cdots, 3d \} \\
	&\Rightarrow \vec{\Gamma} := \left( \abs{\mf{h}_{d+2}^\expl} - \abs{\mf{h}_d^\expl}, \abs{\mf{h}_{d+4}^\expl} - \abs{\mf{h}_{d+2}^\expl}, \cdots, \abs{\mf{h}_{3d}^\expl} - \abs{\mf{h}_{3d-2}^\expl}\right)^\top \in \RR^d.
	\end{split}
	\end{equation}
	Let $\mf{D}$ be the $d$-by-$d$ discrete Laplacian matrix and $\mymathbb{1} := (1, 1, \cdots, 1) \in \RR^d$. The swap angle can be estimated by the statistical estimator
	\begin{equation}
	\hat{\theta}_\mathrm{pd} = \frac{1}{2} \frac{\mymathbb{1}^\top \mf{D}^{-1} \vec{\Gamma}}{\mymathbb{1}^\top \mf{D}^{-1} \mymathbb{1}}. 
	\end{equation}
	The performance guarantee of this estimator is given in the following theorem. We also discuss the case that the a priori is given by the QSPC-F estimator in the next corollary.
	\begin{theorem}\label{thm:bias-prog-diff}
		Assume an unbiased estimator $\hat{\varphi}_\mathrm{pri}$ with variance $\mathrm{Var}\left(\hat{\varphi}_\mathrm{pri}\right)$ is used as a priori. When $d\theta \le \frac{1}{9}$, the estimator $\hat{\theta}_\mathrm{pd}$ is a biased estimator with bounded bias
		\begin{equation}
		\abs{\mathrm{Bias}_\mathrm{pd}} := \abs{\expt{ \hat{\theta}_\mathrm{pd}} - \theta} \le \frac{13}{2} d^2\theta \mathrm{Var}\left( \hat{\varphi} \right) + 37 (d\theta)^3
		\end{equation}
		and variance
		\begin{equation}
		\mathrm{Var}\left( \hat{\theta}_\mathrm{pd} \right) = \frac{3}{4Md(d+1)(d+2)} \approx \frac{3}{4d^3 M}.
		\end{equation}
	\end{theorem}
	\begin{proof}
		Let the amplitude of the reconstructed function be
		\begin{equation}
		\begin{split}
			\mf{f}_d(\omega-\varphi, \theta) := \abs{\mf{h}_d(\omega;\theta,\varphi,\chi)} = \sin\theta\abs{\frac{\sin(d\sigma)}{\sin\sigma}} \sqrt{1 - \sin^2(\theta) \frac{\sin^2(d\sigma)}{\sin^2(\sigma)}},
		\end{split}
		\end{equation}
		which follows \cref{cor:modulus-magnitude-sin2theta} and $\sigma := \arccos\left(\cos(\theta)\cos(\omega-\varphi)\right)$. Furthermore, let
		\begin{equation}
			\wt{\mf{f}}^\circ_d(\omega-\varphi, \theta) := \sin\theta \frac{\sin\left(d(\omega-\varphi)\right)}{\sin(\omega-\varphi)},\quad \mf{f}^\circ_d(\omega-\varphi, \theta) := \abs{\wt{\mf{f}}^\circ_d(\omega-\varphi, \theta)}.
		\end{equation}
		Note that when $\abs{\omega-\varphi} \le \frac{\pi}{d}$, the defined function agrees with the amplitude of itself $\mf{f}^\circ_d(\omega-\varphi, \theta) = \wt{\mf{f}}^\circ_d(\omega-\varphi, \theta)$. Furthermore, for any $\omega$, we have the following bound by using triangle inequality
		\begin{equation}
			\begin{split}
				&\abs{\mf{f}_d^\circ(\omega-\varphi,\theta) - \mf{f}_d(\omega-\varphi,\theta)} \le \abs{\sin\theta \frac{\sin(d\sigma)}{\sin\sigma} \sqrt{1 - \sin^2(\theta) \frac{\sin^2(d\sigma)}{\sin^2(\sigma)}} - \sin\theta \frac{\sin\left(d(\omega-\varphi)\right)}{\sin(\omega-\varphi)}}\\
				&\le \sin\theta \abs{\frac{\sin(d\sigma)}{\sin\sigma}} \left(1 - \sqrt{1 - \sin^2(\theta) \frac{\sin^2(d\sigma)}{\sin^2(\sigma)}} \right) + \sin\theta \abs{ \frac{\sin(d\sigma)}{\sin\sigma} - \frac{\sin\left(d(\omega-\varphi)\right)}{\sin(\omega-\varphi)} }\\
				& := J_1(d) + J_2(d).
			\end{split}
		\end{equation}
		The first term can be further upper bounded by using the fact that $\max_x \abs{\frac{\sin(dx)}{\sin x}} = d$
		\begin{equation}
			J_1(d) = \frac{\sin^3\theta \abs{\frac{\sin(d\sigma)}{\sin\sigma}}^3}{1 + \sqrt{1 - \sin^2(\theta) \frac{\sin^2(d\sigma)}{\sin^2(\sigma)}}}  \le \frac{(d\theta)^3}{1 + \sqrt{1-(d\theta)^2}} \le \frac{(d\theta)^3}{1 + 2\sqrt{2}/3}
		\end{equation}
		where the last inequality uses the condition $3d\theta \le \frac{1}{3}$. The last inequality is established so that it holds for any $J_1(d), \cdots, J_1(3d)$. Note that the Chebyshev polynomial of the second kind is $U_{d-1}(\cos\sigma) = \frac{\sin(d\sigma)}{\sin\sigma}$ and it is related to the derivative of the Chebyshev polynomial of the first kind as $U_{d-1} = \frac{1}{d-1} T_{d-1}^\prime$. Using the intermediate value theorem, there exists $\xi$ in between $\cos\theta\cos(\omega-\varphi)$ and $\cos(\omega-\varphi)$ so that
		\begin{equation}
			\begin{split}
				J_2(d) &= \sin\theta\abs{U_{d-1}\left(\cos\theta\cos(\omega-\varphi)\right) - U_{d-1}\left(\cos(\omega-\varphi)\right)}\\
				&= \sin\theta \abs{U_{d-1}^\prime(\xi)}\abs{\cos(\omega-\varphi)}\left( 1 - \cos\theta \right) \le \frac{\theta^3}{2(d-1)}  \max_{-1\le x \le 1} \abs{T_{d-1}^{\prime\prime}(x)}\\
				&\le \frac{\theta^3}{2(d-1)} \frac{(d-1)^2\left((d-1)^2-1\right)}{3} \max_{-1\le x \le 1} \abs{T_{d-1}(x)} = \frac{d(d-1)(d-2)\theta^3}{6} \le \frac{(d\theta)^3}{6}.
			\end{split}
		\end{equation}
		Here, the Markov brothers' inequality (\cref{thm:Markovs-ineq}) is invoked to bound the second order derivative. Thus, the approximation error is
		\begin{equation}\label{eqn:prog-diff-err1}
			\max_{\omega \in [0,\pi]} \abs{\mf{f}_d^\circ(\omega-\varphi,\theta) - \mf{f}_d(\omega-\varphi,\theta)} \le C (d\theta)^3 \quad \text{where } C = \frac{1}{1 + 2\sqrt{2}/3} + \frac{1}{6} \approx 0.6814.
		\end{equation}
		When $\abs{\omega-\varphi} \le \frac{\pi}{d}$, the absolute value can be discarded and we can consider $\wt{\mf{f}}_d^\circ$ instead. Taking the difference of the function, it yields
		\begin{equation}\label{eqn:prog-diff-err2}
		\wt{\mf{f}}_{d+2}^\circ(\omega-\varphi, \theta) - \wt{\mf{f}}_{d}^\circ(\omega-\varphi, \theta) = 2 \sin\theta \cos\left( (d+1) (\omega-\varphi) \right).
		\end{equation}
		Let the differential signal be
		\begin{equation}
		\Gamma_d(\omega-\varphi, \theta)  := \mf{f}_{d+2}(\omega-\varphi, \theta) - \mf{f}_{d}(\omega-\varphi, \theta) = 2\theta + \delta_d(\omega-\varphi, \theta)
		\end{equation}
		where $\delta_d(\omega-\varphi, \theta)$ is the systematic error raising in the linearization of the model. Using \cref{eqn:prog-diff-err1,eqn:prog-diff-err2}, when $\abs{\omega-\varphi} \le \frac{\pi}{d+2}$, the systematic error is bounded as
		\begin{equation}\label{eqn:systematic-error-bound}
		\begin{split}
		&\abs{\delta_d(\omega-\varphi, \theta)} \le \abs{\wt{\mf{f}}_{d+2}^\circ(\omega-\varphi, \theta) - \wt{\mf{f}}_{d}^\circ(\omega-\varphi, \theta) - 2\theta} + C\theta^3\left( d^3 + (d+2)^3 \right)\\
		&\le 2\abs{\cos\left( (d+1) (\omega-\varphi) \right)}\left(\theta - \sin\theta\right) + 2\theta\left( 1 - \cos\left( (d+1) (\omega-\varphi) \right) \right) + C \theta^3\left( d^3 + (d+2)^3 \right)\\
		&\le \theta \left(d+1\right)^2\left( \omega - \varphi \right)^2 + C \theta^3\left( d^3 + (d+2)^3 \right) + 2\theta^3.
		\end{split}
		\end{equation}
		Furthermore, the differential signal is also bounded
		\begin{equation}
			\begin{split}
				&\abs{\Gamma_d(\omega-\varphi,\theta)} \le \abs{\wt{\mf{f}}_{d+2}^\circ(\omega-\varphi, \theta) - \wt{\mf{f}}_{d}^\circ(\omega-\varphi, \theta)} + C\theta^3\left( d^3 + (d+2)^3 \right)\\
				&= 2\sin\theta \abs{\cos\left( (d+1) (\omega-\varphi) \right)} + C\theta^3\left( d^3 + (d+2)^3 \right) \le 2\theta + C\theta^3\left( d^3 + (d+2)^3 \right).
			\end{split}
		\end{equation}
		In the experimental implementation, we perform the QSPC circuit with $\omega = \hat{\varphi}_\mathrm{pri}$ and degree $d, d+2, d+4, \cdots, 3d$. The resulted dataset contains $\left\{ \mf{f}_j^\expl := \abs{\mf{h}^\expl_j} : j = d, d+2, \cdots, 3d \right\}$ and the differential signal can be computed respectively
		\begin{equation}\label{eqn:prog-diff-lm}
		\Gamma_j^\expl := \mf{f}_{j+2}^\expl - \mf{f}_j^\expl = \Gamma_j(\hat{\varphi}-\varphi, \theta) + w_{j+2} - w_j = 2\theta + \delta_j(\hat{\varphi}-\varphi, \theta) + w_{j+2} - w_j
		\end{equation}
		where $w_j := \mf{f}_j^\expl - \mf{f}_j(\hat{\varphi} - \varphi)$ is the noise of the sampled data. When the SNR is large, Ref. \cite{Tretter1985} suggests the noise can be approximated by the real component of the noise on the complex-valued data $\mf{h}_j^\expl$. Analyzed in the proof of \cref{lma:monte-carlo-error-magnitude}, the variance of the noise concentrates around a constant
		\begin{equation}
		\expt{w_j} = 0 \quad \text{ and } \quad \frac{1}{4M} - \frac{(j\theta)^2}{M} \le \mathrm{Var}(w_j) \le \frac{1}{4M}.
		\end{equation}
		Assume $3d\theta \ll 1$, the covariance matrix of the colored noise $w_{j+2} - w_j$ is well approximated by a constant matrix
		\begin{equation}
		\expt{\left(w_{d + 2(j+1)} - w_{d + 2j}\right)\left(w_{d + 2(k+1)} - w_{d + 2k}\right)} \approx \frac{1}{4M} \mf{D}_{j,k}.
		\end{equation}
		Let the data vector be 
		\begin{equation}
		\vec{\Gamma} = \left( \Gamma_d^\expl, \Gamma_{d+2}^\expl, \cdots, \Gamma_{3d-2}^\expl \right)^\top \in \RR^d
		\end{equation}
		and the systematic error vector be
		\begin{equation}
		\vec{\delta}(\hat{\varphi}-\varphi, \theta) = \left( \delta_d(\hat{\varphi}-\varphi, \theta), \delta_{d+2}(\hat{\varphi}-\varphi, \theta), \cdots, \delta_{3d-2}(\hat{\varphi}-\varphi, \theta) \right)^\top \in \RR^d.
		\end{equation}
		The statistical estimator solving the linearized problem of \cref{eqn:prog-diff-lm} is
		\begin{equation}
		\hat{\theta}_\mathrm{pd} = \frac{1}{2} \frac{\mymathbb{1}^\top \mf{D}^{-1} \vec{\Gamma}}{\mymathbb{1}^\top \mf{D}^{-1} \mymathbb{1}}. 
		\end{equation}
		 According to Ref. \cite{Kay1989}, the matrix-multiplication form can be exactly represented as a convex combination: for any $d$-dimensional vector $\vec{X} = (X_0, \cdots, X_{d-1})^\top$
		\begin{equation}
			\frac{\mymathbb{1}^\top \mf{D}^{-1} \vec{X}}{\mymathbb{1}^\top \mf{D}^{-1} \mymathbb{1}} = \sum_{k=0}^{d-1} \mu_k X_k
		\end{equation}
		where
		\begin{equation}
		\mu_k := \frac{\frac{3}{2} (d+1)}{(d+1)^2-1} \left( 1 - \left( \frac{k - \frac{d-1}{2}}{\frac{d+1}{2}}\right)^2 \right) > 0 \text{ and } \sum_{k=0}^{d-1} \mu_k = 1.
		\end{equation}
		The variance of the estimator is
		\begin{equation}
		\mathrm{Var}\left( \hat{\theta}_\mathrm{pd} \right) = \frac{1}{4} \frac{1}{4M} \frac{1}{\mymathbb{1}^\top \mf{D}^{-1} \mymathbb{1}} = \frac{3}{4Md(d+1)(d+2)} \approx \frac{3}{4d^3 M}.
		\end{equation}
		The conditional mean of the estimator is bounded as
		\begin{equation}
			\abs{\expt{\hat{\theta}_\mathrm{pd} \bigg| \hat{\varphi}_\mathrm{pri}}} = \abs{\frac{1}{2} \sum_{k=0}^{d-1} \mu_k \Gamma_{d+2k}(\hat{\varphi}_\mathrm{pri}-\varphi,\theta)} \le \frac{1}{2} \max_{k=0,\cdots,d-1} \abs{\Gamma_{d+2k}(\hat{\varphi}_\mathrm{pri}-\varphi,\theta)} \le \theta + C(3d\theta)^3.
		\end{equation}
		To make the bound in \cref{eqn:systematic-error-bound} justified, we first assume that $\abs{\hat{\varphi}_\mathrm{pri}-\varphi} \le \frac{\pi}{3d}$. Invoking Chebyshev's inequality, the assumption fails with probability
		\begin{equation}
		\bP\left(\abs{\hat{\varphi}_\mathrm{pri}-\varphi} > \frac{\pi}{3d}\right) \le \frac{9d^2}{\pi^2}\mathrm{Var}\left(\hat{\varphi}_\mathrm{pri}\right) \le d^2 \mathrm{Var}\left(\hat{\varphi}_\mathrm{pri}\right).
		\end{equation}
		When $\abs{\hat{\varphi}_\mathrm{pri}-\varphi} \le \frac{\pi}{3d}$, the conditional expectation of the estimator is
		\begin{equation}
		\expt{\hat{\theta}_\mathrm{pd} \mathds{1}_{\abs{\hat{\varphi}_\mathrm{pri}-\varphi} \le \frac{\pi}{3d}} \bigg| \hat{\varphi}_\mathrm{pri}} = \left(\theta + \frac{1}{2} \frac{\mymathbb{1}^\top \mf{D}^{-1} \vec{\delta}(\hat{\varphi}_\mathrm{pri}-\varphi, \theta)}{\mymathbb{1}^\top \mf{D}^{-1} \mymathbb{1}}\right)\mathds{1}_{\abs{\hat{\varphi}_\mathrm{pri}-\varphi} \le \frac{\pi}{3d}}. 
		\end{equation}
		Invoking \cref{eqn:systematic-error-bound}, when $\abs{\hat{\varphi}_\mathrm{pri}-\varphi} \le \frac{\pi}{3d}$, the bias of the estimator is bounded as
		\begin{equation}
		\begin{split}
		&\abs{\expt{\left(\hat{\theta}_\mathrm{pd} - \theta\right)\mathds{1}_{\abs{\hat{\varphi}_\mathrm{pri}-\varphi} \le \frac{\pi}{3d}}}} = \abs{\expt{\expt{\left(\hat{\theta}_\mathrm{pd} - \theta\right)\mathds{1}_{\abs{\hat{\varphi}_\mathrm{pri}-\varphi} \le \frac{\pi}{3d}} \bigg| \hat{\varphi}_\mathrm{pri}}}}\\
		&= \frac{1}{2}\abs{\sum_{k=0}^{d-1} \mu_k \expt{\delta_{d+2k}(\hat{\varphi}_\mathrm{pri}-\varphi, \theta)\mathds{1}_{\abs{\hat{\varphi}_\mathrm{pri}-\varphi} \le \frac{\pi}{3d}}}} \le \frac{1}{2}\max_{k=0,\cdots,d-1} \expt{\abs{\delta_{d+2k}(\hat{\varphi}_\mathrm{pri}-\varphi, \theta)}} \\
		&\le \frac{1}{2} \theta\left(3d-1\right)^2 \mathrm{Var}\left( \hat{\varphi}_\mathrm{pri} \right) + C\theta^3\left( \frac{1}{C} + \frac{(3d-2)^3 + (3d)^3}{2} \right) \le \frac{1}{2}\theta (3d)^2 \mathrm{Var}\left( \hat{\varphi}_\mathrm{pri} \right) + C(3d\theta)^3.
		\end{split}
		\end{equation}
		On the other hand, when $\abs{\hat{\varphi}_\mathrm{pri}-\varphi} > \frac{\pi}{3d}$, the bias of the estimator is bounded as
		\begin{equation}\label{eqn:bias-upper-bound-outside}
		\begin{split}
			&\abs{\expt{\left(\hat{\theta}_\mathrm{pd} - \theta\right)\mathds{1}_{\abs{\hat{\varphi}_\mathrm{pri}-\varphi} > \frac{\pi}{3d}}}} =  \abs{\expt{\left(\expt{\hat{\theta}_\mathrm{pd} \bigg| \hat{\varphi}_\mathrm{pri}} - \theta\right)\mathds{1}_{\abs{\hat{\varphi}_\mathrm{pri}-\varphi} > \frac{\pi}{3d}}}}\\
			&\le \left(2\theta + C(3d\theta)^3\right) \bP\left(\abs{\hat{\varphi}_\mathrm{pri}-\varphi} > \frac{\pi}{3d}\right) \le 2d^2\theta \mathrm{Var}\left(\hat{\varphi}_\mathrm{pri}\right) + C (3d\theta)^3.
		\end{split}
		\end{equation}
		Combining these two cases and using triangle inequality, the bias is bounded as
		\begin{equation}
		\begin{split}
			\abs{\mathrm{Bias}_\mathrm{pd}} &\le \abs{\expt{\left(\hat{\theta}_\mathrm{pd} - \theta\right)\mathds{1}_{\abs{\hat{\varphi}_\mathrm{pri}-\varphi} \le \frac{\pi}{3d}}}} + \abs{\expt{\left(\hat{\theta}_\mathrm{pd} - \theta\right)\mathds{1}_{\abs{\hat{\varphi}_\mathrm{pri}-\varphi} > \frac{\pi}{3d}}}}\\
			&\le \frac{13}{2} d^2\theta \mathrm{Var}\left( \hat{\varphi}_\mathrm{pri} \right) + 37 (d\theta)^3.
		\end{split}
		\end{equation}
		Here, we use $54C \le 37$ to simplify the preconstant. The proof is completed.
	\end{proof}
	\begin{corollary}\label{cor:bias-prog-diff-QSPC-F}
		When $\hat{\varphi}_\mathrm{pri} = \hat{\varphi}$ is the QSPC-F $\varphi$-estimator in \cref{def:estimator-qsp-pc}, the bias of the estimator is bounded as
		\begin{equation}
			\abs{\mathrm{Bias}_\mathrm{pd}} \le \frac{39}{16 d^2 M \theta} + \frac{7 d\theta}{M} + 19 \left(d\theta\right)^3.
		\end{equation}
	\end{corollary}
	\begin{proof}
		The upper bound follows the substitution $\mathrm{Var}\left(\hat{\varphi}\right) \approx \frac{3}{8 d^4\theta^2 M}$. Furthermore, the second term comes from the refinement in the upper bound in \cref{eqn:bias-upper-bound-outside}
		\begin{equation}
			C (3d\theta)^3 \bP\left(\abs{\hat{\varphi}-\varphi} > \frac{\pi}{3d}\right) \le C (3d\theta)^3 d^2 \mathrm{Var}\left(\hat{\varphi}\right) \le \frac{81 C d \theta}{8 M} \le \frac{7d\theta}{M}.
		\end{equation}
	\end{proof}
	the analysis in this section indicates that trusting the a priori phase as the ``peak'' location and estimating $\theta$ from the differential signal at the ``peak'' will unavoidably introduce bias to the $\theta$-estimator. Unless the a priori is deterministic and is exactly equal to $\varphi$, the ``peak'' is not the exact peak even subjected to the controllable statistical fluctuation of $\hat{\varphi}_\mathrm{pri}$. Hence, it suggests that we need to interpret the a priori $\hat{\varphi}_\mathrm{pri}$ as an estimated peak location which is close to the exact peak location $\varphi$. This gives rise to the regression-based methods in the next subsection.
	
	\subsection{Peak regression and peak fitting}
	In order to circumvent the over-confident reliance on the a priori guess of $\varphi$, the method can be improved by regressing distinct samples with respect to analytical expressions on the unknown angle parameters. Suppose $n$ samples are made with $\{(\omega_j, d_j, \mf{h}^{\expl, j}) : j = 1, \cdots, n\}$. One can consider perform a nonlinear regression on the data to infer the unknown parameters, which is given by the following minimization problem
	\begin{equation}
	    \hat{\theta}_\mathrm{pr}, \hat{\varphi}_\mathrm{pr}, \hat{\chi}_\mathrm{pr} = \myargmin_{\theta, \varphi, \chi} \sum_{j = 1}^n \abs{\mf{h}_{d_j}(\omega_j; \theta,\varphi,\chi) - \mf{h}^{\expl, j}}^2.
	\end{equation}
	When the number of additional samples $n$ is large enough, the estimator derived from the minimization problem is expected to be unbiased and the variance scales as $\Or(1/(d^2nM))$ according to the M-estimation theory \cite{KeenerTheoreticalStatistics2010}. However, the practical implementation of these estimators is easily affected by the complex landscape of the minimization problem. Meanwhile, the sub-optimality and the run time of black-box optimization algorithms also limits the use of these estimator.
	
	To overcome the difficulty due to the complex landscape of nonlinear regression, we propose another technique to improve the accuracy of the swap-angle estimator by fitting the peak of the amplitude function $\mf{f}_d(\omega-\varphi, \theta)$. We observe that the amplitude function is well captured by a parabola on the interval $\mc{I} := \left[ \varphi - \frac{\pi}{2d}, \varphi + \frac{\pi}{2d} \right]$. Consider $n_\mathrm{pf}$ equally spaced sample points on the interval $\mc{I}$: $\omega_j^\mathrm{(pf)} = \hat{\varphi}_\mathrm{pri} + \frac{\pi}{d} \left(\frac{j}{n_\mathrm{pf}-1} - \frac{1}{2}\right)$ where $j = 0, 1, \cdots, n_\mathrm{pf}-1$. We find the best parabola fitting the sampled data $\mf{f}_d^\expl\left(\omega_j^\mathrm{(pf)} \right)$ whose maximum $\mf{f}_d^\mathrm{(pf\ max)}$ attains at $\omega^\mathrm{(pf\ max)}$. Given that $\hat{\varphi}_\mathrm{pri}$ is an accurate estimator of the angle $\varphi$, we accept the parabolic fitting result if the peak location does not deviate $\hat{\varphi}_\mathrm{pri}$ beyond some threshold $\varepsilon^\mathrm{thr}$, namely, the fitting is accepted if $\abs{\omega^\mathrm{(pf\ max)} - \hat{\varphi}_\mathrm{pri}} < \varepsilon^\mathrm{thr}$. Upon the acceptance, the estimator is $\hat{\theta}_\mathrm{pf} := \mf{f}_d^\mathrm{(pf\ max)} / d$. Ignoring the systematic bias caused by the overshooting of $\hat{\varphi} \neq \varphi$, the variance of the estimator is approximately $\Or\left( \frac{1}{d^2 n_\mathrm{pf}} \right)$. The detailed procedure is given in \cref{alg:qspc-peak-fitting}.
	
	\subsection{Numerical performance of QSPC against Monte Carlo sampling error}\label{subsec:num-result-MC}
\begin{figure}[htbp]
		\centering
		\includegraphics[width=\textwidth]{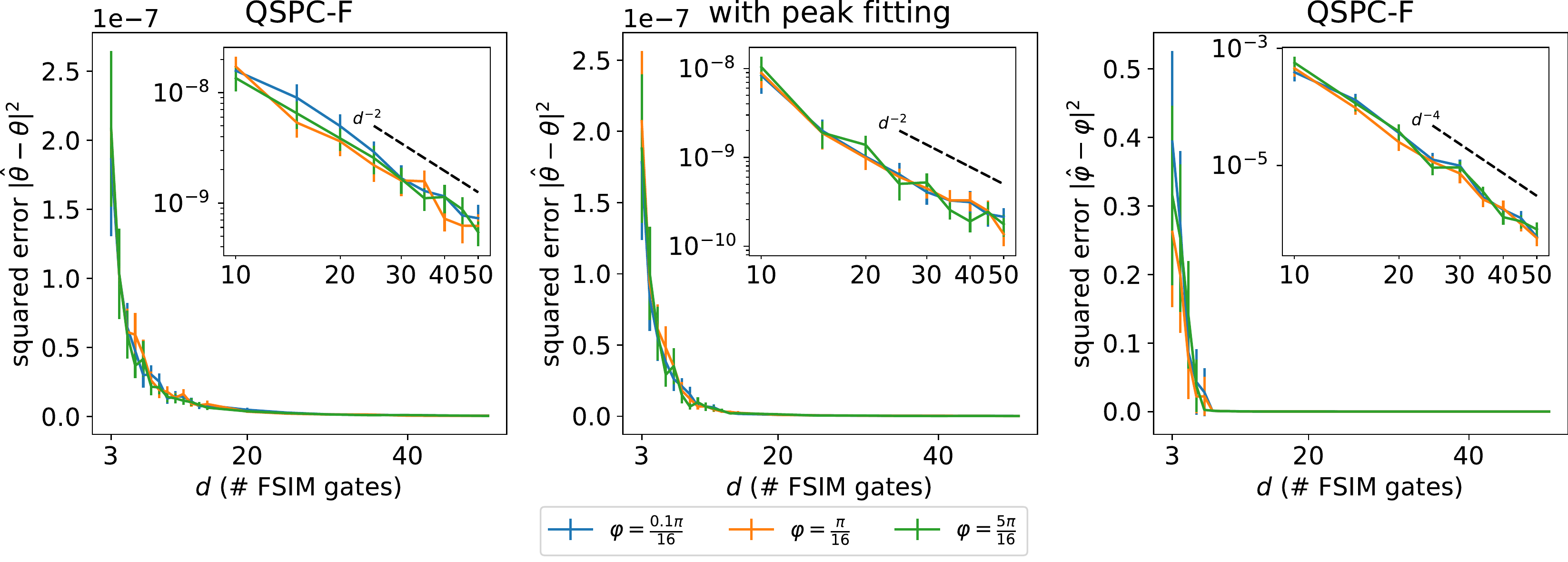}
		\caption{Squared error of estimators as a function of the number of \fsim s. The only source of noise in the numerical experiments is Monte Carlo sampling error. The number of measurement samples is $M = 1\times 10^5$, and $n_\mathrm{pf} = 15$ is used in the peak fitting. The swap angle is set to $\theta = 1 \times 10^{-3}$ and the  phase parameter is set to $\chi = 5\pi/32$. The error bar of each point stands for the confidence interval derived from $96$ independent repetitions.}
		\label{fig:degree_mc}
	\end{figure}

	To numerically test the performance of QSPC and justify the analysis in the presence of Monte Carlo sampling error, we simulate the quantum circuit and perform the inference. In \cref{fig:degree_mc}, we plot the squared error of each estimator as a function of the number of \fsim s $d$ in each quantum circuit. Consequentially, each data point is the mean squared error (MSE), which is a metric of the performance according to the bias-variance decomposition $\mathrm{MSE} = \mathrm{Var} + \mathrm{bias}^2$. The numerical results in \cref{fig:degree_mc} indicates that although $\theta = 1\times 10^{-3}$ is small, QSPC-F estimators achieve an accurate estimation with a very small $d$. The numerical results also justify that the performance of the estimator does not significantly depend on the value of the single-qubit phase $\varphi$. Meanwhile, using the peak fitting in \cref{alg:qspc-peak-fitting}, the variance in $\theta$-estimation is improved so that the MSE curve is lowered. Zooming the MSE curve in log-log scale, the curve scales as a function of $d$ as the theoretically derived variance scaling in \cref{prop:variance-qsp-pc-fourier}. We will discuss the scaling of the variance in \cref{sec:lower-bound-qspc-metrology} in more details.
	
\begin{figure}[htbp]
		\centering
		\includegraphics[width=\textwidth]{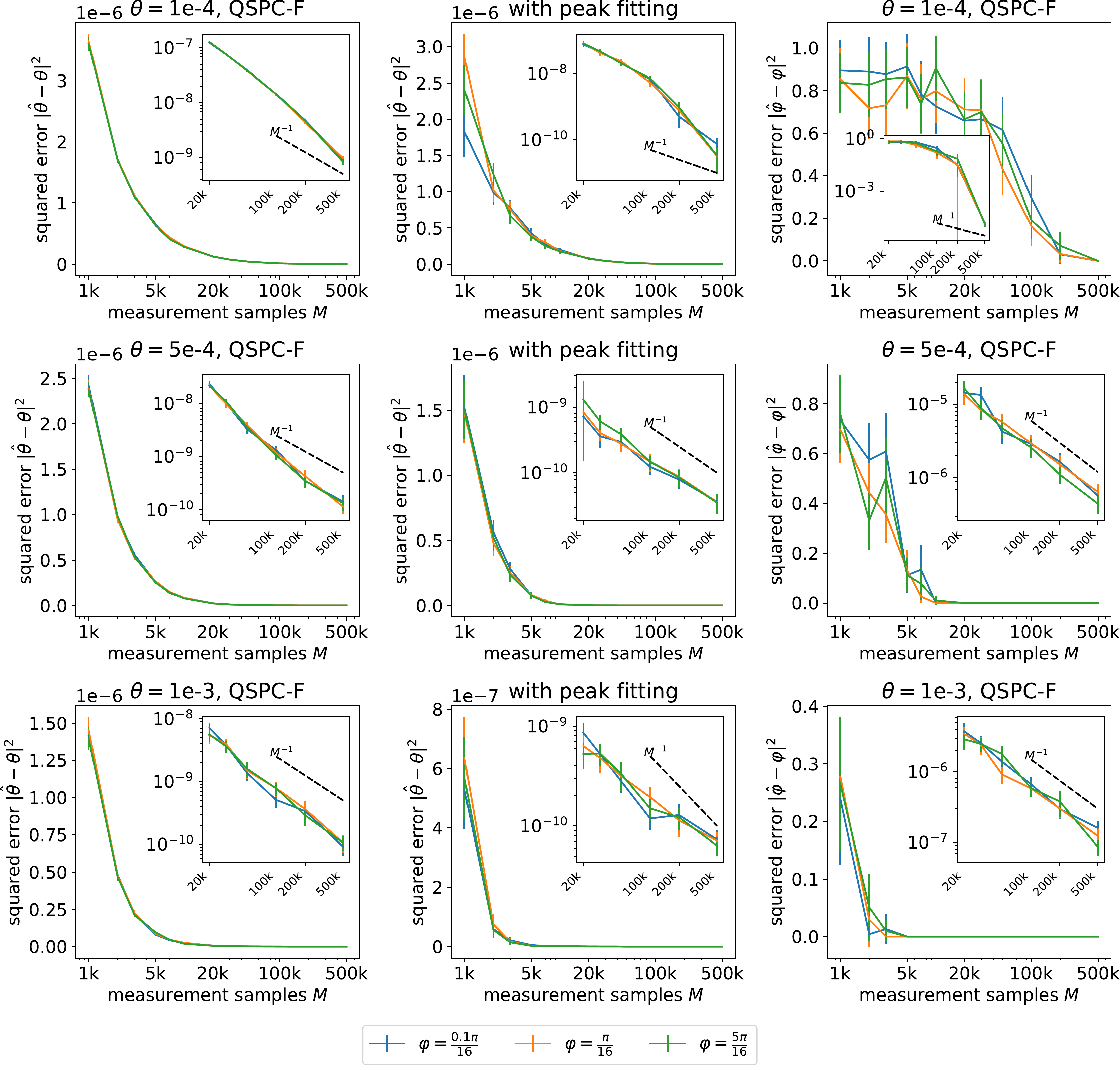}
		\caption{Squared error of estimators as a function of the number of measurement samples. The only source of noise in the numerical experiments is Monte Carlo sampling error. The circuit degree is set to $d = 50$ and the \fsim\ phase parameter is set to $\chi = 5\pi/32$. $n_\mathrm{pf} = 15$ is used in the peak fitting. The error bar of each point stands for the confidence interval derived from $96$ independent repetitions.}
		\label{fig:meas_mc}
	\end{figure}

	In \cref{fig:meas_mc}, we perform the numerical simulation with variable swap angle $\theta$ and number of measurement samples $M$. The numerical results show that the accuracy of $\varphi$-estimation is more vulnerable to decreasing $\theta$. This is explainable from the theoretically derived variance in \cref{prop:variance-qsp-pc-fourier} which depends on the swap angle as $1/\theta^2$. Although the theoretical variance of $\theta$ is expected to be invariant for different $\theta$ values, the numerical results show that the MSE of $\theta$-estimation gets larger when smaller $\theta$ is used, and the scaling of the curve differs from the classical scaling $1/M$. The reason is that when $\theta \le 5\times 10^{-4}$, the SNR is not large enough so that the theoretical derivation can be justified. When using a bigger $d$ or $M$, the curve will converge to the theoretical derivation. When $\theta = 1\times 10^{-3}$, the setting of the experiments is enough to get a large enough SNR. Hence, the scaling of the MSE curves in the bottom panels in \cref{fig:meas_mc} agrees with the classical scaling $1/M$ of Monte Carlo sampling error.
	
	\section{Lower bounding the performance of quantum metrology for QSPC}\label{sec:lower-bound-qspc-metrology}
	In the designed calibration algorithm, gate parameters are estimated from experimental data by running $2(2d-1)$ quantum circuits whose depths are $\Theta(d)$. If we simply think under the philosophy of the Heisenberg limit of quantum metrology in Ref. \cite{Lloyd2006}, we would expect the variance of statistical estimators bounded from below as 
	$$\Omega\left(1/\left((\text{classical repetition})\times(\text{quantum repetition})^2)\right)\right) = \Omega(1/d^3)$$
	when $d$ is large enough. However, theoretical analysis in \cref{prop:variance-qsp-pc-fourier} and numerical simulation in \cref{fig:degree_mc} show that the variance of the $\varphi$-estimator in QSPC-F depends on the parameter $d$ as $\mathrm{Var}\left(\hat{\varphi}\right) \sim 1/d^4$. In this section, we will analyze this nontrivial counterintuitive result. In the end, we prove that for a fixed unknown \fsim, the $1/d^4$-dependency only appears in the pre-asymptotic regime where the condition of the theorems holds, i.e., $d \theta \ll 1$. When passing to the limit of large enough $d$, the variances of statistical estimators agree with that suggested by the Heisenberg limit. Although such faster than Heisenberg limit scaling only applies in a finite range of circuit depth~($d \theta \ll 1$), it has drastically increased our metrology performance in practice against time-dependent errors, and thus deserves further investigation in its generalization to other domains of noise learning.
	
	\subsection{Pre-asymptotic regime $d \ll 1/\theta$}\label{subsec:CRLB-pre-asymptotic}
	We derive the optimal variance scaling permitted using our metrology method in finite circuit depth, i.e. pre-asymptotic regime in this subsection. More particularly, we require that for a given range of gate parameter $\theta \in [\theta_{\text{min}},\theta_{\text{max}}] $, our metrology circuit depth obeys: $d \ll 1/\theta_{\text{min}}$ in the pre-asymptotic regime. This also implies that for any $\theta$ under the consideration we have $d\theta \ll 1$.
	
	The quantum circuits in QSPC form a class of parametrized quantum circuits whose measurement probabilities are trigonometric polynomials in a tunable variable $\omega$. For simplicicty, the gate parameters of the unknown \fsim\ is denoted as $\Xi = (\xi_k) = (\theta, \varphi, \chi)$. According to the modeling of Monte Carlo sampling error in \cref{lma:monte-carlo-error-magnitude}, the experimentally estimated probabilities are approximately normal distributed. Given the normality and assuming the limit $M \gg 1$, the element of the Fisher information matrix is
	\begin{equation}\label{eqn:Fisher-information-element-definition}
		I_{kk^\prime}(\Xi) = \sum_{j=0}^{2d-2} \Sigma_{X, j}^{-2} \frac{\partial p_X(\omega_j; \Xi)}{\partial \xi_k}\frac{\partial p_X(\omega_j; \Xi)}{\partial \xi_{k^\prime}} + \sum_{j=0}^{2d-2} \Sigma_{Y, j}^{-2} \frac{\partial p_Y(\omega_j; \Xi)}{\partial \xi_k}\frac{\partial p_Y(\omega_j; \Xi)}{\partial \xi_{k^\prime}}.
	\end{equation}
	According to \cref{eqn:variance-mc-error-concentration}, the variance of the Monte Carlo sampling error concentrates near a constant. Hence
	\begin{equation}
	I_{kk^\prime}(\Xi) = 4M\left(1 + \Or(d^2\theta^2)\right) \sum_{j=0}^{2d-2} \left(\frac{\partial p_X(\omega_j; \Xi)}{\partial \xi_k}\frac{\partial p_X(\omega_j; \Xi)}{\partial \xi_{k^\prime}} + \frac{\partial p_Y(\omega_j; \Xi)}{\partial \xi_k}\frac{\partial p_Y(\omega_j; \Xi)}{\partial \xi_{k^\prime}}\right).
	\end{equation}
	Using the reconstructed function, the element of the Fisher information matrix can be expressed as
	\begin{align}
		I_{kk^\prime}(\Xi) &= 4M\left(1 + \Or(d^2\theta^2)\right) \Re\left(\sum_{j=0}^{2d-2} \frac{\partial \mf{h}(\omega_j; \Xi)}{\partial \xi_k}\frac{\partial \overline{\mf{h}(\omega_j; \Xi)}}{\partial \xi_{k^\prime}}\right)\\
		&= 4M(2d-1)\left(1 + \Or(d^2\theta^2)\right) \Re\left(\sum_{j=-d+1}^{d-1} \frac{\partial c_j(\Xi)}{\partial \xi_k}\frac{\partial \overline{c_j(\Xi)}}{\partial \xi_{k^\prime}}\right) \label{eqn:Fisher-information-Fourier-coefficients}\\
		&= \frac{4M(2d-1)}{\pi} \left(1 + \Or(d^2\theta^2)\right) \Re\left(\int_{-\pi/2}^{\pi/2} \frac{\partial \mf{h}(\omega; \Xi)}{\partial \xi_k}\frac{\partial \overline{\mf{h}(\omega; \Xi)}}{\partial \xi_{k^\prime}} \rd \omega\right). \label{eqn:Fisher-information-integral}
	\end{align}
	Here, we use the construction of QSPC in which the tunable angles are equally spaced in one period of the reconstructed function. The second equality (\cref{eqn:Fisher-information-Fourier-coefficients}) invokes \cref{thm:structure-of-qsp-pc} and the discrete orthogonality of Fourier factors. The last equality (\cref{eqn:Fisher-information-integral}) is due to the Parseval's identity. 
	
	When $d\theta \ll 1$ and $\theta \ll 1$, the Fourier coefficients are well captured by the approximation in \cref{thm:structure-of-qsp-pc} which gives $c_j(\Xi) \approx \I e^{-\I \chi} e^{-\I(2j+1)\varphi} \theta \bI_{j \ge 0}$. Consequentially, using \cref{eqn:Fisher-information-Fourier-coefficients}, in the pre-asymptotic regime $d \ll 1/\theta$, the Fisher information matrix is approximately
	\begin{equation}
	    I(\Xi) \approx 4M(2d-1) \left(\begin{array}{ccc}
	        d & 0 & 0 \\
	        0 & \frac{d(4d^2-1)}{3}\theta^2 & d^2\theta^2\\
	        0 & d^2\theta^2 & d \theta^2
	    \end{array}\right).
	\end{equation}
	Invoking Cram\'{e}r-Rao bound, the covariance matrix of any statistical estimator is lower bounded as
	\begin{equation}
	   \mathrm{Cov}\left(\hat{\theta}_\text{any}, \hat{\varphi}_\text{any}, \hat{\chi}_\text{any}\right) \succeq I^{-1}(\Xi) \approx \frac{1}{4Md(2d-1)} \left(\begin{array}{ccc}
	        1 & 0 & 0 \\
	        0 & \frac{3}{(d^2-1)\theta^2} & -\frac{3d}{(d^2-1)\theta^2}\\
	        0 & -\frac{3d}{(d^2-1)\theta^2} & \frac{4d^2-1}{(d^2-1)\theta^2}
	    \end{array}\right).
	\end{equation}
	Consequentially, in the pre-asymptotic regime, the optimal variances of the statistical estimator are 
	\begin{align}
	    &\mathrm{Var}\left(\hat{\theta}_\text{opt}\right) = \frac{1}{4Md(2d-1)} \approx \frac{1}{8 M d^2},\label{eqn:opt-theta-var}\\
	    &\mathrm{Var}\left(\hat{\varphi}_\text{opt}\right) = \frac{3}{4Md(2d-1)(d^2-1)\theta^2} \approx \frac{3}{8Md^4\theta^2},\label{eqn:opt-varphi-var}\\
	    &\mathrm{Var}\left(\hat{\chi}_\text{opt}\right) = \frac{1}{4Md(2d-1)\theta^2} \frac{4d^2-1}{d^2-1} \approx \frac{1}{2Md^2\theta^2}.\label{eqn:opt-chi-var}
	\end{align}
	
	Remarkably, the variances of QSPC-F estimators in \cref{prop:variance-qsp-pc-fourier} exactly match the optimality given in \cref{eqn:opt-theta-var,eqn:opt-varphi-var}. We thus proves the optimality of our QSPC-F estimator for inferring gate parameter $\theta$ and $\varphi$. Moreover, we like to point out that the faster than Heisenberg-limit scaling of parameter $\varphi$ in this asymptotic regime is critical to the successful  experimental deployment of our methods. This is because the dominant time-dependent error results in a   time-dependent drift error in $\varphi$, and a faster convergence in circuit depth provides faster metrology runtime to minimize such drift error during the measurements. 
	
	\subsection{Asymptotic regime $d \to \infty$}
	Thinking under the framework of Heisenberg limit in Ref. \cite{Lloyd2006}, for a fixed $\theta$, the optimal variances of $\theta$ and $\varphi$ estimators are expected to scale as $1/d^3$ while that of $\chi$ estimator scales as $1/d$ due to the absence of amplification in the quantum circuit. In contrast to these scalings, we show in the last subsection that the scalings of $\varphi$ and $\chi$ estimators can achieve $1/d^4$ and $1/d^2$ in the pre-asymptotic regime $d \ll 1/\theta$. In this subsection, we will argue that the scalings predicted by the Heisenberg scaling hold if further passing to the asymptotic limit $d \to \infty$. As a consequence, there is a nontrivial transition of variance scalings of QSPC-F estimators in pre-asymptotic regime and the asymptotic regime. We demonstrate such subtle transition in the fundamental   efficiency allowed for the given metrology protocol with both numerical simulation and analytic reasoning in this  section.
	
	As $d \to \infty$, the measurement probabilities no longer admit the property of concentration around constants. Using the variance derived in \cref{eqn:variance-mc-error-concentration}, the diagonal element of Fisher information matrix is exactly equal to
	\begin{equation}
	\begin{split}
	    &I_{kk}(\Xi) = M \sum_{j=0}^{2d-2} \left(\frac{1}{p_X(\omega_j; \Xi)\left(1-p_X(\omega_j; \Xi)\right)} \frac{\partial p_X(\omega_j; \Xi)}{\partial \xi_k}\frac{\partial p_X(\omega_j; \Xi)}{\partial \xi_k}\right.\\
	    &\hspace*{5em}\ignorespaces\left.+ \frac{1}{p_Y(\omega_j; \Xi)\left(1-p_Y(\omega_j; \Xi)\right)} \frac{\partial p_Y(\omega_j; \Xi)}{\partial \xi_k}\frac{\partial p_Y(\omega_j; \Xi)}{\partial \xi_k}\right)\\
	    &= M \sum_{j=0}^{2d-2} \left(- \frac{\partial \log p_X(\omega_j; \Xi)}{\partial \xi_k}\frac{\partial \log\left(1- p_X(\omega_j; \Xi)\right)}{\partial \xi_k} - \frac{\partial \log p_Y(\omega_j; \Xi)}{\partial \xi_k}\frac{\partial \log\left(1- p_Y(\omega_j; \Xi)\right)}{\partial \xi_k} \right).
	\end{split}
	\end{equation}
Moreover $p_X(\omega_j; \Xi)$ and $p_Y(\omega_j; \Xi)$ are trigonometric polynomials in $\theta$ and $\varphi$ of degree at most $d$ while in $\chi$ of degree $1$ due to the absence of amplification. Therefore the log-derivatives of $\theta$ and $\varphi$ are $\Or(d)$ in most regular cases while they are $\Or(1)$ for $\chi$. Hence, we expect from the Cram\'{e}r-Rao bound that
	\begin{equation}\label{eqn:crlb-asymptotic}
	    \mathrm{Var}\left(\hat{\theta}_\mathrm{opt}\right), \mathrm{Var}\left(\hat{\varphi}_\mathrm{opt}\right) = \Omega\left(\frac{1}{d^3}\right),\quad \text{and } \mathrm{Var}\left(\hat{\chi}_\mathrm{opt}\right) = \Omega\left(\frac{1}{d}\right) \quad \text{as } d \to \infty.
	\end{equation}
	These results match the scalings predicted by the Heisenberg limit which holds in the asymptotic limit $d \to \infty$.
	
	\subsection{Numerical results}
	
	We compute the Cram\'{e}r-Rao lower bound (CRLB) of the statistical inference problem defined by QSPC. The lower bound is given by the diagonal element of inverse Fisher information matrix
	\begin{equation}
	    \mathrm{CRLB}\left(\hat{\xi}_k\right) = \left(I^{-1}(\Xi)\right)_{kk}
	\end{equation}
	where the Fisher information matrix is element-wisely defined in \cref{eqn:Fisher-information-element-definition}. At the same time, we also compute the approximation to the optimal variance in the pre-asymptotic regime $d \ll 1/\theta$ derived in \cref{eqn:opt-theta-var,eqn:opt-varphi-var,eqn:opt-chi-var}. The numerical results are given in \cref{fig:exact_crlb-qspcf}. It can be seen that the approximated optimal variance agrees very well with the exact CRLB. In the asymptotic regime with large enough $d$, the optimal variance scaling given by the CRLB is as predicted in \cref{eqn:crlb-asymptotic}. Furthermore, the numerical results justify that there exists a nontrivial transition around $d \approx 1/\theta$ making the optimal variance scalings completely different in the pre-asymptotic and asymptotic regime.
	
	\begin{figure}[htbp]
	    \centering
	    \includegraphics[width=\textwidth]{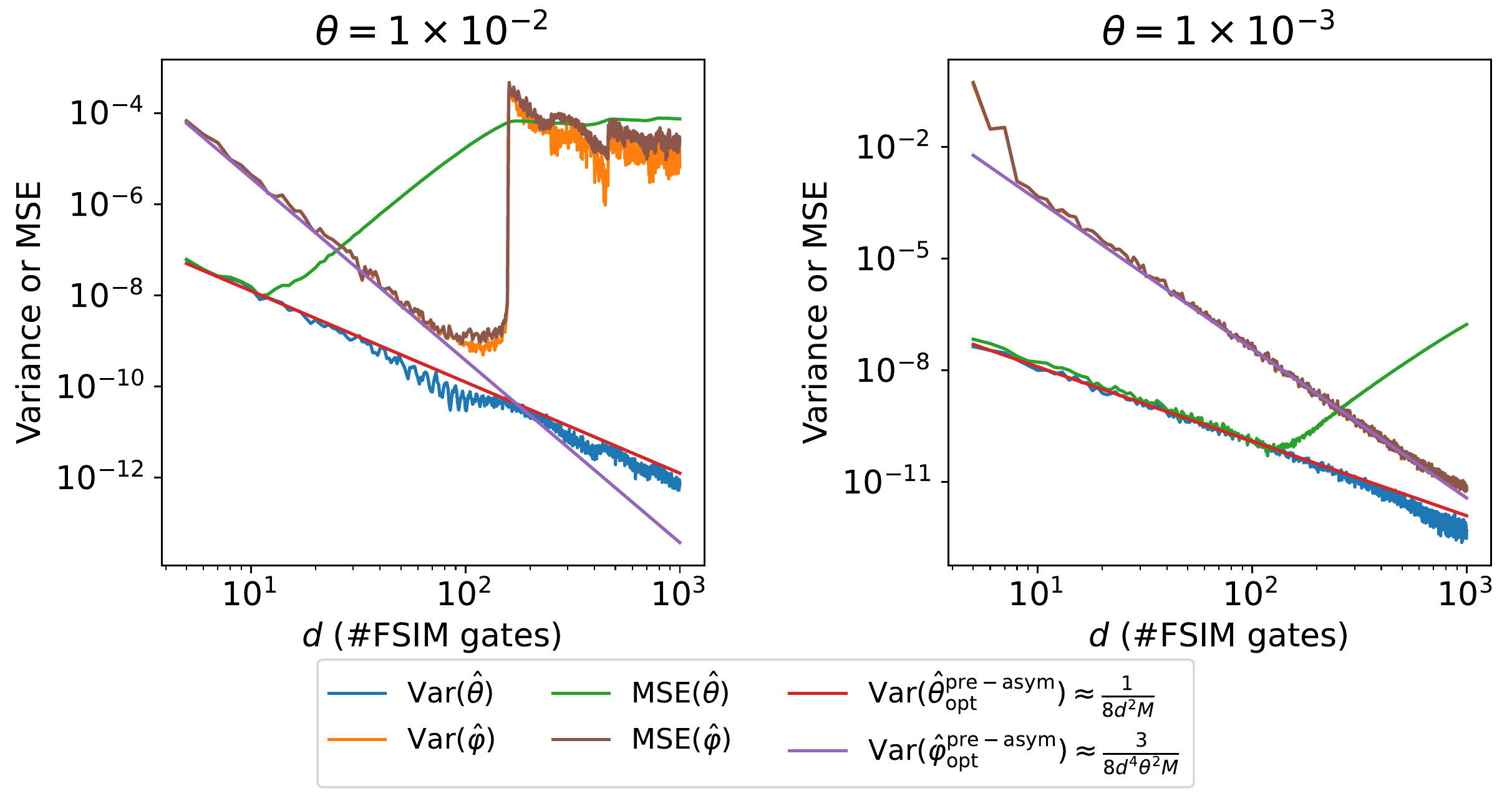}
	    \caption{Variance and mean-square error (MSE) of QSPC-F estimators. The left panel corresponds to the case where $\theta$ is relatively large and $d\theta \ll 1$ condition fails quickly at around $d=10$, beyond which bias dominates the estimator's  MSE since our inference model assumption~($d\theta \ll 1$) fails. The right panel corresponds to the case where $d\theta \ll 1$ condition holds all the way to around $d=100$.  The single-qubit phases are set to $\varphi = \pi/16$ and $\chi = 5\pi/32$. The number of measurement samples is set to $M = 1\times10^5$. Each data point is derived from $100$ independent repetitions.}
	    \label{fig:variance-qspcf}
	\end{figure}
	
	To justify the optimality of QSPC-F and investigate the situation where the conditions for deriving QSPC-F hold, we numerically estimate the variances of QSPC-F estimators and compare them with the derived optimal variances in the pre-asymptotic regime in \cref{eqn:opt-theta-var,eqn:opt-varphi-var}. The QSPC-F estimators are derived by approximating the original statistical inference problem by a linear model. When $d$ gets large, the model violation due to the approximation contributes to the bias of QSPC-F estimators. We compute the mean-square error (MSE) and using the bias-variance decomposition $\mathrm{MSE} = \mathrm{Var} + \mathrm{bias}^2$ to quantify the bias. The numerical results are displayed in \cref{fig:variance-qspcf}. Our simulation shows that the bias of $\theta$-estimator dominates the MSE and contaminates the inference accuracy after $d$ becomes larger than a threshold determined by the pre-asymptotic regime $d\theta \ll1$. Despite the bias due to the model violation, the MSE of the $\theta$-estimator still achieves some accuracy of order $\theta^2$ which suggests that the $\theta$-estimator might give a reasonable estimation of a similar order with model violation in larger $d$. The numerical results show that the $\varphi$-estimator is more robust where the MSE deviates significantly from the theoretical scaling in the pre-asymptotic regime after $d \ge 1/\theta$ is large enough to pass to the asymptotic regime. Furthermore, the MSE well matches the variance which implies that the bias in $\varphi$-estimator is always small. The difference in the robustness of the $\theta$- and $\varphi$-estimators is credited to the construction of QSPC-F in which the inferences of $\theta$ and $\varphi$ are completely decoupled due to the data post-processing using FFT. 
	
	\cref{fig:exact_crlb-qspcf,fig:variance-qspcf} suggest the following. (1) In the pre-asymptotic regime, QSPC-F estimators achieve the optimality in the sense of saturating the Cram\'{e}r-Rao lower bound and exhibit   robustness against time-dependent errors in $\varphi$ in both simulation and experimental deployments. Furthermore, the construction of QSPC-F estimators only involves direct algebraic operations rather than iterative optimization, and the reduced inference problems in Fourier space are linear statistical models whose global optimum is unique for each realization. This not only  enables the fast and reliable data post-processing but also allows us to analyze its performance analytically. (2) Passing to the asymptotic regime, given the significant bias of $\theta$-estimator and the sharp transition of the variance of $\varphi$-estimator, one has to use other estimators to saturate the optimal variance scaling and unbiasness, for example, maximum-likelihood estimators (MLE). Furthermore, we remark that the analysis based on the Cram\'{e}r-Rao lower bound is made by fixing the data generation (measuring quantum circuits) but varying data post-processing.

	\section{Analysis of realistic error}\label{sec:realistic-error}
	Although QSPC-F estimators are derived from modeling Monte Carlo sampling error, we numerically show their robustness against realistic errors in this section. This section is organized as follows. We discuss the sources of realistic errors including depolarizing error, time-dependent error, and readout error in each subsection. We study the methods for correcting some realistic errors by analyzing experimental data. Furthermore, we perform numerical experiments to justify the robustness of our proposed quantum metrology scheme.
	\subsection{Depolarizing error}\label{sec:depolarizing}
	The quantum error largely contaminates the signal. In the two-qubit system, we assume the quantum error is captured by a depolarizing quantum channel, where the density matrix is transformed to the convex combination of the correctly implemented density matrix and that of the uniform distribution on bit-strings. Therefore, assuming the infinite number of measurement samples (vanishing Monte Carlo sampling error), the measurement probability is 
	\begin{equation}
	p_{X(Y)|\alpha}(\omega; \theta, \varphi, \chi) = \alpha p_{X(Y)}(\omega; \theta, \varphi, \chi) + \frac{1-\alpha}{4}
	\end{equation}
	where $\alpha \in [0, 1]$ is referred to as the circuit fidelity. Then, the sampled reconstructed function is also shifted and scaled accordingly $\mf{h}_\alpha(\omega; \theta, \varphi, \chi) = \alpha \mf{h}(\omega; \theta, \varphi, \chi) - \frac{1-\alpha}{4}(1+\I)$. Consequentially, the Fourier coefficients are expected to be scaled by $\alpha$ simultaneously and the constant shift only contributes to the zero-indexed Fourier coefficient, namely
	\begin{equation}
	\abs{c^\expl_{0|\alpha}} = \abs{\alpha c^\expl_0 - \frac{1-\alpha}{4}(1+\I)} \approx \alpha\theta + \frac{1-\alpha}{2\sqrt{2}},\quad \abs{c^\expl_{k|\alpha}} \approx \alpha \theta, \ \forall k = 1, \cdots, d-1.
	\end{equation}
	We remark that the approximation of $\abs{c^\expl_{0|\alpha}}$ holds when the circuit fidelity is not close to one , namely, $\theta \ll 1-\alpha$. Yet when the circuit fidelity is close to one, the depolarizing error can be neglected as higher order effect. Using this feature, the circuit fidelity can be estimated from the difference between the Fourier coefficient of zero index and those of nonzero indices. Then, the estimators of the circuit fidelity and the swap angle are given by
	\begin{equation}\label{eqn:estimate-alpha}
	\begin{split}
	& \hat{\alpha} = 1 - 2\sqrt{2}\left(\abs{c^\expl_{0|\alpha}} - \frac{1}{d-1} \sum_{k=1}^{d-1} \abs{c^\expl_{k|\alpha}}\right),\\
	& \hat{\theta} = \frac{1}{\hat{\alpha}} \times \frac{1}{d-1} \sum_{k=1}^{d-1} \abs{c^\expl_{k|\alpha}}.
	\end{split}
	\end{equation}
	We numerically test the accuracy of these estimators in \cref{sec:additional-numerical}.

	\subsection{Time-dependent error}\label{coherentdriftSec}
The dominant time-dependent noise in superconducting qubits two-qubit control is in the frequency of the qubits. It can be modeled by time-dependent Z phase error in \fsim.  Observed from  experimental data, the magnitude of the time-dependent drift error increases when more gates are applied to the circuit. To emulate the realistic time-dependent noise, we model the noise by introducing a random deviation in angle parameters, which is referred to as the coherent angle uncertainty. Given a perfect \fsim\ parametrized as $U_\fsim(\theta,\varphi,\chi,*)$, the erroneous quantum gate due to the coherent angle uncertainty is another \fsim\ parametrized as $U_\fsim(\theta_\mathrm{unc},\varphi_\mathrm{unc},\chi_\mathrm{unc},*)$. Here, angle parameters subjected to the uncertainty are distributed uniformly at random around the perfect value
	\begin{equation}\label{eqn:coherent-noise-1}
	\theta_\mathrm{unc} \in [\theta - D_\theta, \theta + D_\theta],\ \varphi_\mathrm{unc} \in [\varphi - D_\varphi, \varphi + D_\varphi],\ \chi_\mathrm{unc} \in [\chi - D_\chi, \chi + D_\chi]
	\end{equation}
	where $D_\theta, D_\varphi, D_\chi$ stand for the maximal deviations of uncertain parameters. Inspired by experimental results, maxmal deviations of phase angles are increasing when more \fsim's are applied. Moreover, there is a Gaussian noise~\cite{niu2019universal} in the analog pulse realizations causing small  fluctuations on all gate parameters. To capture this feature and the rough estimate from the experimental data, we set the uncertainty model when the $j$-th \fsim\ is applied as
	\begin{equation}\label{eqn:coherent-noise-2}
	D_\theta^\scp{j} = 0.1 \times \theta,\ D_\varphi^\scp{j} = D_\chi^\scp{j} = 0.3 \times \frac{j}{d}.
	\end{equation}
	
	We would like to remark that the proposed model has already taken the phase drift in $Z$-rotation gates into account, which is effectively factored in the random phase drift in the single-qubit phase $\varphi$ and $\chi$ in the \fsim.
	
	\subsection{Numerical performance of the calibration against depolarizing error and time-dependent drift error}\label{sec:additional-numerical}
	In the numerical simulation, we add a depolarizing error channel after each individual gate. In terms of the quantum channel, it is quantified as
	\begin{equation}
	\begin{split}
	\mc{E}_{A_0}\left(\varrho\right) =& \left(1-\frac{3}{4}r\right) \varrho + \frac{r}{4} \left( \left(X_{A_0}\otimes I_{A_1}\right)\varrho\left(X_{A_0}\otimes I_{A_1}\right)\right.\\
	&\left.+ \left(Y_{A_0}\otimes I_{A_1}\right)\varrho\left(Y_{A_0}\otimes I_{A_1}\right) + \left(Z_{A_0}\otimes I_{A_1}\right)\varrho\left(Z_{A_0}\otimes I_{A_1}\right)\right),\\
	\mc{E}_{A_0,A_1}\left(\varrho\right) =& (1-r)\varrho + r \frac{I_{A_0,A_1}}{4}
	\end{split}
	\end{equation}
	where $r$ is the error rate. At the same time, the quantum circuit subjects to  drift error according to \cref{eqn:coherent-noise-1,eqn:coherent-noise-2}. 
	
\begin{figure}[htbp]
		\centering
		\includegraphics[width=.75\textwidth]{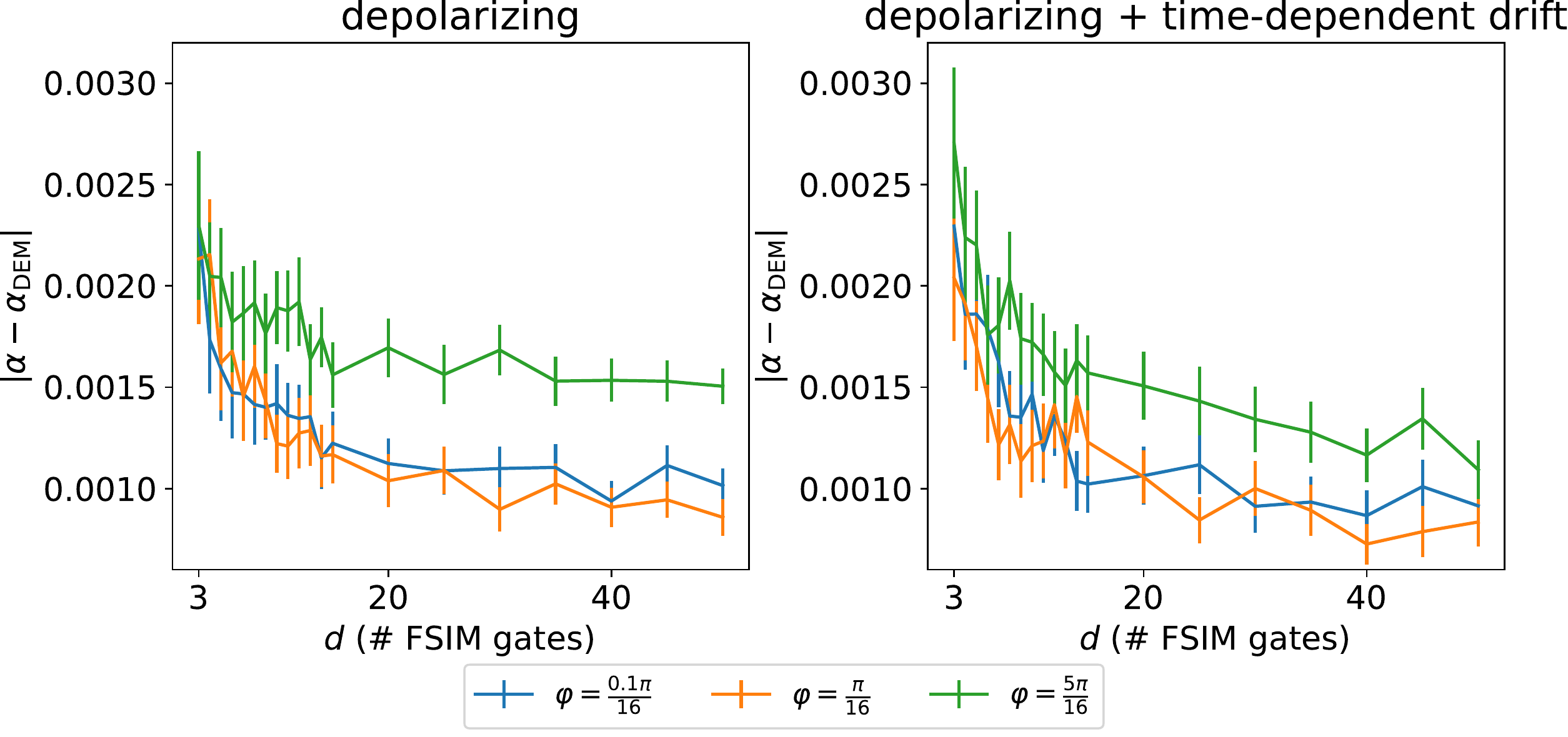}
		\caption{Estimating circuit fidelity using QSPC. The reference value $\alpha_\mathrm{DEM}$ is the circuit fidelity estimated from the digital error model. The sources of noise in the numerical experiments are Monte Carlo sampling error, depolarizing error and  drift error. The depolarizing error rate is set to $r = 1 \times 10^{-3}$ and the number of measurement samples is set to $M = 1 \times 10^{5}$. The parameters of \fsim\ are set to $\theta = 1\times 10^{-3}$ and $\chi = 5\pi/32$. The error bar of each point stands for the confidence interval derived from $96$ independent repetitions.}
		\label{fig:alpha_degree}
	\end{figure}

	In \cref{fig:alpha_degree}, we numerically test the accuracy of estimating the circuit fidelity using the Fourier space data according to the estimator in \cref{eqn:estimate-alpha}. The reference value of the circuit fidelity is computed from the digital error model (DEM) \cite{BoixoIsakovSmelyanskiyEtAl2018} with 
	\begin{equation}
	    \alpha_\text{DEM} := (1-r)^{n_\text{gates}} \approx (1-r)^{2d+5} + \Or(r).
	\end{equation}
	Here, $n_\text{gates}$ stands for the number of total gates in the quantum circuit. Because of the additional phase gate used in the Bell-state preparation, the quantum circuit for computing $p_Y$ uses $n_\text{gates} = 2d+6$ gates while that for $p_X$ uses $n_\text{gates} = 2d+5$ gates. This ambiguity in a gate makes the left-hand side approximates the circuit fidelity up to $\Or(r)$. In \cref{fig:alpha_degree}, the performance of the circuit fidelity estimation is quantified by the deviation $\abs{\hat{\alpha} - \alpha_\text{DEM}}$. As the circuit depth of QSPC increases, it turns out that the deviation decreases to $\sim 0.001$ which is equal to the error rate $r$. The decreasing deviation is due to the improvement of the SNR when increasing the circuit depth. Furthermore, the plateau near $0.001$ is due to the ambiguity discussed in the reference $\alpha_\text{DEM}$. In the left panel, we turn off the time-dependent drift error and the quantum circuit is only subject to Monte Carlo sampling error and depolarizing error. However, the performance of the circuit fidelity estimation does not differ significantly after turning on the time-dependent drift error. The numerical results suggest that the depolarizing error can be inferred with considerable accuracy even in the presence of more complex time-dependent error. 
	
\begin{figure}[htbp]
		\centering
		\includegraphics[width=\textwidth]{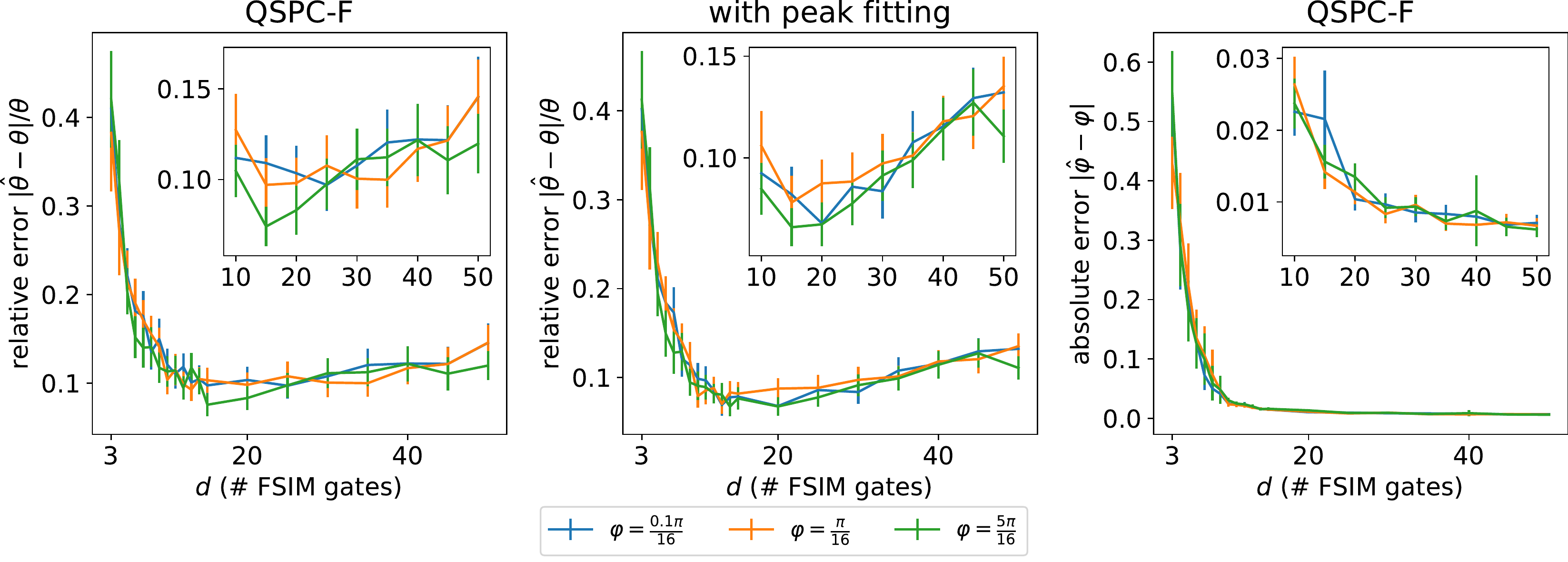}
		\caption{Accuracy of estimators as a function of the number of \fsim s. The sources of noise in the numerical experiments are Monte Carlo sampling error, depolarizing error and time-dependent drift error. The depolarizing error rate is set to $r = 1 \times 10^{-3}$ and the number of measurement samples is set to $M = 1 \times 10^{5}$. The swap angle is set to $\theta = 1 \times 10^{-3}$ and the phase parameter is set to $\chi = 5\pi/32$. The error bar of each point stands for the confidence interval derived from $96$ independent repetitions.}
		\label{fig:degree_cu}
	\end{figure}

\begin{figure}[htbp]
		\centering
		\includegraphics[width=\textwidth]{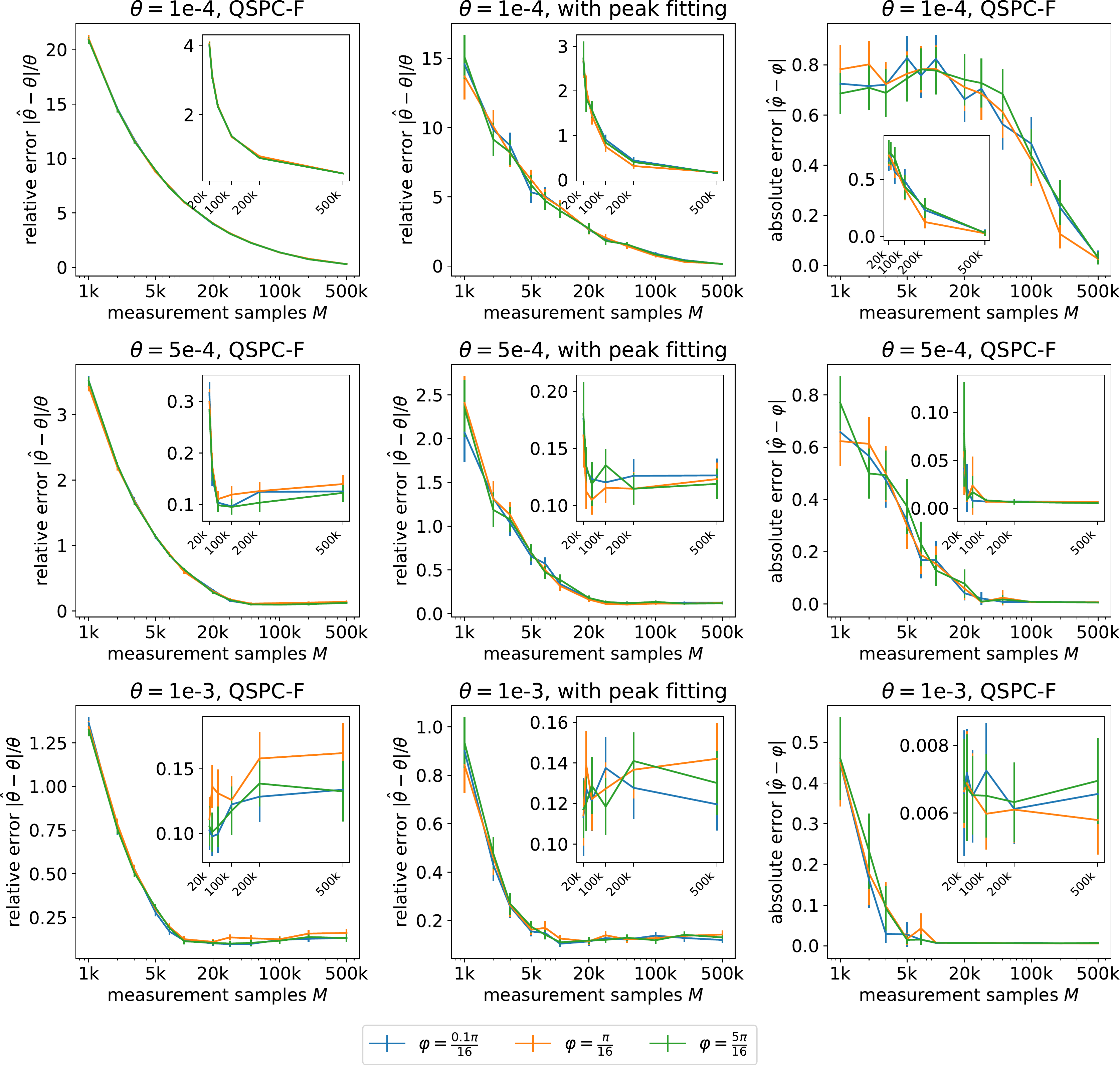}
		\caption{Accuracy of estimators as a function of the number of measurement samples. The sources of noise in the numerical experiments are Monte Carlo sampling error, depolarizing error and time-dependent drift error. The depolarizing error rate is set to $r = 1 \times 10^{-3}$. The circuit degree is set to $d = 50$ and the \fsim\ phase parameter is set to $\chi = 5\pi/32$.The error bar of each point stands for the confidence interval derived from $96$ independent repetitions.}
		\label{fig:meas_cu}
	\end{figure}

	In \cref{fig:degree_cu,fig:meas_cu}, we test our proposed metrology scheme in the presence of Monte Carlo sampling error, depolarizing error and time-dependent error. Although the system is subjected to realistic errors, the numerical results suggest that the QSPC-F estimators show some robustness against errors and they can give reasonable estimation results with one or two correct digits. Furthermore, the accuracy of $\varphi$-estimation is also not fully contaminated by the time-dependent error on it. The improvement due to the peak fitting becomes less significant under realistic errors because the structure of the highest peak is heavily distorted in the presence of realistic errors. More interestingly, the numerical results show the accuracy of $\theta$-estimation does not decay and even increases after some $d^*$. This transition is due to a tradeoff. When $d$ becomes larger, the inference is expected to be more accurate because the gate parameters are more amplified. However, in the presence of realistic error, the \fsim\ is subjected to both time-independent errors and time-dependent drift error. A quantum circuit with more \fsim s violates the model derived from the noiseless setting more. The competition between these two opposite effects makes the estimation error attains some minimum at $d^*$. This observation also suggests that in the experimental deployment, one can consider using a moderate $d$ with respect to the tradeoff.
	
	In \cref{fig:meas_cu}, we perform the numerical simulation with variable swap angle and number of measurement samples. Similar to the case of Monte Carlo sampling error, the estimation results are less accurate when $\theta$ is small because of the insufficient SNR. The numerical results indicate that the estimation accuracy cannot be further improved after the number of measurement samples is greater than some $M^*$. That is because increasing $M$ can only mitigate Monte Carlo sampling error. When $M$ is large enough, the sources of errors are dominated by depolarizing error and time-dependent drift error which cannot be sufficiently mitigated by large $M$. Combing with the discussion on $d^*$, the numerical results suggest that the experimental deployment does not require an extremely large $d$ and $M$, and using a moderate choice of $d^*$ and $M^*$ suffices to get some accurate estimation.
	
	\subsection{Readout error}\label{sec:readout}
	The readout error is modeled by a stochastic matrix whose entry is interpreted as a conditional probability. This matrix is referred to as the confusion matrix in the readout. For a two-qubit system, it takes the form
	\begin{equation}
	R := [\bPP(\mathrm{binary}(j) | \mathrm{binary}(i))]_{i, j = 0}^3 = \left( \begin{array}{*4{c}}
	\bPP(00 | 00) & \bPP(01 | 00) & \bPP(10 | 00) & \bPP(11 | 00)\\
	\bPP(00 | 01) & \bPP(01 | 01) & \bPP(10 | 01) & \bPP(11 | 01)\\
	\bPP(00 | 10) & \bPP(01 | 10) & \bPP(10 | 10) & \bPP(11 | 10)\\
	\bPP(00 | 11) & \bPP(01 | 11) & \bPP(10 | 11) & \bPP(11 | 11)
	\end{array} \right)
	\end{equation}
	where $\bPP(\mathrm{binary}(j) | \mathrm{binary}(j))$ is the conditional probability of measuring the qubits with the bit-string $\mathrm{binary}(j)$ given that the quantum state is $\ket{\mathrm{binary}(i)}$. The sum of each row of the confusion matrix is equal to one due to the normalization of probability. The confusion matrix can be determined by performing additional quantum experiments in which $I \otimes I$, $I \otimes X$, $X \otimes I$ and $X \otimes X$ are measured to determine each row respectively. If the probability vector from the measurement with readout error is $\vec{q}^\expl = (q^\expl(00), q^\expl(01), q^\expl(10), q^\expl(11))^\top$, the probability vector after correcting the readout error is given by inverting the confusion matrix
	\begin{equation}
	\vec{p}^\expl = (p^\expl(00), p^\expl(01), p^\expl(10), p^\expl(11))^\top = \left( R^\top \right)^{-1} \vec{q}^\expl.
	\end{equation}
	
	In practice, the confusion matrix is determined by finite measurement samples which could introduce error to the confusion matrix due to the statistical fluctuation. We analyze the error and its effect in \cref{thm:confmat}. As a consequence, the theorem indicates a minimal requirement on the measurement sample size so that the readout error can be accurately corrected.
	\begin{theorem}\label{thm:confmat}
		Let $\vec{p}^\expl_\mathrm{fs}$ be the probability vector computed by inverting the confusion matrix estimated by finite samples. To achieve the bounded error $\norm{\vec{p}^\expl - \vec{p}^\expl_\mathrm{fs}}_2 \le \epsilon$ with confidence level $1 - \alpha$, it suffices to set the number of measurement samples in each experiment determining the confusion matrix as
		\begin{equation}
		M_\mathrm{cmt} = \ceil{\frac{2\kappa^2(\kappa+\epsilon)^2 \ln\left(32/\alpha\right)}{\epsilon^2}}
		\end{equation}
		where
		\begin{equation}
		\kappa = \max_{i = 0, \cdots, 3} \frac{1}{2R_{ii} - 1}.
		\end{equation}
	\end{theorem}
	
	\subsection{Calibration with experimental data}\label{sec:calibrate-experiment}
\begin{figure}[htbp]
		\centering
		\includegraphics[width=\textwidth]{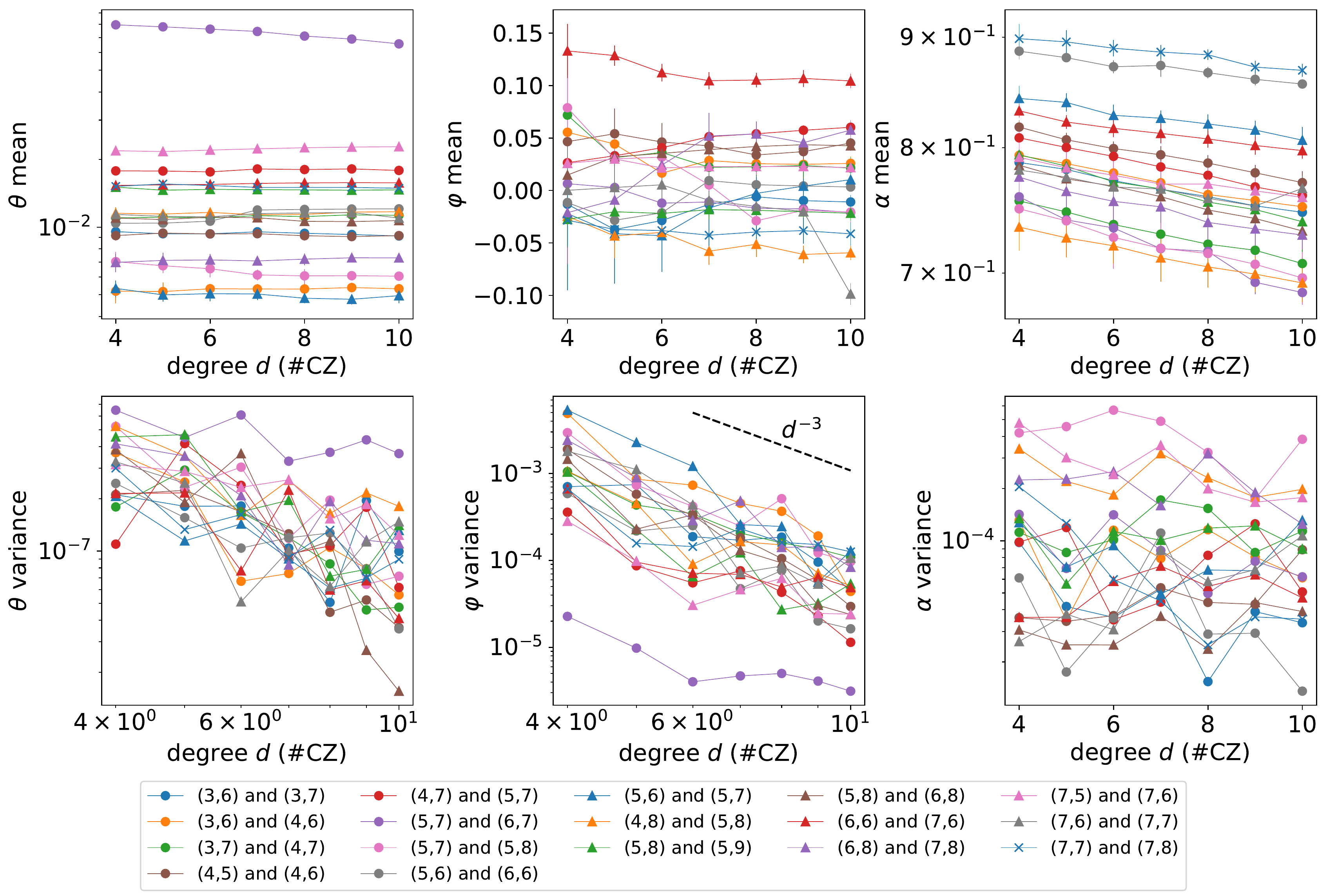}
		\caption{Calibration of CZ with extremely small unwanted swap angle. Each data point is the average of $10$ independent repetitions and the error bars in the top panels stand for the standard deviation across those repetitions. The number of measurement samples is set to $M = 1\times 10^4$.}
		\label{fig:cz_calibrate_exp}
	\end{figure}

In this subsection we review the experimental deployment of our metrology method	and compare it against the leading alternative methods in learning extremely small swap angle in \fsim. We use Google Quantum AI superconducting qubits~\cite{GoogleQuantumSupremacy2019} platform to conduct the experiments described in Algorithm \ref{alg:qspc-peak-fitting}   and Fig.~\ref{fig:qspc}. We apply our QSPC method to calibrate $\theta$ and $\varphi$ angles of seventeen pairs of CZ gates. Each CZ gate qubit pair are labeled by $(x_1, y_1)$ and $(x_2, y_2)$, indices of the both qubits on a grid architecture, e.g. $(3,6)$ and $(3,7)$ qubits. We plot the statistics of the learned gate angle parameters in \cref{fig:cz_calibrate_exp}. As shown in the figure, the unwanted swap angle for almost qubits are small, of order below $10^{-2}$. In comparison, the leading alternative methods are unable to achieve the learning accuracy comparable to such small magnitudes of the gate angle parameters. In \cref{tab:qubit-pair-error-rate}, we list the effective depolarizing error rate on the single-excitation subspace inferred from the exponential decay of circuit fidelities derived from QSPC-F methods.

\begin{figure}
    \centering
    \includegraphics{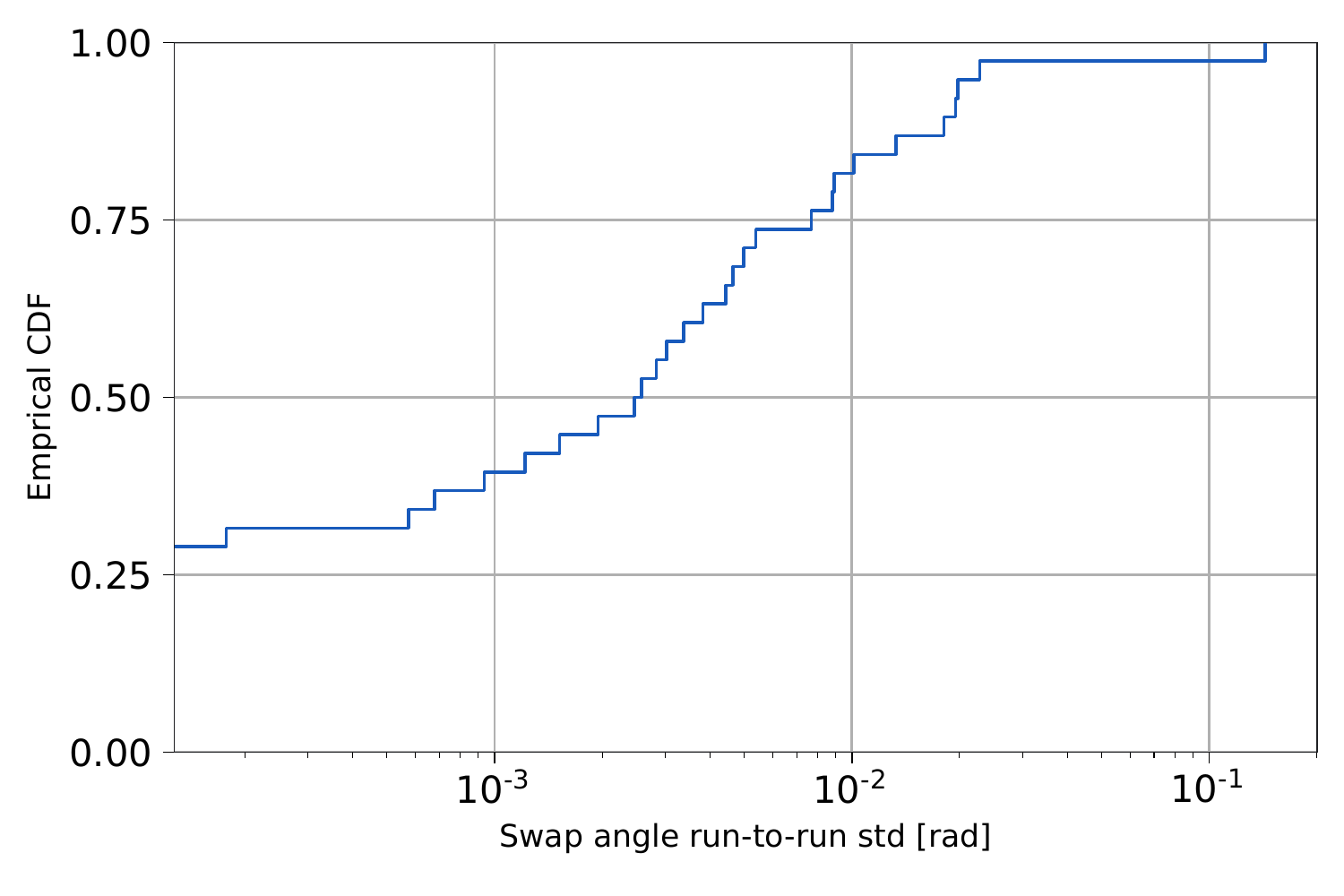}
    \caption{Distribution of run-to-run variation of swap-angle estimation across a device.
    The swap angles were estimated using Floquet Calibration~\cite{neill_accurately_2021} on four independent datasets for each CZ gate, with 10,000 samples per circuit and maximum depth 30.
    Due to the behavior of the Floquet estimator for particularly small swap angles, a substantial fraction of swap angles were estimated to be identically 0, leading to the portion of the cumulative distribution function that extends off the plot to the left.}
    \label{fig:periodic-calibration-theta-std}
\end{figure}

The performance advantage behind our QSPC-F method over prior art lies in its robustness against time-dependent noise in gate parameter $\varphi$. In traditional methods, both XEB and Floquet Calibration, the measurement observables is a nonlinear function of both $\varphi$ and $\theta$. So if there is time-dependent drift in $\varphi$ during each experiment, or over different repetitions of the same experiment routine, the value of inferred $\theta$ will be directly affected. For example, as shown in \cref{fig:periodic-calibration-theta-std}, the existing leading calibration method, Floquet calibration~\cite{neill_accurately_2021} will give a large range of different value of $\theta$ inference for the same pair of CZ gate over different runs. We know from the design of our superconducting qubit two-qubit gate~\cite{foxen2020}, such drift in $\theta$ is not physical, and is direct consequence of time-dependent drift in value $\varphi$. In comparison, QSPC-F is tolerant to realistic time-dependent error in $\varphi$ when inferring swap angle $\theta$ due to the analytic separation between the two parameter through QSP transformation combined with Fourier analysis. Notice, if the error in $\varphi$ is sufficiently large to invalidate the assumptions made in the analysis according \cref{alg:qspc-peak-fitting}, QSPC-F method will fail as well. But given the current device drift values in experiments, QSPC-F method offers a significantly improved performance in stability of $\theta$ estimation over both Floquet calibration and XEB by one magnitude in STD.

\begin{figure}[htbp]
		\centering
		\includegraphics[width=0.7\textwidth]{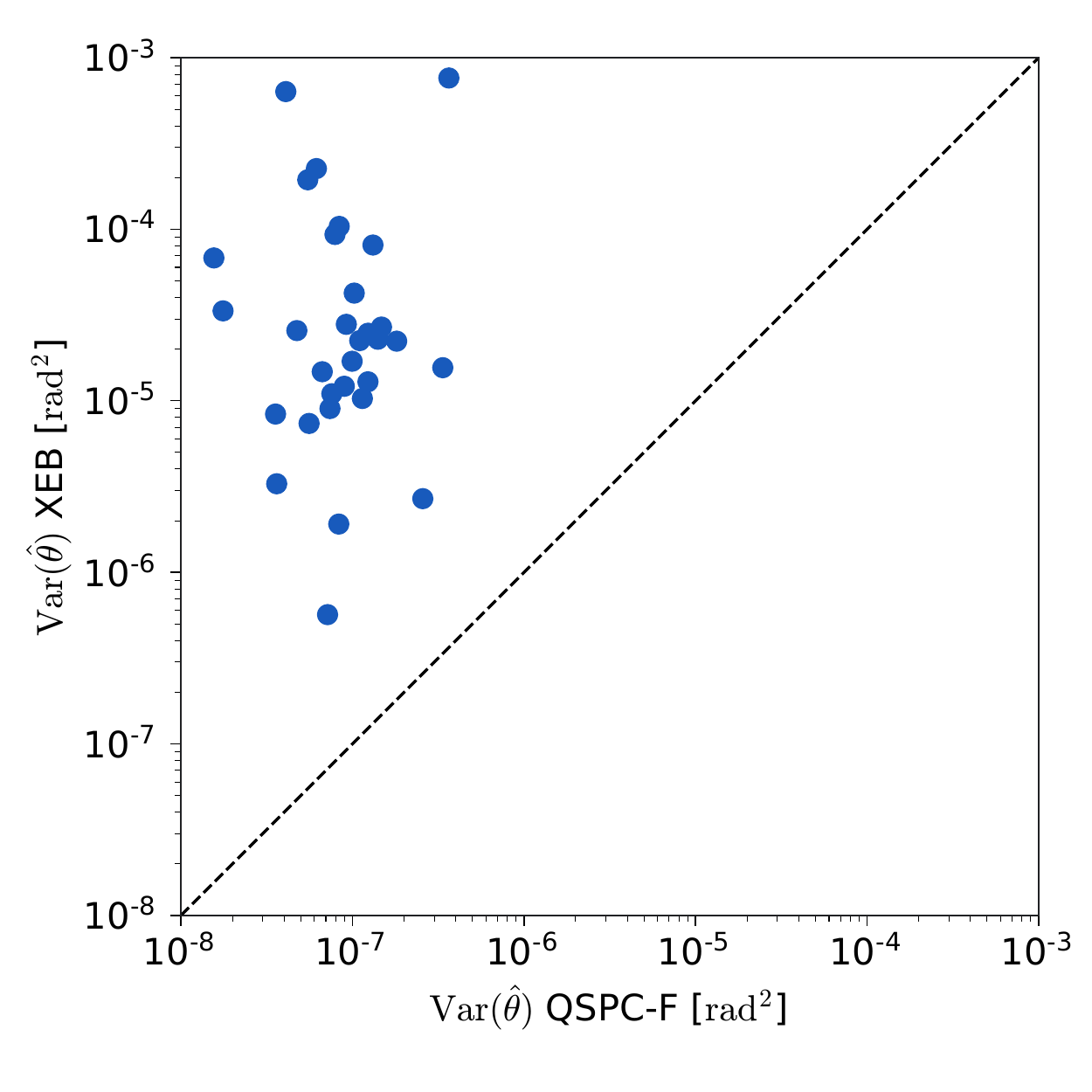}
		\caption{Comparison of the accuracy in learning swap angle $\theta$ of CZ gates over seventeen pairs of qubits between QSPC-F  and XEB.  }
		\label{fig:cz_calibrate_compare_var}
	\end{figure}

To validate the stability of QSPC-F calibration methods, we repeat the same calibration routine on each CZ gate pair over 10 independent repetitions. This allows us to bootstrap the variance of the QSPC estimator on $\theta$ and $\varphi$. We show the results on both the variance, and value of the estimated $\theta$ and $\varphi$ on seventeen pairs of CZ gate over different circuit depth $d$ used in QSPC-F in \cref{fig:periodic-calibration-theta-std}. We show that on average the learned variance on $\theta$ is around $10^{-7}$ for a depth-10 QSPC-F experiment. This corresponds to $3\times 10^{-4}$ in STD, which is one to two magnitudes lower than the value of $\theta$ itself. In comparison, we also performed the same set of experiments using XEB, see result in \cref{fig:cz_calibrate_compare_var}. The variance of $\theta$ infered by XEB is of order $10^{-4}$~(three orders of magnitudes larger than QSPC-F). Consequently, we show that XEB is insufficient to learn the value of $\theta$ in our experiments  with larger than unity signal-to-noise ratio.

\begin{table}[htbp]
\centering
\begin{tabular}{@{} *{5}{c} @{}}\midrule
(3,6) and (3,7) & (3,6) and (4,6) & (3,7) and (4,7) & (4,5) and (4,6) & (4,7) and (5,7) \\
$4.52\times 10^{-3}$ & $4.73\times 10^{-3}$ & $5.39\times 10^{-3}$ & $4.69\times 10^{-3}$ & $5.15\times 10^{-3}$ \\\midrule
(5,7) and (6,7) & (5,7) and (5,8) & (5,6) and (6,6) & (5,6) and (5,7) & (4,8) and (5,8) \\
$8.25\times 10^{-3}$ & $5.89\times 10^{-3}$ & $2.81\times 10^{-3}$ & $3.59\times 10^{-3}$ & $4.96\times 10^{-3}$ \\\midrule
(5,8) and (5,9) & (5,8) and (6,8) & (6,6) and (7,6) & (6,8) and (7,8) & (7,5) and (7,6) \\
$5.84\times 10^{-3}$ & $5.70\times 10^{-3}$ & $3.36\times 10^{-3}$ & $5.13\times 10^{-3}$ & $3.32\times 10^{-3}$ \\\midrule
(7,6) and (7,7) & (7,7) and (7,8) &  &  &  \\
$2.36\times 10^{-3}$ & $2.89\times 10^{-3}$ &  &  &  \\\midrule
\end{tabular}
\caption{Qubit pairs and the inferred effective error rate on the single-excitation subspace. The error rate is estimated by the regression with respect to the exponential decay. The regression data are the circuit fidelity estimated from QSPC-F in \cref{fig:cz_calibrate_exp} (top-right panel).}
\label{tab:qubit-pair-error-rate}
\end{table}


	\newpage
	\appendix

	\section{Computing the polynomial representation on a special set of points}
\begin{lemma}\label{lma:poly-rep-special-pts}
	Let $d = 2^j$ for some $j = 0, 1, 2, \cdots$. Then 
	\begin{equation}
	P_\omega^\scp{d}(x) = e^{\I\omega} \left(\cos\left(d \sigma\right) + \I \frac{\sin\left(d \sigma\right)}{\sin\sigma} \left(\sin \omega\right) x\right) \text{ and } Q_\omega^\scp{d}(x) = \frac{\sin\left(d \sigma\right)}{\sin\sigma}
	\end{equation}
	where $\sigma = \arccos\left(\left(\cos \omega\right) x \right)$.
\end{lemma}
\begin{proof}
	A system of recurrence relations can be established by inserting the resolution of identity in the matrix multiplication:
	\begin{equation}\label{eqn:recurrence-Q}
	\begin{split}
	\I \sqrt{1-x^2} Q^\scp{d}(x) &= \braket{0|U^\scp{d}(\omega,\theta)|1} = \braket{0|U^\scp{d/2}(\omega,\theta) e^{-\I\omega Z} \left(\ket{0}\bra{0} + \ket{1}\bra{1}\right) U^\scp{d/2}(\omega,\theta) |1}\\
	&= e^{-\I\omega} P_\omega^\scp{d/2}(x) \I \sqrt{1-x^2} Q_\omega^\scp{d/2}(x) + e^{\I\omega} \I\sqrt{1-x^2} Q_\omega^\scp{d/2}(x) P_\omega^{\scp{d/2} *}(x)\\
	&= \I \sqrt{1-x^2} Q_\omega^\scp{d/2}(x) 2 \Re\left(e^{-\I\omega} P_\omega^\scp{d/2}(x)\right)\\
	\Rightarrow Q_\omega^\scp{d}(x) &= 2 Q_\omega^\scp{d/2}(x) \Re\left(e^{-\I\omega} P_\omega^\scp{d/2}(x)\right),
	\end{split}
	\end{equation}
	and
	\begin{equation}\label{eqn:recurrence-P}
	\begin{split}
	P_\omega^\scp{d}(x) &= \braket{0|U^\scp{d}(\omega,\theta)|0} = \braket{0|U^\scp{d/2}(\omega,\theta) e^{-\I\omega Z} \left(\ket{0}\bra{0} + \ket{1}\bra{1}\right) U^\scp{d/2}(\omega,\theta) |0}\\
	&= e^{-\I\omega} \left(P_\omega^\scp{d/2}(x)\right)^2 - e^{\I\omega} (1-x^2) \left(Q_\omega^\scp{d/2}(x)\right)^2\\
	&\stackrel{(\star)}{=} -e^{\I\omega} + 2 P_\omega^\scp{d/2}(x) \Re\left(e^{-\I\omega} P_\omega^\scp{d/2}(x) \right)\\
	\Rightarrow &\Re\left(e^{-\I\omega} P_\omega^\scp{d}(x) \right) = -1 + 2 \Re^2\left(e^{-\I\omega} P_\omega^\scp{d/2}(x) \right),\\
	\text{and } & \Im\left(e^{-\I\omega} P_\omega^\scp{d}(x) \right) = 2 \Im\left(e^{-\I\omega} P_\omega^\scp{d/2}(x) \right) \Re\left(e^{-\I\omega} P_\omega^\scp{d/2}(x) \right).
	\end{split}
	\end{equation}
	Here, equation $(\star)$ uses the special unitarity of $U^\scp{d/2}(\omega,\theta)$ which yields $P_\omega^\scp{d/2}(x) P_\omega^{\scp{d/2} *}(x) + (1-x^2) \left(Q_\omega^\scp{d/2}(x)\right)^2 = 1$ by taking determinant. We will first solve the nonlinear recurrence relation for $\Re\left(e^{-\I\omega} P_\omega^\scp{d}\right)$ in \cref{eqn:recurrence-P}. Note that the second-order Chebyshev polynomial of the first kind is $T_2(x) = 2x^2 - 1$. Then,
	\begin{equation}
	\Re\left(e^{-\I\omega} P_\omega^\scp{d}(x) \right) = T_2\left(\Re\left(e^{-\I\omega} P_\omega^\scp{d/2}(x) \right)\right) = \cdots = \underbrace{T_2\circ \cdots \circ T_2}_{\log_2(d)}\left(\Re\left(e^{-\I\omega} P_\omega^\scp{1}(x) \right)\right)
	\end{equation}
	Using the composition identity of the Chebyshev polynomials $T_n \circ T_m = T_{nm}$, we have $\underbrace{T_2\circ \cdots \circ T_2}_{\log_2(d)} = T_{d}$. On the other hand, when $d=1$, we have
	\begin{equation}
	\begin{split}
	U^\scp{1}(\omega, \arccos(x)) &= e^{\I\omega Z} e^{\I \arccos(x) X} e^{\I\omega Z} = \left(
	\begin{array}{cc}
	e^{2\I\omega} x & \I\sqrt{1-x^2} \\
	\I \sqrt{1-x^2} & e^{-2\I\omega} x
	\end{array}
	\right)\\
	\Rightarrow e^{-\I\omega} P_\omega^\scp{1}(x) &= e^{\I\omega} x,\ Q_\omega^\scp{1}(x) = 1.
	\end{split}
	\end{equation}
	Therefore
	\begin{equation}
	\Re\left(e^{-\I\omega} P_\omega^\scp{d}(x) \right) = T_{d}\left(\left(\cos\omega\right) x\right).
	\end{equation}
	Furthermore, $Q_\omega^\scp{d}$ and $\Im\left(e^{-\I\omega} P_\omega^\scp{d} \right)$ can be determined from the recurrence relation in \cref{eqn:recurrence-Q,eqn:recurrence-P}
	\begin{equation}
	Q_\omega^\scp{d}(x) = d \prod_{j=0}^{\log_2(d)-1} T_{2^j}\left(\left(\cos\omega\right) x\right),\ \Im\left(e^{-\I\omega} P_\omega^\scp{d}(x) \right) = Q_\omega^\scp{d}(x) \left(\sin\omega\right) x.
	\end{equation}
	For convenience, let $\cos \sigma := \left(\cos \omega\right) x = \cos\omega \cos\theta$. Then
	\begin{equation}
	\begin{split}
	Q_\omega^\scp{d}(x) \sin\sigma &= \left(\frac{d}{2} \prod_{j=1}^{\log_2(d)-1}\right) 2 \cos \sigma \sin\sigma = \left(\frac{d}{4} \prod_{j=2}^{\log_2(d)-1}\right) 2 \cos(2\sigma) \sin(2\sigma)\\
	&= \cdots = 2 \cos\left(\frac{d}{2} \sigma\right) \sin\left(\frac{d}{2} \sigma\right) = \sin\left(d \sigma\right).
	\end{split}
	\end{equation}
	Therefore
	\begin{equation}
	P_\omega^\scp{d}(x) = e^{\I\omega} \left(\cos\left(d \sigma\right) + \I \frac{\sin\left(d \sigma\right)}{\sin\sigma} \left(\sin \omega\right) x\right), \text{ and } Q_\omega^\scp{d}(x) = \frac{\sin\left(d \sigma\right)}{\sin\sigma}.
	\end{equation}
\end{proof}

	\section{Estimating measurement sizes to accurately determine the confusion matrix}\label{sec:analysis}
	\begin{proof}[Proof of \cref{thm:confmat}]
		In each experiment given the exact outcome $u \in \{0,1\}^2$ without readout error and exact measurement probability vector $\vp^{(u)} := \left(p(00|u), p(01|u), p(10|u), p(11|u)\right)$ taking readout error into account, the number of measurement outcomes corresponding to each bit-string is multinomial distributed
		\begin{equation}
		\vk^{(u)} := \left(k(00|u), k(01|u), k(10|u), k(11|u)\right) \sim \mathrm{Multinomial}(M_\mathrm{cmt}, \vp^{(u)})
		\end{equation}
		where $k(s|u) := \#(\text{outcome is } s \text{ in }M_\mathrm{cmt} \text{ samples})$. The bit-string frequency
		\begin{equation}
		\vq^{(u)} = \left(q(00|u), q(01|u), q(10|u), q(11|u)\right) := \left(\frac{k(00|u)}{M_\mathrm{cmt}}, \frac{k(01|u)}{M_\mathrm{cmt}}, \frac{k(10|u)}{M_\mathrm{cmt}}, \frac{k(11|u)}{M_\mathrm{cmt}}\right)
		\end{equation}
		is therefore an estimate to the measurement probability since $\expt{\vq^{(u)}} = \vp^{(u)}$. However, the statistical fluctuation makes the estimate deviates the exact probability. Applying Hoeffding's inequality, we have
		\begin{equation}
		\bP\left(\abs{q(s|u) - p(s|u)} > \frac{\wt{\epsilon}}{4} \right) = \bP\left(\abs{k(s|u) - M_\mathrm{cmt} p(s|u)} > \frac{\wt{\epsilon} M_\mathrm{cmt}}{4}\right) \le 2 e^{- \frac{\wt{\epsilon}^2 M_\mathrm{cmt}}{8}}.
		\end{equation}
		Let the confusion matrix determined by finite samples be $R_\mathrm{fs}$ where $\left(R_\mathrm{fs}\right)_{ij} = q\left(\mathrm{binary}(j) | \mathrm{binary}(i)\right)$ and the subscript ``fs'' abbreviates ``finite sample''. Then, the deviation can be bounded as
		\begin{equation}
		\begin{split}
		&\bP\left(\norm{R_\mathrm{fs} - R}_2 > \wt{\epsilon}\right) \le \bP\left(\norm{R_\mathrm{fs} - R}_F > \wt{\epsilon}\right) = \bP\left(\sum_{s, u \in \{0,1\}^2} \abs{q(s|u) - p(s|u)}^2 > \wt{\epsilon}^2\right)\\
		&\le \bP\left(\bigcup_{s, u \in \{0,1\}^2} \left\{ \abs{q(s|u) - p(s|u)} > \frac{\wt{\epsilon}}{4} \right\}\right) \le \sum_{s, u \in \{0,1\}^2} \bP\left(\abs{q(s|u) - p(s|u)} > \frac{\wt{\epsilon}}{4} \right)\\
		&\le 32 e^{- \frac{\wt{\epsilon}^2 M_\mathrm{cmt}}{8}}.
		\end{split}
		\end{equation}
		Therefore, to achieve $\norm{R_\mathrm{fs} - R}_2 \le \wt{\epsilon}$ with confidence level $1 - \alpha$, it suffices to set the number of measurement samples in each experiment as
		\begin{equation}
		M_\mathrm{cmt} = \ceil{\frac{8 \ln\left(32/\alpha\right)}{\wt{\epsilon}^2}}.
		\end{equation}
		Expanding the matrix inverse in terms of power series and denoting $\Delta_\mathrm{fs} := R_\mathrm{fs} - R$ for convenience, we have
		\begin{equation}
		R_\mathrm{fs}^{-1} = \left(R + \Delta_\mathrm{fs}\right)^{-1} = R^{-1} \left(I + \Delta_\mathrm{fs} R^{-1}\right)^{-1} = R^{-1} + \sum_{j=1}^\infty R^{-1} \left(\Delta_\mathrm{fs} R^{-1}\right)^j.
		\end{equation}
		Furthermore, we get
		\begin{equation}
		\norm{R_\mathrm{fs}^{-1} - R^{-1}}_2 \le \norm{R^{-1}}_2 \sum_{j=1}^\infty \norm{\Delta_\mathrm{fs} R^{-1}}_2^j \le \frac{\norm{\Delta_\mathrm{fs}}_2 \norm{R^{-1}}_2^2}{1 - \norm{\Delta_\mathrm{fs}}_2 \norm{R^{-1}}_2}.
		\end{equation}
		Note that $\norm{R^{-1}}_2 = \lambda_\mathrm{min}^{-1}(R)$. To proceed, we have to lower bound the smallest eigenvalue of the confusion matrix. Note that all eigenvalues of the confusion matrix are real as a property of stochastic matrix. Applying Gershgorin circle theorem, all eigenvalues of the confusion matrix are contained in the union of intervals
		\begin{equation}
		\bigcup_{i = 0}^3 \left[R_{ii} - \sum_{j \ne i} R_{ij}, R_{ii} + \sum_{j \ne i} R_{ij}\right].
		\end{equation}
		Consequentially, the smallest eigenvalue of the confusion matrix is lower bounded
		\begin{equation}
		\lambda_\mathrm{min}(R) \ge \min_{i=0, \cdots, 3} \left(R_{ii} - \sum_{j \ne i} R_{ij}\right) = \min_{i = 0, \cdots, 3} \left(2 R_{ii} - 1\right) =: \kappa^{-1}.
		\end{equation}
		Thus, by properly choosing the number of measurement samples, with confidence level $1-\alpha$, we can bound the inverse confusion matrix as
		\begin{equation}
		\norm{R_\mathrm{fs}^{-1} - R^{-1}}_2 \le \frac{\wt{\epsilon} \kappa^2}{1 - \wt{\epsilon} \kappa}.
		\end{equation}
		When computing the probability vector by inverting the confusion matrix dermined by finite measurement samples, the error is bounded as
		\begin{equation}
		\norm{\vec{p}^\expl - \vec{p}^\expl_\mathrm{fs}}_2 \le \norm{R_\mathrm{fs}^{-1} - R^{-1}}_2 \norm{\vec{q}^\expl}_2 \le \norm{R_\mathrm{fs}^{-1} - R^{-1}}_2 \norm{\vec{q}^\expl}_1 \le \frac{\wt{\epsilon} \kappa^2}{1 - \wt{\epsilon} \kappa}.
		\end{equation}
		Let 
		\begin{equation}
		\frac{\wt{\epsilon} \kappa^2}{1 - \wt{\epsilon} \kappa} = \epsilon \Rightarrow \wt{\epsilon} = \frac{\epsilon}{\kappa(\kappa+\epsilon)}
		\end{equation}
		Thus, to achieve the bounded error $\norm{\vec{p}^\expl - \vec{p}^\expl_\mathrm{fs}}_2 \le \epsilon$ with confidence level $1 - \alpha$, it suffices to set the number of measurement samples in each experiment determining the confusion matrix as
		\begin{equation}
		M_\mathrm{cmt} = \ceil{\frac{8\kappa^2(\kappa+\epsilon)^2 \ln\left(32/\alpha\right)}{\epsilon^2}}.
		\end{equation}
		The proof is completed.
	\end{proof}
	
\section{Upper bounding the derivative of polynomials}\label{app:poly-ineq}
	In the analysis in the paper, we sometimes upper bound the error by the derivative of some polynomials. The following theorem is useful to get a further upper bound.
	\begin{theorem}[Markov brothers' inequality \cite{Markov1890}]\label{thm:Markovs-ineq}
	    Let $P \in \RR_d[x]$ be any algebraic polynomial of degree at most $d$. For any nonnegative integer $k$, it holds that
	    \begin{equation}
	        \max_{x \in [-1,1]} \abs{P^{(k)}(x)} \le \max_{x \in [-1,1]} \abs{P(x)} \prod_{j=0}^{k-1} \frac{d^2-j^2}{2j+1}.
	    \end{equation}
	    The equality is attained for Chebyshev polynomial of the first kind $T_d(x)$.
	\end{theorem}
	
\end{document}